\documentclass[]{aa}
\usepackage{amsmath,graphicx,amssymb}
\usepackage{txfonts}
\usepackage{natbib}
\usepackage{hyperref}
\usepackage{subfigure}
\usepackage[dvipsnames]{xcolor}

\newcommand{\Teff}{$T_\mathrm{eff}$}
\newcommand{\logg}{$\log g$}

\begin{document}

\title{Chemically peculiar stars on the pre-main sequence}

\author{L. Kue{\ss}\inst{1}, E. Paunzen\inst{2}, N. Faltov{\' a}\inst{2}, D. Jadlovsk{\' y}\inst{2}, M. Labaj\inst{2},
M. Mesar{\v c}\inst{2}, P. Mondal\inst{2}, M.~Pri{\v s}egen\inst{3},
T. Ramezani\inst{2}, J. Sup{\' i}kov{\' a}\inst{2,4},
K. Sva{\v c}inkov{\' a}\inst{2}, M. V{\' i}tkov{\' a}\inst{2}, C. Xia\inst{2}, K.~Bernhard\inst{5,6}, 
S.~H{\"u}mmerich\inst{5,6}}

\institute{Department of Astrophysics, Vienna University, T{\"u}rkenschanzstraße 17, 1180 Vienna, Austria
\and Department of Theoretical Physics and Astrophysics, Masaryk University,
Kotl\'a\v{r}sk\'a 2, 611\,37 Brno, Czechia \\
\email{epaunzen@physics.muni.cz}
\and Advanced Technologies Research Institute, Faculty of Materials Science and Technology in Trnava, Slovak University of Technology in Bratislava, Bottova 25, 917 24 Trnava, Slovakia
\and Institute of Computer Science, Masaryk University, Brno, Czechia
\and Bundesdeutsche Arbeitsgemeinschaft f{\"u}r Ver{\"a}nderliche Sterne e.V. (BAV), Berlin,
              Germany 
          \and
              American Association of Variable Star Observers (AAVSO), Cambridge, USA
}

\date{}

\abstract
{The chemically peculiar (CP) stars of the upper main sequence are defined by 
spectral peculiarities that indicate unusual elemental abundance patterns in the presence
of diffusion in the calm, stellar atmospheres. Some of them have a stable local magnetic field of up to several
kiloGauss. The pre-main-sequence evolution of these objects is still a mystery and contains many open questions.}
{We identify CP stars on the pre-main sequence to determine possible mechanisms that lead to the occurrence of chemical peculiarities in the (very) early stages of stellar evolution.}
{We identified likely pre-main-sequence stars by fitting the spectral energy distributions. The subsequent analysis using stellar spectra and photometric time series helped us to distinguish between CP and non-CP stars. Additionally, we compared our results to the literature to provide the best possible quality assessment.}
{Out of 45 candidates, about 70 \% seem to be true CP stars or CP candidates. Furthermore, 9 sources appear to be CP stars on the pre-main sequence, and all are magnetic. We finally report a possible CP2 star that is also a pre-main-sequence star and was not previously in the literature.}
{The evolution of the peculiarities seems to be related to the (strong) magnetic fields in these CP2 stars.}


\titlerunning{CP stars on the PMS}
\authorrunning{L. Kue{\ss} et al.}

\maketitle

\section{Introduction} \label{introduction}

An interesting phenomenon of the stars on the upper main sequence are the so-called chemically peculiar (CP) stars. First discovered by \cite{1897AnHar..28....1M}, they still puzzle ustoday.  This group of stars not only exhibits over- or underabundances of certain elements, but it was discovered that they could show spectroscopic and photometric variability due to the presence of a magnetic field \citep{1947ApJ...105..283D, 1950MNRAS.110..395S}.
\cite{1974ARA&A..12..257P} assembled a classification scheme that is still valid today. He introduced four peculiarity subgroups.

The first subgroup, the classical Am (metallic lined, CP1) stars, contains A- to F-type stars whose spectra are different from the typical spectra of similar spectral types in that iron and similar elements are overabundant, while other elements such as calcium and scandium are underabundant. Many of the classical Am stars are found in binary systems, which may contribute to the occurrence of peculiarities that are due to tidal braking. Tidal breaking allows the helium convection zone to settle, which in turn gives way to atomic diffusion to the surface layers \citep{2005A&A...443..627T,2009AJ....138...28A}.

The second (CP2 or Ap) group contains late B- to early F- stars with overabundances of Sr, Si, Cr and Eu. Additionally, many objects exhibit strong magnetic fields up to some dozen kiloGauss \citep{2021A&A...652A..31B}. Those strong cause chemical spots around the magnetic poles, and the spots lead to the photometric (the so-called ACV variables) and spectroscopic variability (with periods comparable to the rotation period of the star) that has been described for a plethora of stars (see e.g. \cite{2015A&A...581A.138B,2015AN....336..981B,2018A&A...619A..98H}). The origin of these magnetic fields is still not entirely clear, but it has been suggested that the fields can be frozen-in during the formation of the star \citep{2003A&A...403..693M} and that a field like this can be sustained after the transition from a convective to a radiative core \citep{2023arXiv230617131S}.

The mercury-manganese stars that comprise the third (CP3) group show overabundances of mercury and/or manganese, hence the name. They exhibit no detected magnetic fields, and other than overabundances in iron-peak elements, their atmospheres seem stable. They can be seen as the hot analogues of the CP1 group \citep{2018MNRAS.480.2953G}. However, while there seem to be similarities between the two types, a satisfying model that describes the existence of these peculiarities in the CP3 group remains to be found \citep{2003A&A...397..267A}. 

The fourth group (CP4 or He-weak/He-strong) of stars are B-type stars with an overabundance or lack of helium compared to (apparently) normal stars in the same effective temperature domain. Similarly to the CP2 group, these stars show large-scale magnetic fields, and thus also spectral and photometric variability \citep{1977A&AS...30...11P}.

Additionally, one more group is defined in the literature, the so-called $\lambda$ B{\"o}otis stars (e.g. \cite{1996Ap&SS.237...77S,2014A&A...567A..67P}). The stars in this category show broad hydrogen lines and weak to no lines of heavier elements such as Mg when compared to normal stars at similar temperatures. Many theories to explain these phenomena such as interplay between mass loss and diffusion \citep{1986ApJ...311..326M} or interplay between diffusion and accretion \citep{1990ApJ...363..234V} have been proposed.

After the initial collapse of a molecular coud, a protostar is formed that accretes matter onto its core. When the protostar becomes visible, it arrives at its birthline (e.g. \cite{1983ApJ...274..822S, 1990ApJ...360L..47P, 2005PThPS.158....1L}). After this, a phase of contraction follows until the core of the star reaches temperatures of $\sim 10^7$K, where the hydrogen burning begins. This stage in the evolution is called the pre-main-sequence (PMS) phase. Here, we commonly differ between low-mass T Tauri stars and the medium-mass Herbig Ae/Be stars. When talking about possible PMS-CP stars, we need to focus on the second category. 

Herbig Ae/Be stars are intermediate-mass ($\sim$2 - $10 M_\odot$) PMS stars of late-O to early-F spectral type that show typical signs of pre-main-sequence evolution: emission in Balmer lines due to accretion and infrared-excess due to the disk of dust and debris around the young star (see \cite{2023SSRv..219....7B} for an extensive recent review of this matter). These stars are thought to be linked to the CP phenomenon since similar characteristics have already been found in these stars \citep{2012MNRAS.422.2072F}.

The mechanisms that cause the CP phenomenon, such as diffusion, rotation, and
mass loss strongly depend on time.
Therefore, one of the main questions is at which point during the stellar evolution the peculiarities arise. It is therefore natural to search for the youngest objects that already show signs of the phenomena described above.

Only a few pre-main sequence CP (PMS-CP) stars or candidates have been mentioned in the literature so far for example \cite{2014MNRAS.442.3761N} presented an intermediate-mass PMS star that already showed Am peculiarities in the open cluster Stock 16. A similar study was performed by \cite{2018NewA...58....1C} for the open cluster Hogg 16, resulting in three PMS-CP candidates. Additionally, \cite{2023MNRAS.520.1296P} detected a young He-weak star in the star-forming region NGC 1333. Lastly, \cite{2023Univ....9..210P} reported the discovery of a young CP2 star with a magnetic field with a strength of 3.5\,kG.

The recent third data release (DR3) of the measurements from $Gaia$ \citep{2021A&A...649A...1G} with its precise astrometry is an excellent source for detecting open clusters. \cite{2023A&A...673A.114H} published an extensive catalogue of more than 7000 aggregates based on this dataset.

In this paper, we use the memberships of this catalogue to present data for 45  
PMS-CP candidates in star clusters.

\section{Target selection} \label{target_selection}

We selected CP1 to CP4 stars from eight references and excluded no
objects. The stars were first identified in the latest $Gaia$ release. We found several duplicates, which
were then eliminated from the sample. This left us with about 29\,300 objects from the following
papers (ordered chronologically).

{\it \citet{2009A&A...498..961R}}: The authors started to collect and publish CP stars and candidates 
in the late 1980s. The last edition used for our purpose includes 8205 known or suspected CP stars.
This catalogue remains the basis for almost all studies in this research field. However, the authors included 
all stellar objects listed at least once as peculiar. It is a rather uncritical compilation that must be 
treated cautiously. 

{\it \citet{2013PhDT.........1N}}: The thesis includes magnetic CP stars in open clusters with membership
probabilities deduced before the $Gaia$ data became available. The high quality of this analysis is
proven by the consistency of the conclusions when comparing the newest astrometric data.

{\it \citet{2019ApJS..242...13Q}}: Using spectra from the LAMOST DR5 
\citep{2012RAA....12.1197C} with a signal-to-noise ratio higher than 50,
9372 CP1 stars and 1131 CP2 candidates were compiled into a catalogue. The authors used six machine-learning
algorithms incorporating known CP1 spectra. They concluded that the random forest (RF) algorithm  selected CP stars best. Furthermore, they also manually identified them
based on the spectral features derived from the RF algorithm.

{\it \citet{2020A&A...640A..40H}}: They presented 1002 magnetic CP stars selected by searching LAMOST 
DR4 spectra for the characteristic 5200\,\AA\ flux depression. 
The spectral classification was made with a modified version of the MKCLASS code
\citep{2016AJ....151...13G}. As the final step,
the accuracy of the automatic classifications was estimated by comparison with results from manual 
classification and the literature. This guaranteed the best possible spectral type. 

{\it \citet{2020MNRAS.496..832C}}: They reported 260 newly identified CP3 stars based 
on $H$-band spectra obtained via the Sloan Digital Sky Survey (SDSS) Apache Point Observatory Galactic Evolution Experiment (APOGEE) survey \citep{2017AJ....154...94M}.
The CP3 stars were identified among the telluric standard stars as those
whose metallic absorption content is limited to or dominated by the $H$-band \ion{Mn}{II} lines. 

{\it \citet{2021A&A...645A..34P}}: As done in \citet{2020A&A...640A..40H}, they searched 
for among pre-selected early-type spectra from LAMOST DR4 using a modified version of the 
MKCLASS code which probes several \ion{Hg}{II} and \ion{Mn}{II} features. The spectra of the
resulting 332 candidates were visually inspected. 

{\it \citet{2022ApJS..259...63S}}: Similar to \citet{2019ApJS..242...13Q}, they applied three 
machine-learning algorithms to search for CP1 and CP2 stars within LAMOST DR8 spectra. However, they
selected the XGBoost algorithm \citep{2019AcASn..60...16L} as the most efficient. Their
catalogue comprises 6917 and 1652 newly discovered CP1 and CP2 stars, respectively.

{\it \citet{2023ApJ...943..147S}}: They applied the same techniques as \citet{2020A&A...640A..40H} 
for the LAMOST DR9 but without any critical assessment of the spectral types.

\begin{figure}
    \centering
    \includegraphics[width = \columnwidth]{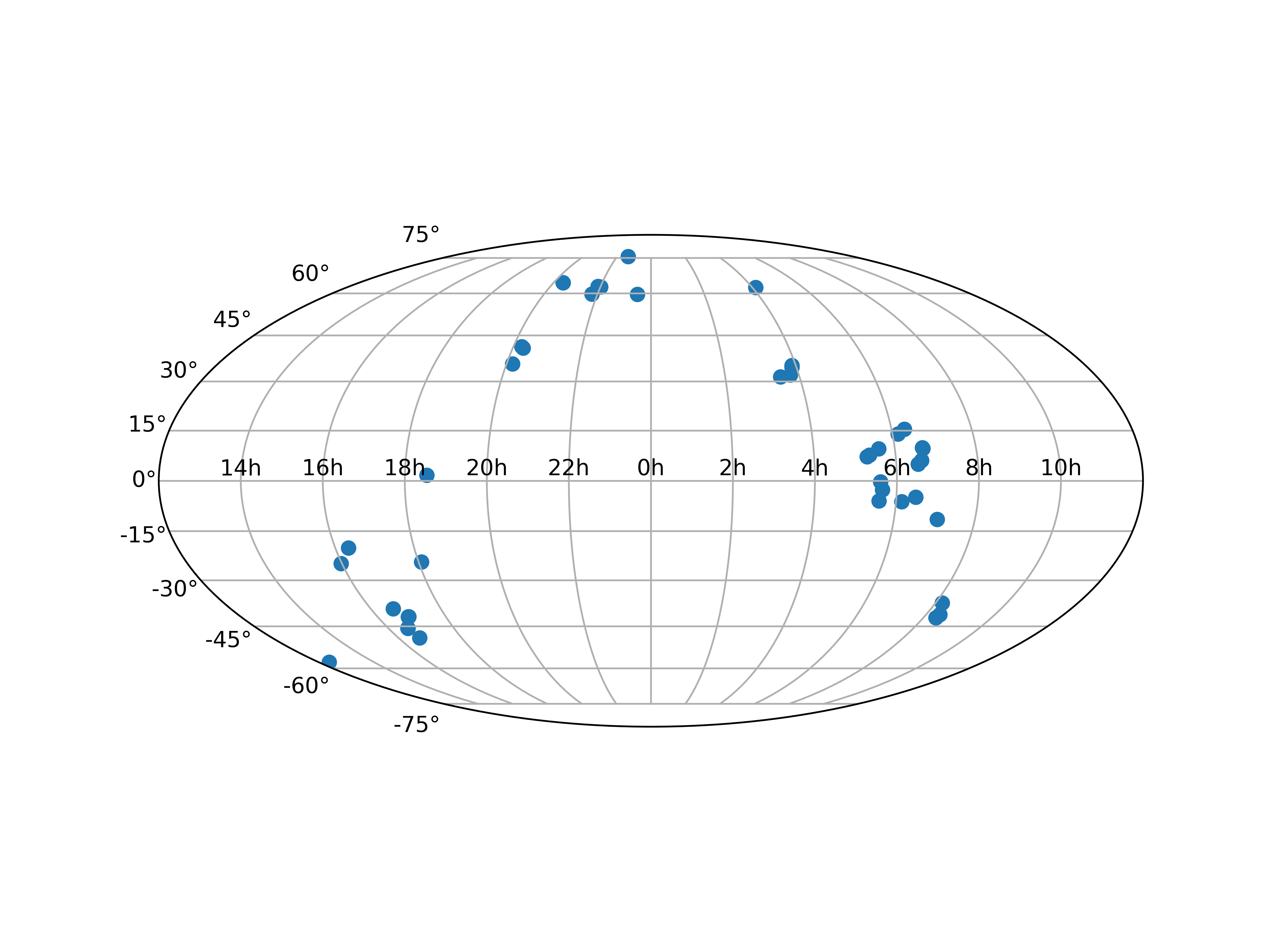}
    \caption{Sky distribution of the final 45 targets we used for the further analysis in this work.}
    \label{fig:sky_distribution}
\end{figure}

\section{Matching with members of open clusters} \label{matching_CPs}

The matching was made via $Gaia$ IDs, which had to be deduced for our CP star sample. 
The published coordinates and identifications were cross-checked within a certain matching radius.
Components of binary systems were all verified manually.

We used the clusters member list and membership probabilities 
by \citet{2023A&A...673A.114H} to match our list of bona-fide CP stars.
They applied the algorithm called Hierarchical Density-Based Spatial Clustering 
of Applications with Noise (HDBSCAN, \cite{McInnes2017}) to recover star clusters. They validated the 
aggregates they found using a statistical density test and a Bayesian convolutional neural network 
for classification in a colour-magnitude diagram. In addition, 
this catalogue contains the parameters (age, reddening, and distance) of 7167 star clusters, 
with more than 700 newly discovered high-confidence star clusters. 
We must emphasize that determining the cluster parameters is 
still challenging, although a reasonable distance estimate can be obtained from the $Gaia$ datasets \citep{2015A&A...582A..19N,2021MNRAS.504..356D}.
To check their results, we compared them with the catalogues by \citet{2020A&A...640A...1C} 
and \citet{2021MNRAS.504..356D}. The results generally agree. Some outliers have been
described in \citet{2023A&A...673A.114H}.

We cross-matched our sample with the catalogue of star clusters and identified
682 high-confidence ($P>0.7$) members within 460 aggregates. 
As the next step, we checked for all members whether they might be PMS objects or close to the zero-age 
main sequence (ZAMS). To do this, we used appropriate isochrones \citep{2012MNRAS.427..127B} and the
location of the stars in the corresponding colour-magnitude diagrams.  Initially, only star clusters with 
ages younger than 50\,Myr were chosen.
Finally, we identified 45 high-confidence ($P>0.7$) members within 39 young star clusters.

\section{$\Delta$a photometry} \label{Da_photometry}

The $\Delta$a photometry tool is potent in investigating CP stars \citep{2005A&A...441..631P}.
It investigates the flux depression at 5200\,\AA\, a spectral feature
that only occurs in CP stars. The photometric system samples
the depth of this feature by comparing the flux at the centre with the adjacent regions using bandwidths of 
110 to 230\,\AA. 
The flux depression is caused by line blanketing of Cr, Fe, and Si in this region, which is enhanced by a magnetic
field \citep{2003MNRAS.341..849K,2007A&A...469.1083K}. This is most significantly visible in magnetic CP stars.
However, not all these objects show this flux depression, 
probably for observational reasons (an unfavourable inclination) and magnetic 
field characteristics \citep{2005A&A...441..631P}. In addition, some (non- or only weakly magnetic)
CP1 and CP3 also show detectable positive $\Delta$a values but with a far lower significance.

 Galactic field stars and open cluster fields alone were surveyed so far \citep{2007A&A...462..591N,2014A&A...564A..42P}.
Another approach to the 
$\Delta$a photometry was presented by \citet{2022A&A...667L..10P}, who used $Gaia$ BP/RP spectra for the
synthesis. They found a detection level of more
than 85\% for almost the entire investigated spectral range of the upper main sequence. 
In total, 597 of the cluster members have an available BP/RP spectrum.

\begin{figure}
    \centering
    \includegraphics[width = \columnwidth]{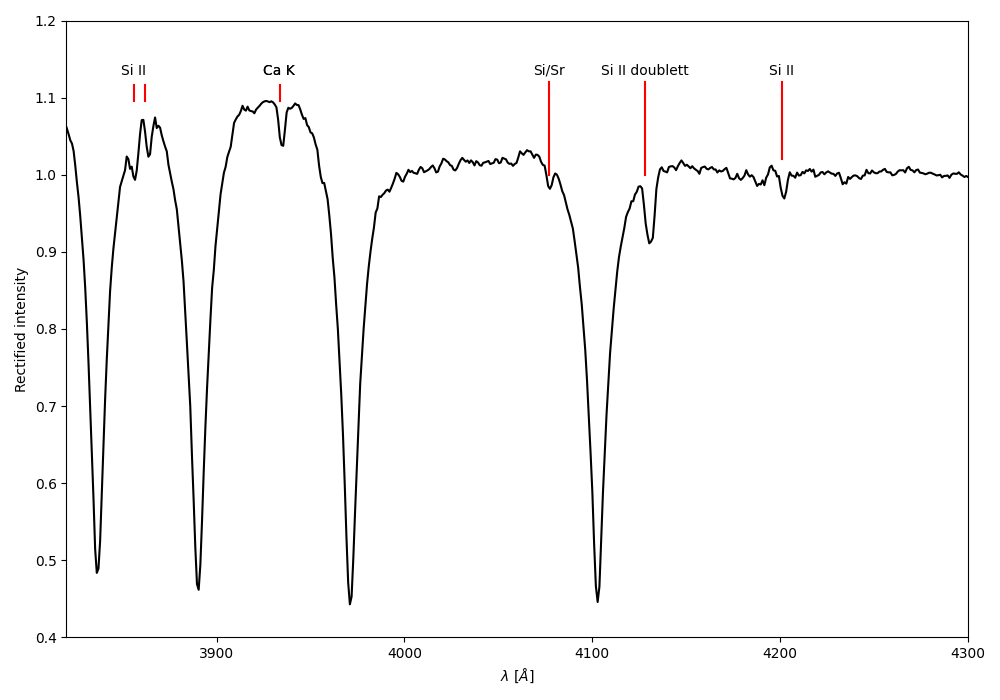}
    \caption{Spectrum of the B8 IV Si (\textit{libr18}) star Gaia DR3 121406905707934464
/ LAMOST J032919.94+312457.0. The Si II lines are shown.}
    \label{fig:example_spec}
\end{figure}

\section{Astrophysical parameters}

 For the statistical analysis, we need the astrophysical parameters of the stars (\Teff, \logg\ or luminosity, and mass). The homogeneity of these parameters is most important for draw the correct conclusions. 
We used the correlations taken from \cite{2024A&A...683L...7P}, who
analysed \Teff\ and \logg\ values of different automatic pipelines 
\citep{2019A&A...628A..94A,2019AJ....158..138S,2022A&A...658A..91A,2023A&A...674A..28F,2023MNRAS.524.1855Z}
for the four CP star subgroups. He derived mean uncertainties between 3 and 12\%. 

We used the isochrones by \cite{2012MNRAS.427..127B} to do this. They include 
a PMS phase starting with $\log t$\,$\ge$\,6.6. 

\section{Light curves}\label{tess_lc}

\subsection{TESS}

Since many stars in the CP2/4 subgroups show variability in their brightness and spectra, a phenomenon explained by the oblique rotator theory \citep{1950MNRAS.110..395S}, it is only natural to search for variabilities like this in our stars. The main classes to search for are the  $\alpha^2$ Canum Venaticorum (ACV), SX Arietis (SXARI), and the rapidly oscillating Ap (roAp) stars.

We used the Python package \texttt{eleanor}\footnote{\url{https://adina.feinste.in/eleanor/}} \citep{2019PASP..131i4502F} to download light curves collected by the Transiting Exoplanet Survey Satellite (TESS) mission \citep{2015JATIS...1a4003R}.The light curves of 33 of our 45 stars were accessible via this method, some even in multiple sectors. After extraction, outliers were removed using a $3\sigma$ cut on the flux data. Subsequently, the flux differences were converted into relative magnitudes using the well-known relation

\begin{equation}
    \Delta mag = -2.5log_{10}\left(\frac{I}{I_0}\right)
\end{equation}

However, these light curves have to be used with caution because in TESS, a relatively large portion of the sky ($21"\times21"$) falls onto each pixel, which may lead to blending with other sources in this area. Thus, the signal cannot be entirely assumed to come from the target star and not a nearby source. An example is the star Gaia DR3 3131891973309856640, whose signal likely comes from the nearby star V640 Mon (HD 47088). The light curve is shown in Fig. \ref{fig:blended_star}

Finally, frequency analysis was performed using a Lomb-Scargle periodogram \cite{1976Ap&SS..39..447L,1982ApJ...263..835S}. The light curves were then phase-folded at the most prominent period.

The light curve data from this method had some issues regarding quality and consistency, thus we could not find a good solution in the frequency analysis for all stars. However, the data were good enough for 13 stars to allow for more or less periodic signals to be seen in the light curves. The results of this analysis are presented in Appendix \ref{appendix_lc_tess}.

Based on the method, the light curve data are processed by \texttt{eleanor}, and some long-term variability might be lost. This is prominent in the case of Gaia DR3 121406905707934464, which has a rotation period of 123.3d, but probably due to the detrending of the light curve, this signal was lost for our analysis (Fig. \ref{fig:lc_gaia_121406905707934464}).

\subsection{CoRoT}

The light curve of one star (Gaia DR3 3326696507149141120) is reported in the database of the COnvection, ROtation and planetary Transits (CoRoT) satellite. The observations span roughly three weeks.
The light curve was treated in the same way as the data obtained from the TESS satellite. 

\section{Spectral classification}

Tthe CP nature of the stars is best confirmed using spectroscopy. We searched the latest public data release (DR9 \footnote{\url{http://www.lamost.org/dr9/}} of the  Large Sky Area Multi-Object Fiber Spectroscopic Telescope (LAMOST, \cite{2012RAA....12.1197C, 2012RAA....12..723Z}) for spectral data. Spectra were available for four stars and could therefore be used for classification. To do this, we used the code MKCLASS \citep{2014AJ....147...80G}, which is an automatic routine that compares the spectra to spectral libraries and determines the best-fitting spectral type according to the quality of the spectra.

On the rectified spectra, we used the spectral libraries \textit{libr18\_225}, \textit{libr18} and \textit{libnor36} to classify the spectra.

Additionally, we searched the ESO database\footnote{\url{https://archive.eso.org/scienceportal/}} for spectra of our stars. Spectral data in a suitable wavelength range for spectral classification were found for seven sources. They either had spectra from the Ultraviolet and Visual Echelle Spectrograph (UVES \cite{2000SPIE.4008..534D}, four stars) or from X-shooter (\cite{2011A&A...536A.105V}, two stars) and one star had a spectrum available from the High Accuracy Radial velocity Planet Searcher (HARPS \cite{2003Msngr.114...20M}). 

We note that the signal-to-noise ratio in the g-band of the LAMOST spectra is not the best for the fainter sources. The automatic classification can therfore vary depending on which library of standard stars is used. Since the number of stars with available spectra is small, we also inspected the spectra by eye to verify the automatic classification and to provide a more accurate result based on the data.

\section{Spectral energy distributions}

To confirm the PMS status of our target stars,that is, to search for infrared excess, and to obtain proper estimates of the stellar parameters, we fitted spectral energy distributions to available photometric data. We used the Python package \texttt{astroARIADNE} \citep{2022MNRAS.513.2719V}\footnote{\url{https://github.com/jvines/astroARIADNE}} to search for photometric data which is available online and fit an atmospheric model (in this case the models from \cite{1993yCat.6039....0K, 2003IAUS..210P.A20C}) to the photometry. After fitting the models, the package uses Bayesian model averaging (BMA) to find the best-fitting stellar parameters, such as the effective temperature, stellar radius, surface gravity, and luminosity. Additionally, the age of the star can be determined by the use of MIST isochrones \citep{2011ApJS..192....3P,2013ApJS..208....4P,2015ApJS..220...15P,2016ApJS..222....8D,2016ApJ...823..102C}. However, since these isochrones do not cover the pre-main-sequence evolution, the ages determined by the fitting routine are not suitable for our case.

\begin{figure}
    \centering
    \includegraphics[width = \columnwidth]{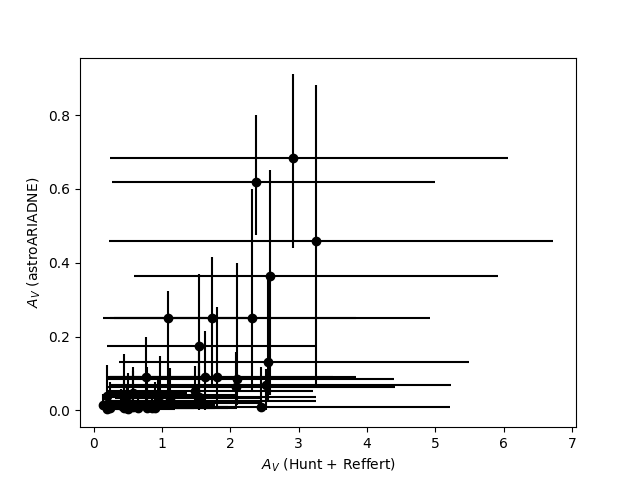}
    \caption{Comparison of the extinction values given by \cite{2023A&A...673A.114H} and those estimated by \texttt{astroARIADNE}.}
    \label{fig:extinction_comp}
\end{figure}

\texttt{astroARIADNE} was unable to find any infrared excess. This is probably a result of the photometric data acquired by the package and the subsequent fitting process. It also provides only the raw SED of the measurements, so no extinction correction is done to infer the actual stellar parameters. Some stars, especially those with higher $A_V$ values, therefore have cooler temperatures than they truly have. One notable example is the star SHI261 (Gaia DR3 121406905707934464), a B-type star with an effective temperature of $11175 \pm 130$K according to \citep{2023Univ....9..210P} who used high-resolution spectroscopy in addition to an SED fit, and determined an effective temperature of only $5587^{+589}_{-463}$K by \texttt{astroAriadne}. However, the final $A_V$ values determined by \texttt{astroARIADNE} are all close to zero (see Fig. \ref{fig:extinction_comp}). The results therefore have to be interpreted with caution.

The results of the SED fitting were compared with the output from the Virtual Observatory SED Analyser\footnote{\url{http://svo2.cab.inta-csic.es/theory/vosa/}} (VOSA) \citep{2008A&A...492..277B}.

\begin{figure}
    \centering
    \includegraphics[width = \columnwidth]{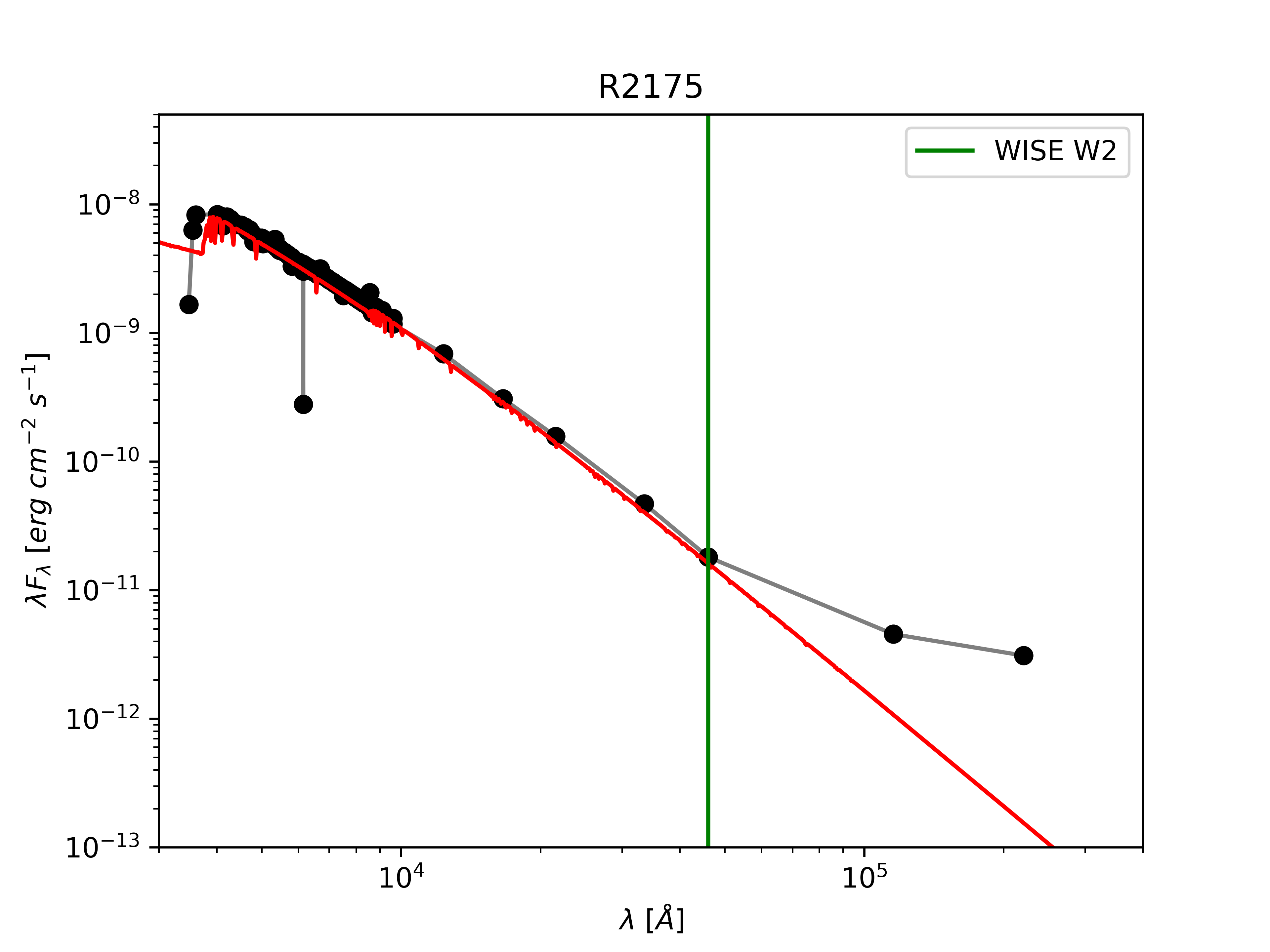}
    \caption{Spectral energy distribution for the star R2175 (Gaia DR3 3326717260430731648) as measured by VOSA (black dots). The red spectrum is a model from \cite{2003IAUS..210P.A20C} with the parameters $T_{eff} = 13000$K, $logg = 4.0$ and $[Fe/H]=-0.5$ dex. The vertical green line denotes the effective wavelength of the WISE W2 band from the point at which the measured SED deviates from the model spectrum, and the IR excess starts to show.}
    \label{fig:sed_example}
\end{figure}

According to the results from VOSA, 15 of our 45 stars had enough data in the near and mid-infrared region to find an IR excess that is commonly associated with debris disks around newly formed stars. An example can is shown in Fig. \ref{fig:sed_example}.

\begin{figure}
    \centering
    \includegraphics[width = \columnwidth]{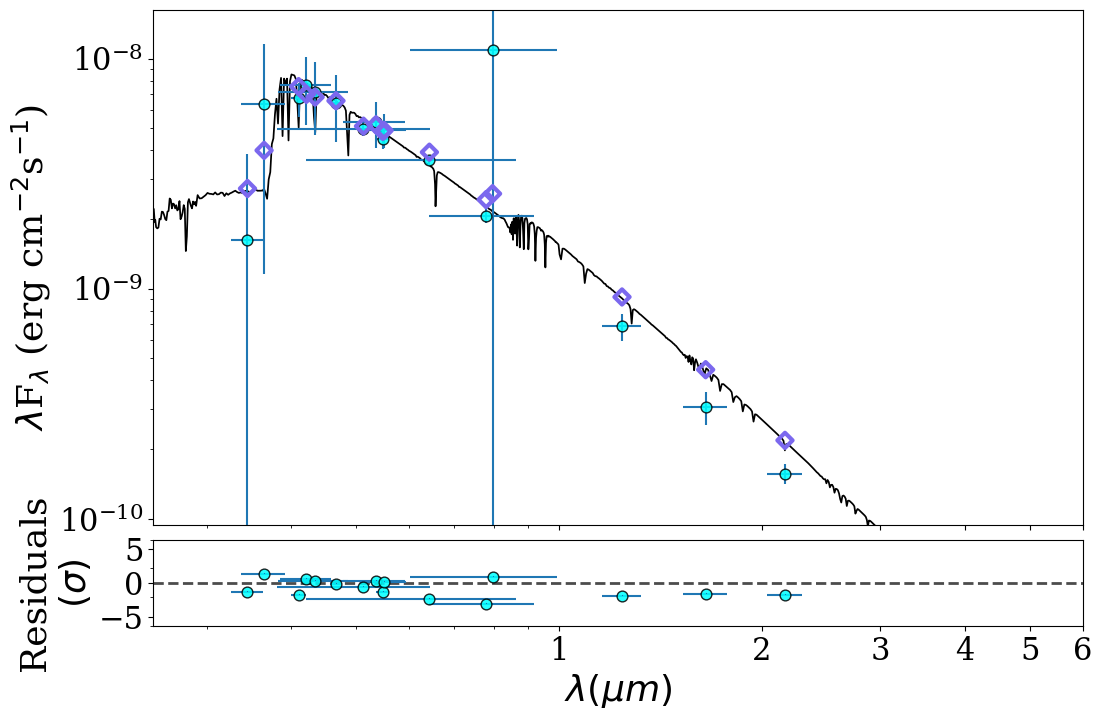}
    \caption{Same as Fig. \ref{fig:sed_example} but here we show the output plot generated by \texttt{astroARIADNE}. The synthetic photometry is shown via the cyan points; the purple diamonds show the fit points. The synthetic spectrum is a spectrum with the parameters $T_{eff} = 9500$K, $logg=3.5$ and $[Fe/H] = -0.1$dex taken from \citep{1993yCat.6039....0K}. This method probes a smaller wavelength range so that no IR excess can be seen here.}
    \label{fig:sed_ariadne}
\end{figure}

\section{Results} \label{results}

\subsection{Pre-main-sequence stars vs non-pre-main-sequence stars}

The results of the SED fitting using VOSA show that 15 of our sample stars exhibit an infrared excess. The measurements suggest that the IR excess starts in the mid-infrared region after 3$mu$m, meaning that the debris disk around each of these stars has a temperature below approximately 1000K. Additionally, 2 of our sources show at least one hydrogen line emission in their spectra. One of these 2 sources (Gaia DR3 5943020022195591552) shows no IR excess in its SED.

\subsection{Variability}
Thirteen of the 33 light curves available from TESS showed clear variability in accordance to them being CP stars. We found 6 definitive ACV variables with a double wave (see Appendix \ref{appendix_lc_tess}). Eight can be classified as having rotating features or binarity in their data. The light curve of one star, Gaia DR3 5541472465805985024, shows a periodic pattern with a period of 4.241 days, which likely is due to rotation. It also shows a higher frequency pulsational pattern, which probably is a pattern commonly found in $\gamma$ Doradus variables or PMS stars (Fig. \ref{fig:rot_gdor})\\
Seven of the remaining light curves show no or irregular variability. They are treated as VAR because some variability of unknown origin is visible. These patterns in variability, for instance the irregular ones, could also be signs of the PMS evolution of the star. The last group, suspected binary stars, consists of three sources, of which for one, Gaia DR3 3131891973309856640, the signal is more likely to come from the nearby known variable V649 Mon (Fig. \ref{fig:blended_star}; see also Tab. \ref{tab:tess_vars} for the variability classification of the whole sample based on TESS light curves).

\subsection{Spectral Types}

\subsubsection{LAMOST}

The spectral types of the four availaable LAMOST spectra were determined by MKCLASS. The result is listedx in Tab. \ref{tab:lamost_spt}. We note that the signal-to-noise ratios of the spectra are not particularly good, so the automatically detected spectral types are somewhat unsure. However, a manual inspection can help in these cases. One of the spectra (Gaia DR3 3368982075084757632) shows emission in the H$\alpha$ line, which is a clear sign of the PMS status of the star. This star is also a CP star candidate without previous mention in the literature. Table \ref{tab:lamost_spt} lists the sample of the stars with LAMOST spectra, the classifications made by MKCLASS, and the visual inspection.

\begin{table*}
    \caption{Spectral types determined by MKCLASS from the LAMOST spectra. The last column lists the latest spectral type we could find in the literature (if available) as a reference.}
    \centering
    \begin{tabular}{c|c|c|c|c|c}

    \textbf{Gaia DR3} & \textit{spt\_libr18} & \textit{libnor36} & \textit{spt\_libr18\_225} & \textbf{SpT manual} & \textbf{SpT Literature}\\
    \hline
    \hline
    121406905707934464  &B8 IV  Si	&B9 III  Si	&F1 mB8 IV-V metal-weak & B9 III-IV Si& B8-9 Si He-wk (1)\\
    3019972890876467968  & Unclassifiable	&B8 III-IV  Si	&F1 mB8 IV-V metal-weak	& B8 IV-V Si: & B6 VI Si (2)\\
    3368982075084757632 & Unclassifiable	&O6 III	&F4 V Fe-1.2 & B9 Ve SrSiEu	& A2 V (3) \\
    3131891973309856640 &B5 IV	&O7 ?	&kB9hF2mG2  Eu  & B7-8 V Eu &  B5 IV (1) \\

    \end{tabular}
    
             \tablebib{
    (1)~\cite{2023Univ....9..210P}, (2)~\cite{2022ApJS..259...63S}, (3)~\cite{2022ApJS..259...38Z}
}  
    \label{tab:lamost_spt}
\end{table*}

\subsubsection{ESO spectra}

Due to the high spectral resolution and/or a spectral range that was unsuitable for MKCLASS, ESO stars were classified manually. Four of the seven stars showed signs of chemical peculiarity. The others have been classified as CP or CP candidates in the literature (see Tab. \ref{tab:manual_spt}).

\begin{table}[]
    \caption{Same as Tab. \ref{tab:lamost_spt} but for the seven stars with spectra available from the ESO archive. }
    \centering
    \begin{tabular}{c|c|c}

    \textbf{Gaia DR3} & \textbf{SpT} & \textbf{SpT Literature}\\
    \hline
    \hline
    3104244792087711360 & B9 Si           &  B9 Si (1)\\
    3217786616242183424 & B3              &  Am (2) \\
    6236109243250504576 & B He-strong?    &  B5 (3) \\
    6071548670839838848 & B7              &  B8 Cr (4) \\
    3345191014284099200 & B9-A0 Si        &  A0 Si (5)\\
    3017188622495555328 & B5              &  B9 He-wk (2)\\
    5943020022195591552 & B2 Ve He-strong    & B2 IV-V (4)\\

    \end{tabular}

    \tablebib{
    (1)~\cite{2022MNRAS.513.1429S}, (2)~\cite{2013AstBu..68..300R},
    (3)~\cite{2022A&A...657A..62K}, (4)~\cite{2021MNRAS.504.3758P},
    (5)~\cite{2017MNRAS.468.2745N}, (6)~\cite{2020A&A...636A..74T}
}  
    \label{tab:manual_spt}
\end{table}

\subsection{Fraction of chemically peculiar stars}

We determined in our sample that are CP stars and those that have the possibility of being normal stars.  The results for the spectral classification are given in tables \ref{tab:lamost_spt} and \ref{tab:manual_spt}. Many of these starsx can be considered CP stars or CP candidates according to our classification and the classification found in the literature. Not every source has a definite CP spectral type in the literature (especially in \cite{2009A&A...498..961R}).
Based on the entire sample and by comparing it to the spectral types found in the literature, 29 of our 45 (64\%) candidates can be considered CP stars. The others either lack a spectral type or have only vague spectral classifications, such as A1-A7 for the star Gaia DR3 2204463918269731072 (see Tab. \ref{tab:spt_all_lit}).

\begin{figure*}
    \centering
    \includegraphics[width = \textwidth]{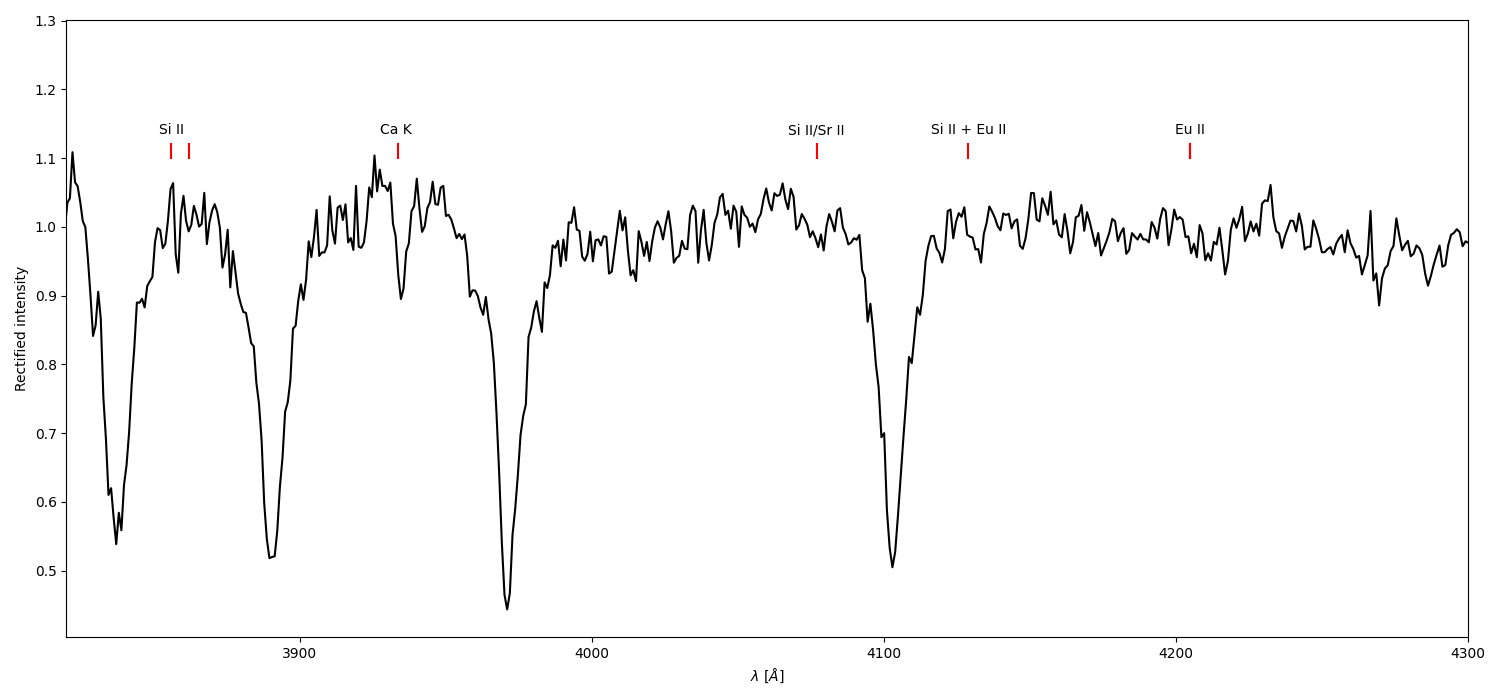}
    \caption{Blue-violet part of the rectified spectrum of the newly found CP candidate Gaia DR3 3368982075084757632. The spectral lines of interest are indicated. However the S/N of the spectrum is a slightly low, so that the exact type remains to be determined.}
    \label{fig:new_cp_cand}
\end{figure*}

\begin{figure*}
    \centering
    \includegraphics[width = \textwidth]{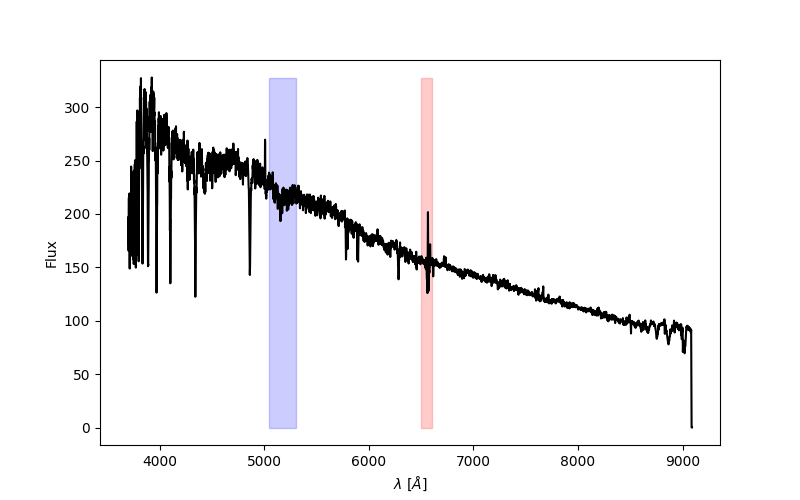}
    \caption{Full spectrum of the newly found PMS CP candidate (see text and Fig. \ref{fig:new_cp_cand} as observed by LAMOST. The 5200\AA\ flux depression is shown in blue and the H$\alpha$ emission is shown in red.}
    \label{fig:new_cp_full}
\end{figure*}

No spectral type is listed for Gaia DR3 3368982075084757632, but it has a LAMOST spectrum showing possible signs of Si and/or Sr/Eu peculiarity (see Fig. \ref{fig:new_cp_cand}) However, the spectrum has a signal-to-noise ratio (S/N) of only 47.77 in the g-band, so the exact spectral type is unclear, whereas the flux depression around 5200\AA\ is relatively clearly visible, so that a magnetic field associated with CP2 stars can be assumed (see Fig. \ref{fig:new_cp_cand} for the part of the spectrum with the lines of interest and Fig. \ref{fig:new_cp_full} for the full LAMOST spectrum of the star). Photometric data from VOSA resulted in an infrared-excess in the mid-infrared region.
When we combine this with the finding that the spectrum also shows emission in the $H\alpha$ line, we can conclude that this is also a Herbig Ae/Be object. This is another indication that the two object types are related.

Combining the spectroscopic results with the results of the TESS light curves, we find that 32 of our 45 candidates are definitely or likely CP stars, resulting in a CP star fraction of 71\%. 

\subsection{Pre-maim-sequence chemically peculiar stars} \label{pms_stars}

Combining all the results from our analysis, which consist of spectral classification, SED fitting, and light curve analysis, we can determine how many of our stars are PMS-CP stars.  Nine probable sources have at least one of these characteristics. Out of these, eight sources have a CP spectral type either in the literature, derived by us or both, and one lacks any CP spectral type. However, the light curve data for this star shows variability associated with ACV variables (Fig. \ref{fig:pms_cp_acv}). The primary data (coordinates, parallaxes and spectral types)of our PMS CP stars are shown in Tab. \ref{tab:pms_cp_final} and the criteria for classifying the stars as pre-main sequence, in our case, IR-excess, are given in  Tab. \ref{tab:ir_ex}. The star Gaia DR3 368982075084757632, also shows emission in H$\alpha$, which is a sign of accretion onto  the star and it thus indicates the PMS status of this source (Fig. \ref{fig:new_cp_full}).

\subsubsection{SHI261 / Gaia DR3 121406905707934464}

This star has already been classified as B8-9 Si He-wk by \cite{2023Univ....9..210P}. The authors also analysed the evolutionary status of this star in detail. They concluded that it is still a PMS object that recently completed the accretion state in its evolution. They derived an IR excess starting in the region of the WISE W3 band. Our analysis based on spectra and SED fitting cenfirms those findings. Our manually determined spectral type is B9 III-IV Si. Additionally, the IR excess determined by VOSA starts around the same wavelengths as in \cite{2023Univ....9..210P}, confirming the PMS status of the star.

\subsubsection{R720 / Gaia DR3 216648703146774016}

The spectral type of this star was given as A0 SiSr in \cite{2009A&A...498..961R}. Our analysis revealed a flux depression around $5200\AA$ in the Gaia BP/RP spectrum, strengthening the argument that it is a CP2 star. The IR excess of the star starts to show in the AKARI WIDE-S passband.

\subsubsection{SHA20504 / Gaia DR3 3019972890876467968}

This star was classified as B6 VI Si by \cite{2022ApJS..259...63S} using machine learning on LAMOST DR8. Our spectral classification of B8 IV-V Si confirms the CP status. The assignment of a PMS-type star comes from the IR excess starting at the WISE W4 passband, as determined by VOSA.

\subsubsection{Z583 / Gaia DR3 3368982075084757632}

This star has not been named a CP star in the literature. However, looking at the spectrum from LAMOST (Fig. \ref{fig:new_cp_full}), we can see the flux depression around $5200\AA$, which, together with our derived spectral type B9 Ve SrSiEu, hints at the star being a CP candidate. However, as mentioned above, the SNR of the spectrum is poor, so we note that the  spectral type is not final and needs a better spectrum to classify the peculiarity or even disprove it accurately. Some spectral lines of interest are denoted in Fig. \ref{fig:new_cp_cand}. The PMS status is made clear by the H$\alpha$ emission in the LAMOST spectra (see the red region in Fig. \ref{fig:new_cp_full}) and the IR excess that was determined from the SED fitting starting at the WISE W3 band.

\subsubsection{R2175 / Gaia DR3 3326717260430731648}

This is another CP star from \cite{2009A&A...498..961R} who reported the spectral type as B6 He var. Its PMS status was determined using the IR excess in the SED starting in the WISE W3 band.

\subsubsection{R5585 / Gaia DR3 6244725050721030528}

This star has been classified as B9 SiCrSr according to \cite{2009A&A...498..961R}. The BP/RP spectrum also revealed a clear flux depression associated with magnetic stars, further confirming the CP nature of this star. Our analysis regarding the PMS status of the source shows IR excess at wavelengths beyond the WISE W3 band.

\subsubsection{R5765 / Gaia DR3 5966515967154648064}

Although we failed to find a classification as a CP star in the literature (\cite{2009A&A...498..961R} determined it to be an A-type star), the positive $\Delta$a value marks it as a candidate for being a magnetic star. Unfortunately, we were unable to obtain a spectrum from the databases we searched that would have allowed a detailed classification. However, the light curve from TESS shows a variability that is commonly associated with ACV variables, which is another strong argument for the CP nature (Fig. \ref{fig:pms_cp_acv}). We conclude that the star is still in the PMS stage based on the SED that shows IR excess starting at the WISE W2 band.

\subsubsection{R7193 / Gaia DR3 2244529022468154880}

This star has also been classified as a CP star in the literature. \cite{2009A&A...498..961R} listed the stars as an A0 Si star, which is accompanied by a positive $\Delta$a value. The CP classification is therefore reliable. It is also a PMS star because we detected an IR excess starting at the WISE W1 band.

\begin{table*}[]
    \centering
    \caption{Basic properties of our PMS CP candidates}.
    \begin{tabular}{c|c|c|c|c|c|c}
    
    \textbf{Gaia DR3} & \textbf{RA ICRS} & \textbf{DEC ICRS} & $\varpi$ & \textbf{Cluster} & \textbf{SpT our} & \textbf{SpT Literature}\\
    \hline
    \hline
    121406905707934464 & 52.3324400	& 31.4158037 & 3.4802$\pm$0.0237 & NGC 1333 & B9 III-V Si & B8-9 Si He-wk (1)\\
    216648703146774016 & 56.4495623	& 32.0162390 & 3.0693$\pm$0.0325 & IC 348   &            & A0 SiSr (2)\\
    3019972890876467968& 92.0529125	& -6.1956720 & 1.2045$\pm$0.0172 & OC 357 & B8 IV-V Si: & B6 IV Si (3)\\
    3368982075084757632& 94.8562983	& 15.3365834 & 0.2395$\pm$0.0409 & UBC 438 &  B9 Ve SrSiEu & A2 V (4)\\
    3326717260430731648& 100.2148105&  9.8637002 & 1.3310$\pm$0.0365 & NGC 2264&           & B6 He var (2)\\
    6244725050721030528& 245.0228316& -20.0565075& 7.6496$\pm$0.0297 & HSC 2931 &          & B9 SiCrSr (2)\\
    5966515967154648064& 253.4669125& -41.8150331& 0.5817$\pm$0.0205 & NGC 6231 &           & A (2)\\
    2244529022468154880& 305.7257939&  64.1840742& 2.1286$\pm$0.0137 & UPK 160 &           & A0 Si (2)\\

    \end{tabular}
    \tablebib{
    (1)~\cite{2023Univ....9..210P}, (2)~\cite{2009A&A...498..961R},
    (3)~\cite{2022ApJS..259...63S}, (4)~\cite{2022ApJS..259...38Z}
    }
    \label{tab:pms_cp_final}
\end{table*}

\begin{table}[h!]
    \centering
    \begin{tabular}{c|c}
    \textbf{Gaia DR3} & \textbf{Begin of IR excess}\\
    \hline
    \hline
    121406905707934464 & IRAS 12$\mu$m	 \\
    216648703146774016 & AKARI WIDE-S	 \\
    3019972890876467968& WISE W4	 \\
    3368982075084757632& WISE W3	 \\
    3326717260430731648& WISE W3 \\
    6244725050721030528& WISE W4 \\
    5966515967154648064& WISE W1 \\
    2244529022468154880& WISE W1 \\
    \end{tabular}
    \caption{List of the filters in which the IR excess stars to show according to VOSA.}
    \label{tab:ir_ex}
\end{table}

\subsection{Non-chemically peculiar stars as members of open clusters} \label{none_CPs}

As described in Sect. \ref{target_selection}, we did not exclude any objects from the various
sources of CP stars. After the matching, we checked all objects to determine whether they belonged to their subgroup. 
For 24 objects from \citet{2009A&A...498..961R} we were unable to find any confirmation
that they would be a CP star. They also have regular $\Delta$a values (Sect. \ref{Da_photometry}).

\subsection{Colour-magnitude diagrams} \label{cmd}

Using the cluster members from \cite{2023A&A...673A.114H}, we plotted a CMD for each of the 39 clusters containing our CP stars and candidates. In the plots, we included the isochrones from \cite{2012MNRAS.427..127B} shifted and adapted by ages, distances, and extinctions given in \cite{2023A&A...673A.114H}. The results are shown in Appendix \ref{cluster_cmds}. One CP candidate has a very red colour index of $BP-RP \approx 3.1$ (Fig. \ref{fig:oc_322}). This star was classified as a CP1 star by \cite{2019ApJS..242...13Q} with the spectral type kA3hA3mA7.\\
However, since some CMDs have relatively high reddening values, we cannot  determine the evolutionary status from the diagram alone. The differential extinction across the clusters would have to be considered as well. However, we can get an idea of where our CP stars and candidates lie.

\section{Conclusions} \label{conclusions}

We studied 45 suspected PMS stars that may also be CP stars to determine which of them are real PMS stars and which are CP stars, or at least candidates, either labelled as such by previous authors or CP candidates that need closer investigation.

We find a CP fraction of 71\% in our sample (including CP stars and CP candidates), of which nine stars are likely still in their PMS phase (Section \ref{pms_stars}). Based on our analysis of the available spectral types in the literature and those we determined, accompanied by photometric variability studies, we conclude that all of our PMS CP sources appear to belong to the subclass of magnetic CP stars.

This result can be interpreted as an argument that the magnetic fields found in these stars are an essential part of the early evolution of CP stars and would at least partially explain the origin of these subtypes. However, the origin and evolution of these magnetic fields remains to be discussed in detail.

However, we note that some of the spectral types are still somewhat ambiguous due to the quality of the data. More and better data are required to accurately determine the correct type of peculiarity, especially for four of our newly found CP candidates.

\begin{acknowledgements}
This work was supported by the grant GA{\v C}R 23-07605S and the European Regional Development Fund, project No. ITMS2014+: 313011W085 (MP).
This work has made use of data from the European Space Agency (ESA) mission 
{\it Gaia} (\url{https://www.cosmos.esa.int/gaia}), processed by the {\it Gaia} Data 
Processing and Analysis Consortium (DPAC, \url{https://www.cosmos.esa.int/web/gaia/dpac/consortium}). 
Funding for the DPAC has been provided by national institutions, in particular, the institutions
participating in the {\it Gaia} Multilateral Agreement. This research has made use of the SIMBAD database,
operated at CDS, Strasbourg, France and 
of the WEBDA database, operated at the Department of Theoretical
Physics and Astrophysics of the Masaryk University.
This publication makes use of VOSA, developed under the Spanish Virtual Observatory (https://svo.cab.inta-csic.es) project funded by MCIN/AEI/10.13039/501100011033/ through grant PID2020-112949GB-I00.
VOSA has been partially updated by using funding from the European Union's Horizon 2020 Research and Innovation Programme, under Grant Agreement nº 776403 (EXOPLANETS-A).
We also thank the anonymous referee for their valuable feedback a, as well as the editor responsible for comments on what can be improved.
\end{acknowledgements}

\bibliographystyle{aa}
\bibliography{PMS_CPs}

\begin{appendix}

\section{Colour-magnitude diagrams of the host clusters}\label{cluster_cmds}

\begin{figure}[ht]
    \centering{\includegraphics[width = \columnwidth]{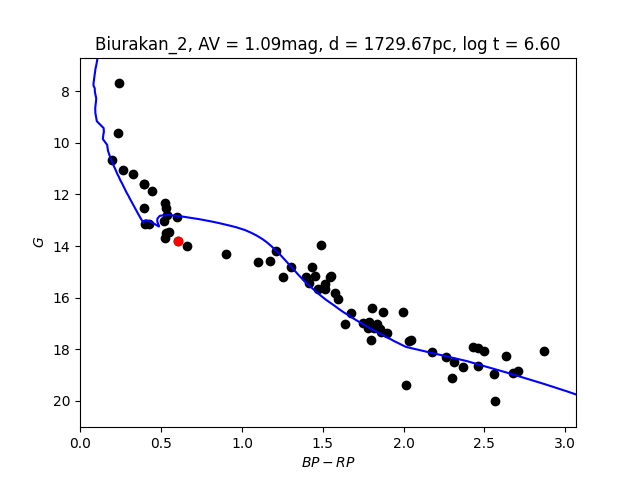}}
    \caption{Colour-magnitude diagram of the cluster Biurakan 2. The black dots are members of the cluster according to \cite{2023A&A...673A.114H}, and the red dots are our CP star candidates. The PARSEC isochrone \citep{2012MNRAS.427..127B} was computed using the $A_V$ and $log\,t$ values from the catalogue of \cite{2023A&A...673A.114H} and using solar metallicity ($Z = 0.0152$). To "fit" the cluster, they were shifted according to the distance of the cluster, which was also taken from \cite{2023A&A...673A.114H}.}
    \label{fig:cmd_cluster}
\end{figure}

\begin{figure}[ht]
    \centering
    \includegraphics[width = \columnwidth]{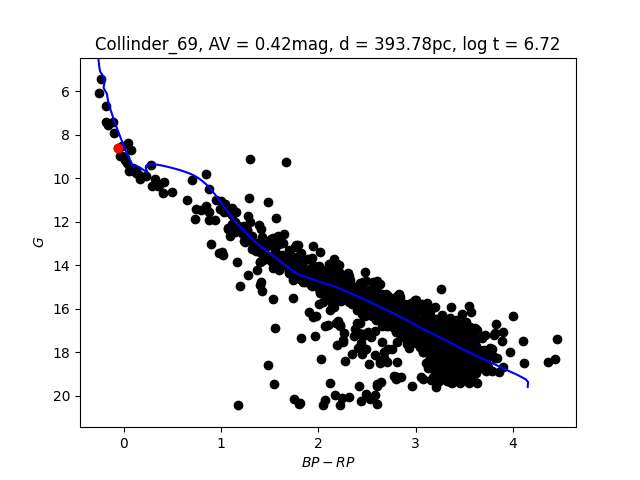}
    \caption{Same as Fig. \ref{fig:cmd_cluster}, but for Collinder 69.}
    \label{fig:enter-label}
\end{figure}

\begin{figure}[ht]
    \centering
    \includegraphics[width = \columnwidth]{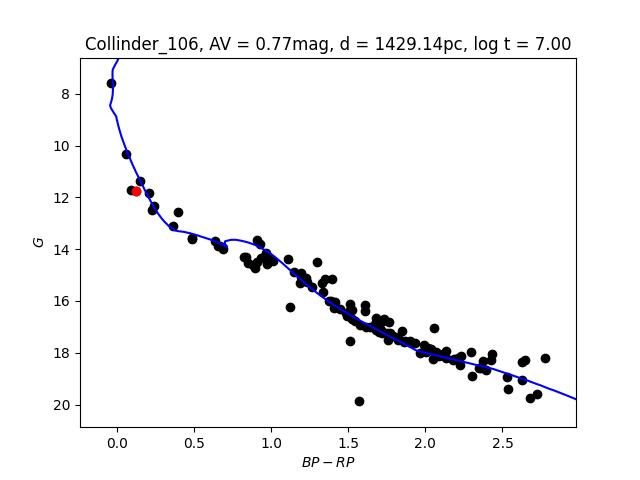}
    \caption{Same as Fig. \ref{fig:cmd_cluster}, but for Collinder 106.}
    \label{fig:enter-label}
\end{figure}

\begin{figure}[ht]
    \centering
    \includegraphics[width = \columnwidth]{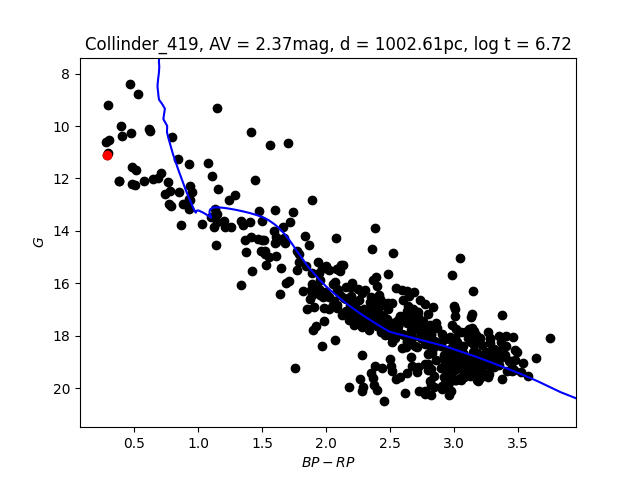}
    \caption{Same as Fig. \ref{fig:cmd_cluster}, but for Collinder 419.}
    \label{fig:enter-label}
\end{figure}

\begin{figure}[ht]
    \centering
    \includegraphics[width = \columnwidth]{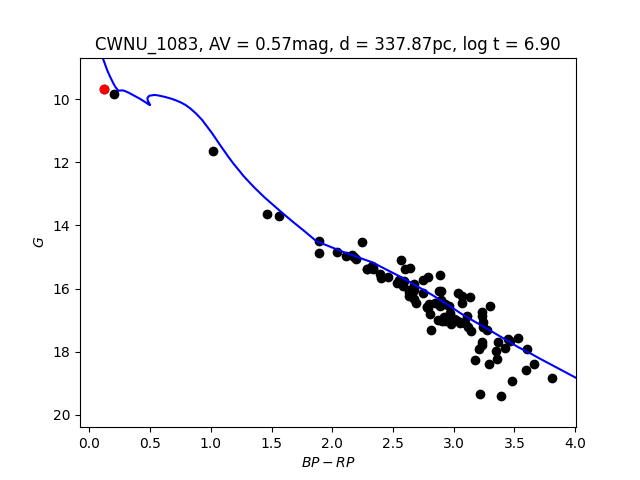}
    \caption{Same as Fig. \ref{fig:cmd_cluster}, but for CWNU 1083.}
    \label{fig:enter-label}
\end{figure}

\begin{figure}[ht]
    \centering
    \includegraphics[width = \columnwidth]{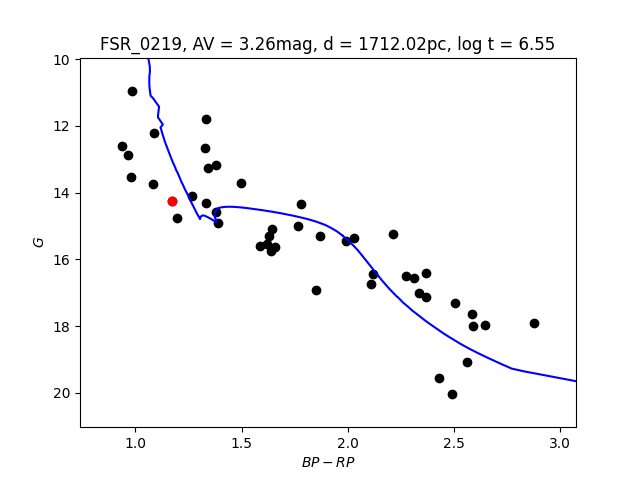}
    \caption{Same as Fig. \ref{fig:cmd_cluster}, but for FSR 0219.}
    \label{fig:enter-label}
\end{figure}

\begin{figure}[ht]
    \centering
    \includegraphics[width = \columnwidth]{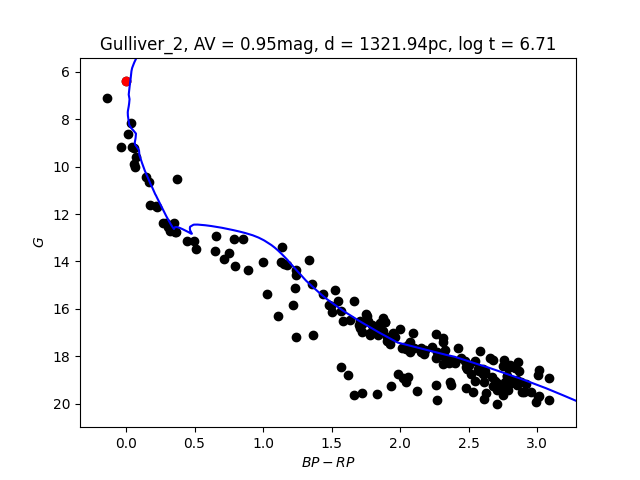}
    \caption{Same as Fig. \ref{fig:cmd_cluster}, but for Gulliver 2.}
    \label{fig:enter-label}
\end{figure}

\begin{figure}[ht]
    \centering
    \includegraphics[width = \columnwidth]{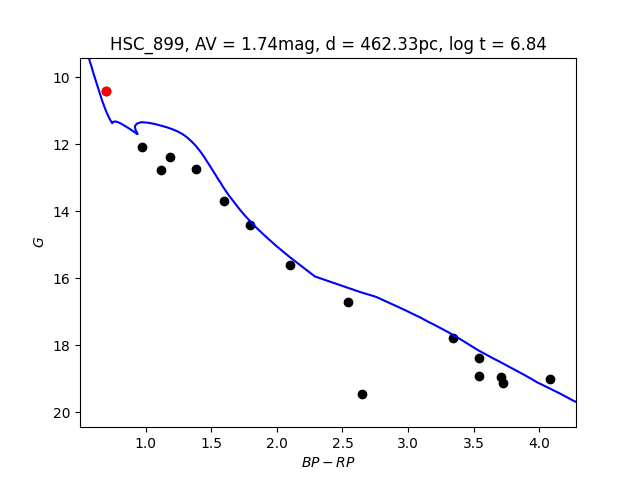}
    \caption{Same as Fig. \ref{fig:cmd_cluster}, but for HSC 899.}
    \label{fig:enter-label}
\end{figure}

\begin{figure}[ht]
    \centering
    \includegraphics[width = \columnwidth]{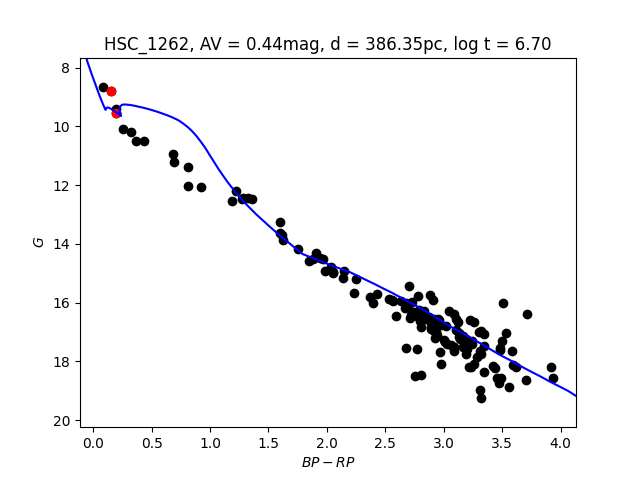}
    \caption{Same as Fig. \ref{fig:cmd_cluster}, but for HSC 1262.}
    \label{fig:enter-label}
\end{figure}

\begin{figure}[ht]
    \centering
    \includegraphics[width = \columnwidth]{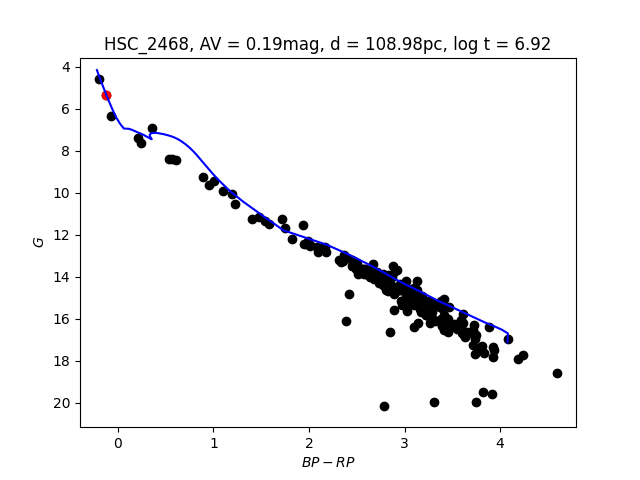}
    \caption{Same as Fig. \ref{fig:cmd_cluster}, but for HSC 2468.}
    \label{fig:enter-label}
\end{figure}

\begin{figure}[ht]
    \centering
    \includegraphics[width = \columnwidth]{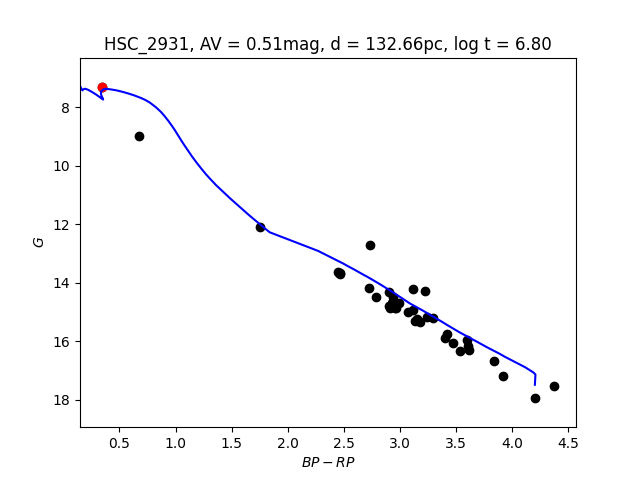}
    \caption{Same as Fig. \ref{fig:cmd_cluster}, but for HSC 2931.}
    \label{fig:enter-label}
\end{figure}

\begin{figure}[ht]
    \centering
    \includegraphics[width = \columnwidth]{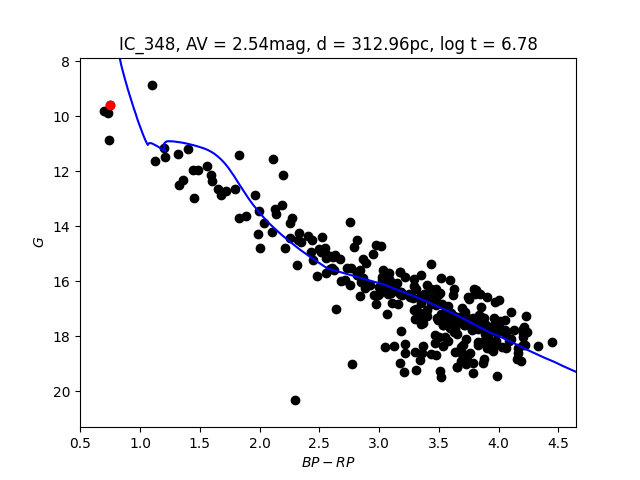}
    \caption{Same as Fig. \ref{fig:cmd_cluster}, but for IC 348.}
    \label{fig:enter-label}
\end{figure}

\begin{figure}[ht]
    \centering
    \includegraphics[width = \columnwidth]{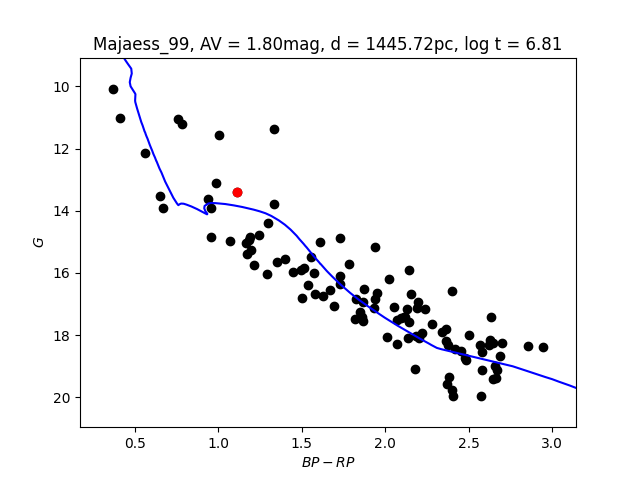}
    \caption{Same as Fig. \ref{fig:cmd_cluster}, but for Majaess 99.}
    \label{fig:enter-label}
\end{figure}

\begin{figure}[ht]
    \centering
    \includegraphics[width = \columnwidth]{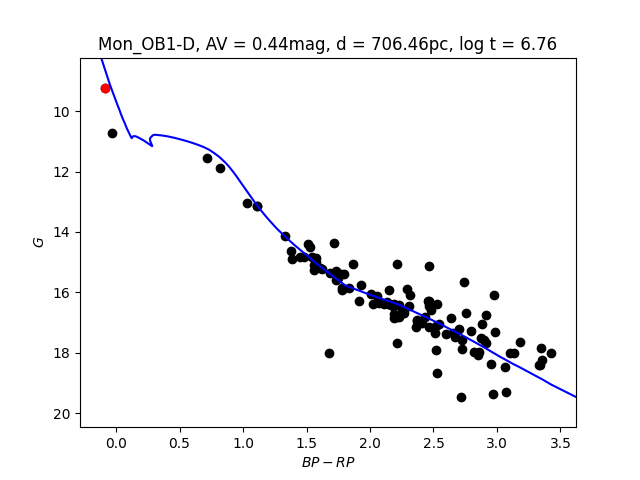}
    \caption{Same as Fig. \ref{fig:cmd_cluster}, but for Mon OB1-D.}
    \label{fig:enter-label}
\end{figure}

\begin{figure}[ht]
    \centering
    \includegraphics[width = \columnwidth]{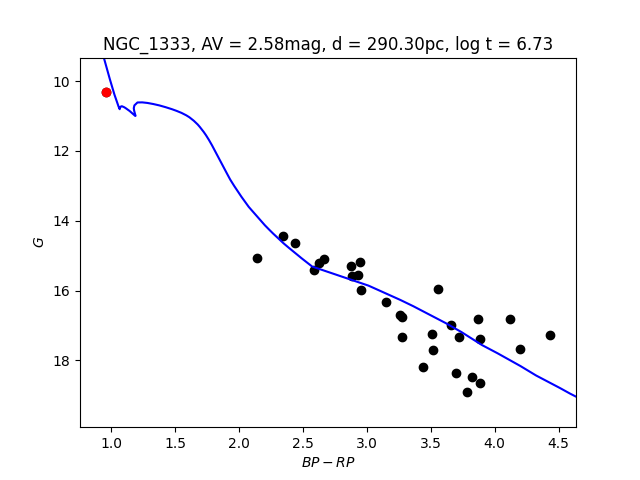}
    \caption{Same as Fig. \ref{fig:cmd_cluster}, but for NGC 1333.}
    \label{fig:enter-label}
\end{figure}

\begin{figure}[ht]
    \centering
    \includegraphics[width = \columnwidth]{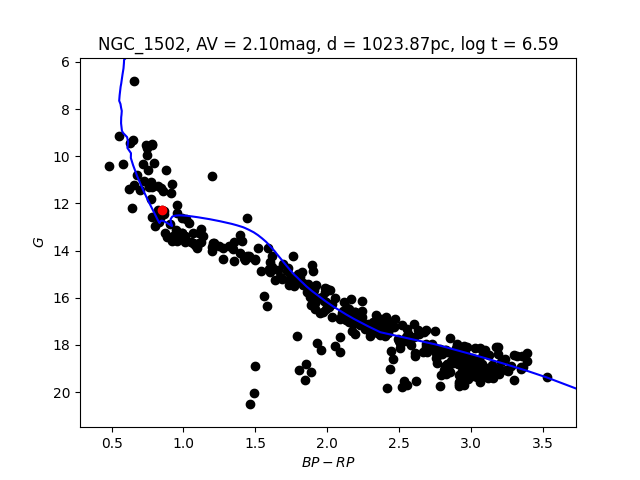}
    \caption{Same as Fig. \ref{fig:cmd_cluster}, but for NGC 1502.}
    \label{fig:enter-label}
\end{figure}

\begin{figure}[ht]
    \centering
    \includegraphics[width = \columnwidth]{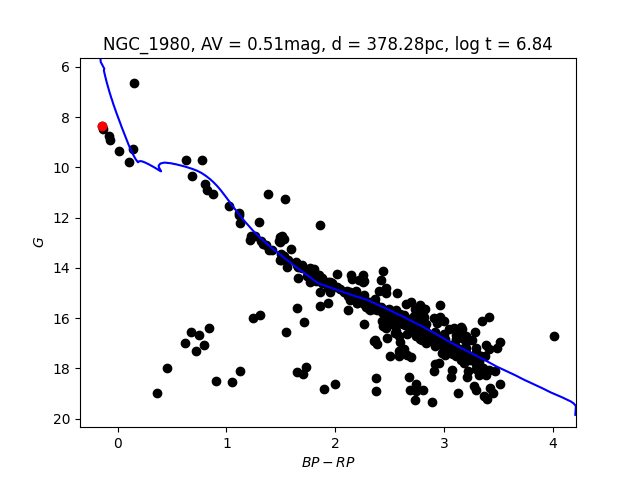}
    \caption{Same as Fig. \ref{fig:cmd_cluster}, but for NGC 1980.}
    \label{fig:enter-label}
\end{figure}

\begin{figure}[ht]
    \centering
    \includegraphics[width = \columnwidth]{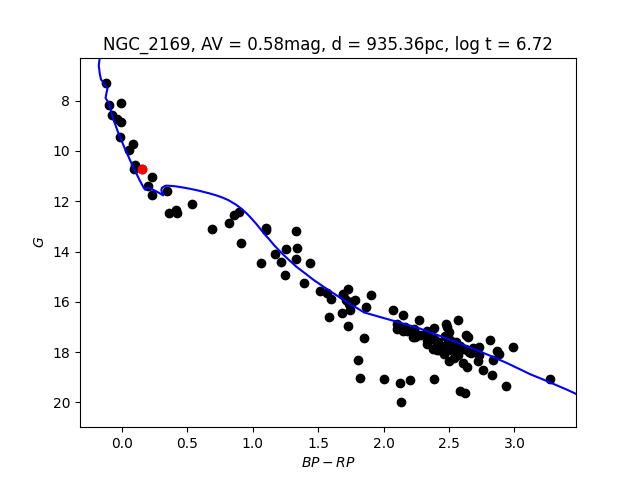}
    \caption{Same as Fig. \ref{fig:cmd_cluster}, but for NGC 2169.}
    \label{fig:enter-label}
\end{figure}

\begin{figure}[ht]
    \centering
    \includegraphics[width = \columnwidth]{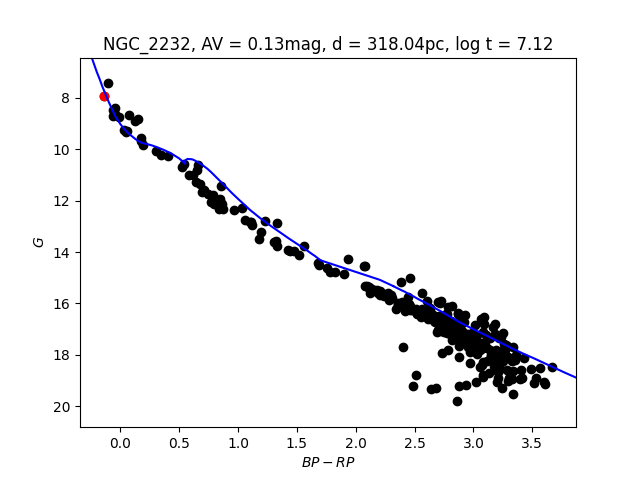}
    \caption{Same as Fig. \ref{fig:cmd_cluster}, but for NGC 2232.}
    \label{fig:enter-label}
\end{figure}

\begin{figure}[ht]
    \centering
    \includegraphics[width = \columnwidth]{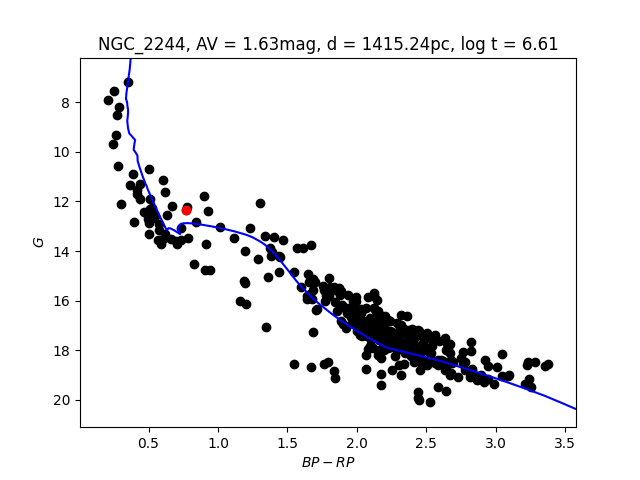}
    \caption{Same as Fig. \ref{fig:cmd_cluster}, but for NGC 2244.}
    \label{fig:enter-label}
\end{figure}

\begin{figure}[ht]
    \centering
    \includegraphics[width = \columnwidth]{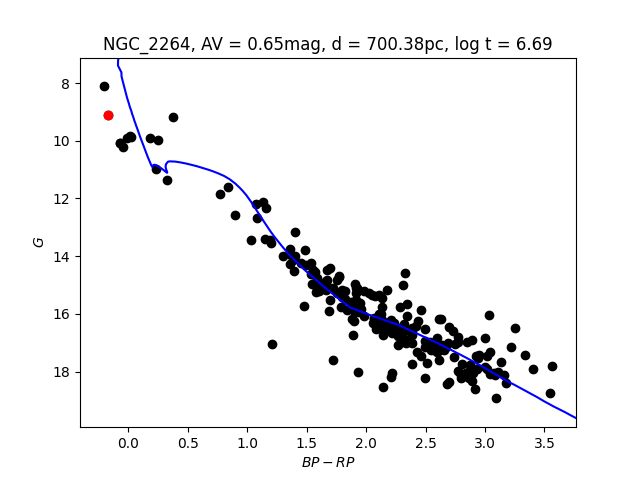}
    \caption{Same as Fig. \ref{fig:cmd_cluster}, but for NGC 2264.}
    \label{fig:enter-label}
\end{figure}

\begin{figure}[ht]
    \centering
    \includegraphics[width = \columnwidth]{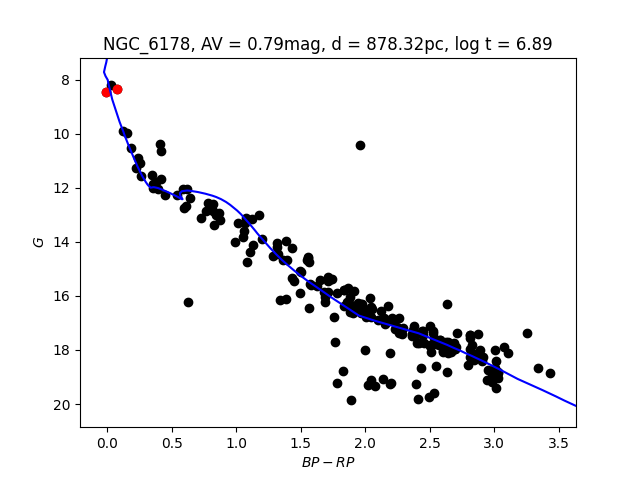}
    \caption{Same as Fig. \ref{fig:cmd_cluster}, but for NGC 6178.}
    \label{fig:enter-label}
\end{figure}

\begin{figure}[ht]
    \centering
    \includegraphics[width = \columnwidth]{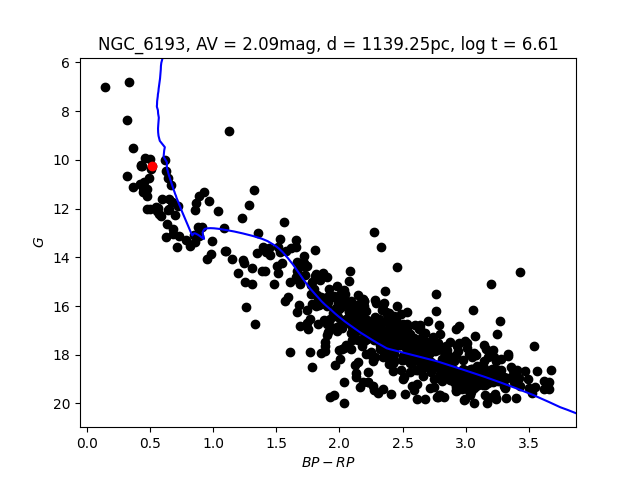}
    \caption{Same as Fig. \ref{fig:cmd_cluster}, but for NGC 6193.}
    \label{fig:enter-label}
\end{figure}

\begin{figure}[ht]
    \centering
    \includegraphics[width = \columnwidth]{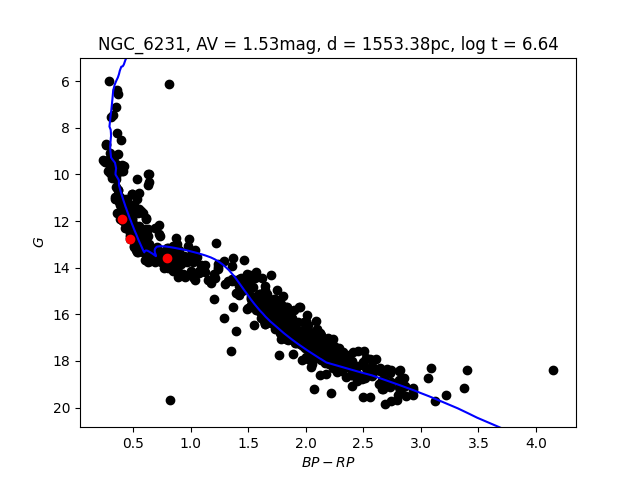}
    \caption{Same as Fig. \ref{fig:cmd_cluster}, but for NGC 6231.}
    \label{fig:enter-label}
\end{figure}

\begin{figure}[ht]
    \centering
    \includegraphics[width = \columnwidth]{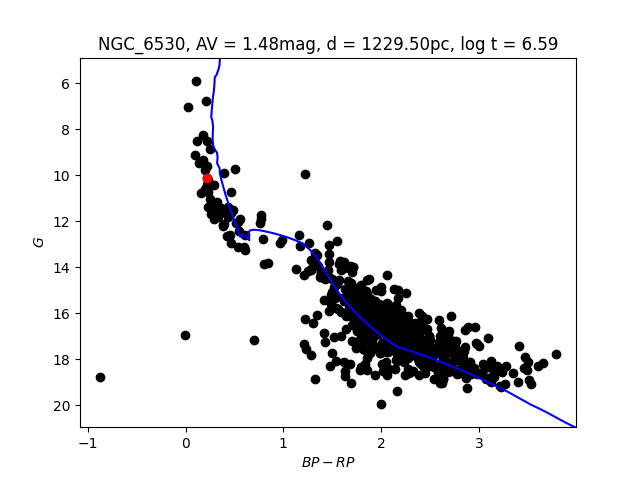}
    \caption{Same as Fig. \ref{fig:cmd_cluster}, but for NGC 6530.}
    \label{fig:enter-label}
\end{figure}

\begin{figure}[ht]
    \centering
    \includegraphics[width = \columnwidth]{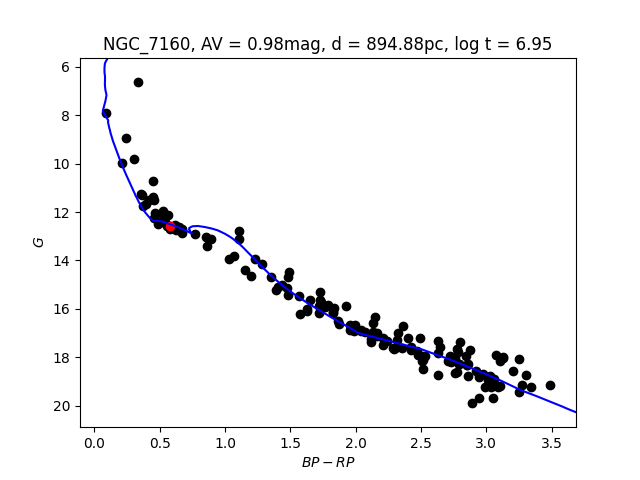}
    \caption{Same as Fig. \ref{fig:cmd_cluster}, but for NGC 7160.}
    \label{fig:enter-label}
\end{figure}

\begin{figure}[ht]
    \centering
    \includegraphics[width = \columnwidth]{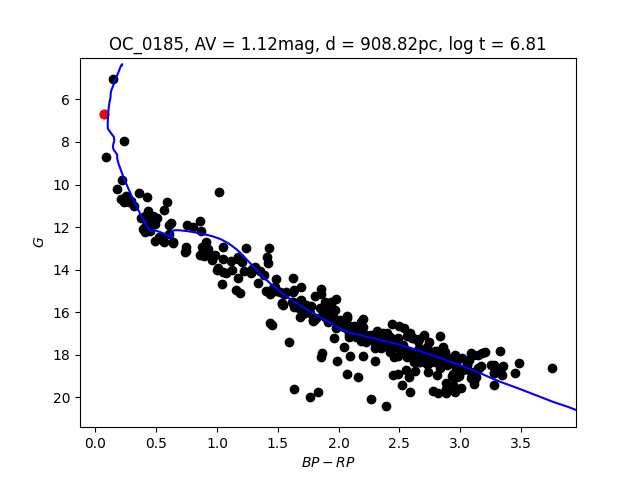}
    \caption{Same as Fig. \ref{fig:cmd_cluster}, but for OC 0185.}
    \label{fig:enter-label}
\end{figure}

\begin{figure}[ht]
    \centering
    \includegraphics[width = \columnwidth]{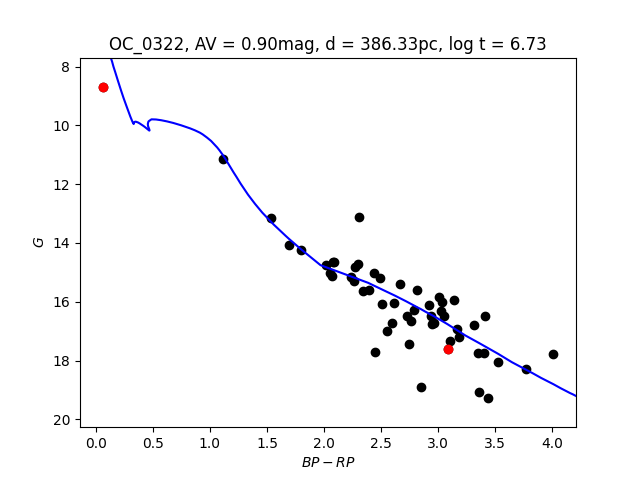}
    \caption{Same as Fig. \ref{fig:cmd_cluster}, but for OC 0322.}
    \label{fig:oc_322}
\end{figure}

\begin{figure}[ht]
    \centering
    \includegraphics[width = \columnwidth]{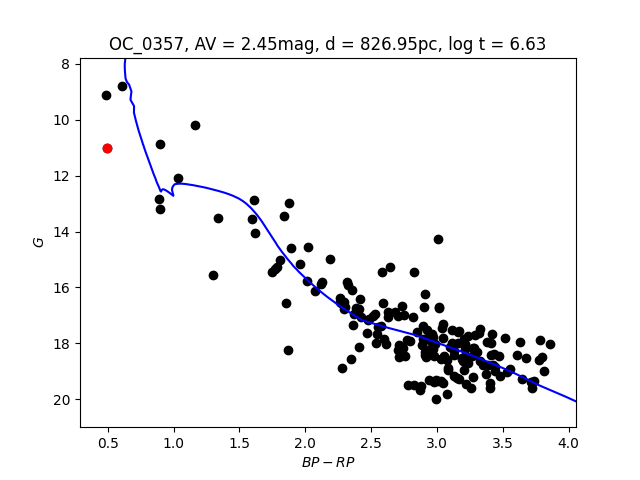}
    \caption{Same as Fig. \ref{fig:cmd_cluster}, but for OC 0357.}
    \label{fig:enter-label}
\end{figure}

\begin{figure}[ht]
    \centering
    \includegraphics[width = \columnwidth]{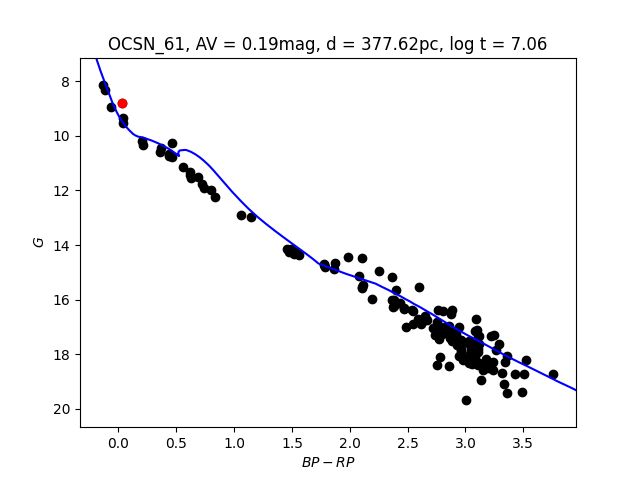}
    \caption{Same as Fig. \ref{fig:cmd_cluster}, but for OCSN 61.}
    \label{fig:enter-label}
\end{figure}

\begin{figure}[ht]
    \centering
    \includegraphics[width = \columnwidth]{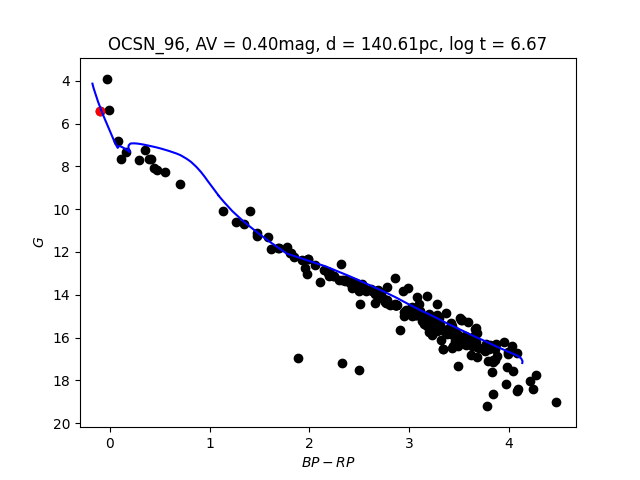}
    \caption{Same as Fig. \ref{fig:cmd_cluster}, but for OCSN 96.}
    \label{fig:enter-label}
\end{figure}

\begin{figure}[ht]
    \centering
    \includegraphics[width = \columnwidth]{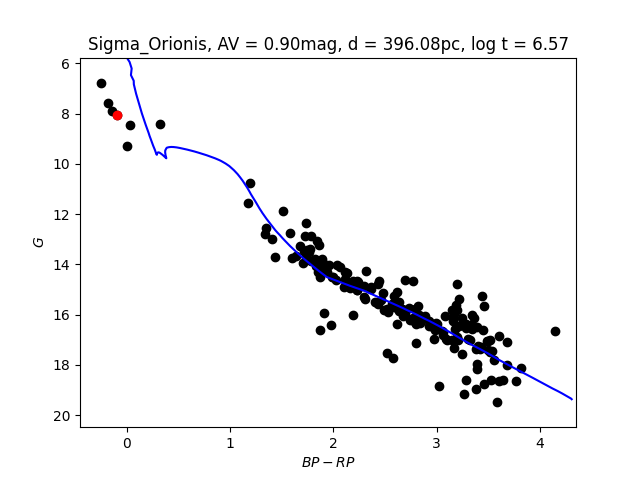}
    \caption{Same as Fig. \ref{fig:cmd_cluster}, but for Sigma Orionis.}
    \label{fig:enter-label}
\end{figure}

\begin{figure}[ht]
    \centering
    \includegraphics[width = \columnwidth]{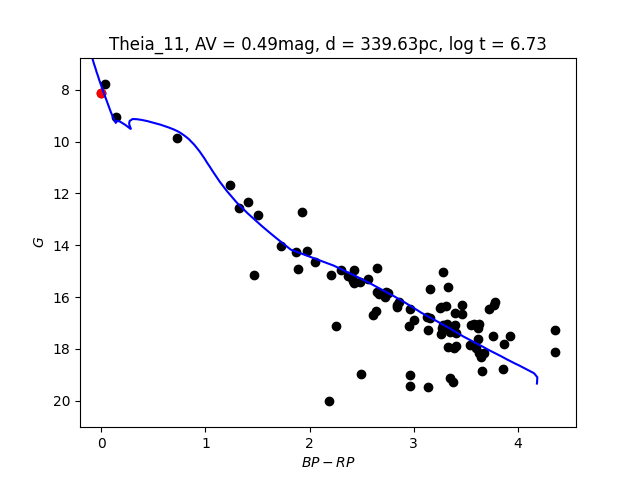}
    \caption{Same as Fig. \ref{fig:cmd_cluster}, but for Theia 11.}
    \label{fig:enter-label}
\end{figure}

\begin{figure}[ht]
    \centering
    \includegraphics[width = \columnwidth]{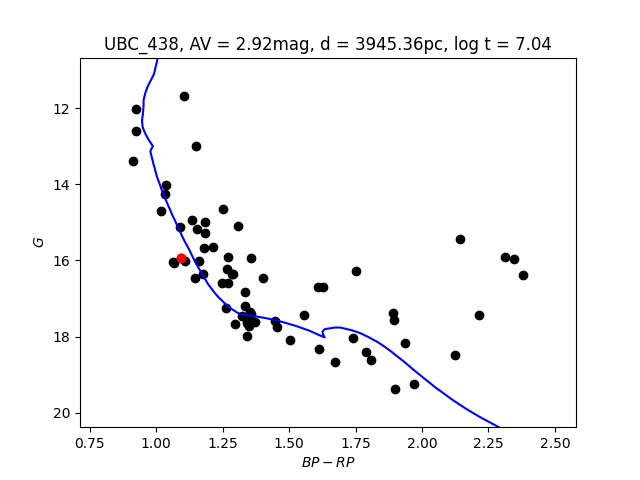}
    \caption{Same as Fig. \ref{fig:cmd_cluster}, but for UBC 438.}
    \label{fig:enter-label}
\end{figure}

\begin{figure}[ht]
    \centering
    \includegraphics[width = \columnwidth]{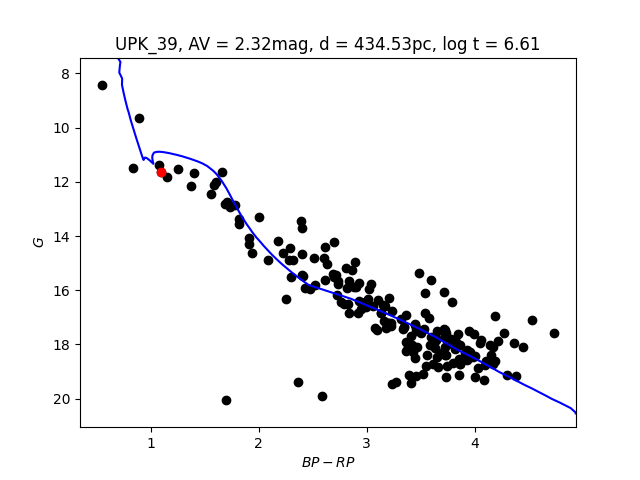}
    \caption{Same as Fig. \ref{fig:cmd_cluster}, but for UPK 39.}
    \label{fig:enter-label}
\end{figure}

\begin{figure}[ht]
    \centering
    \includegraphics[width = \columnwidth]{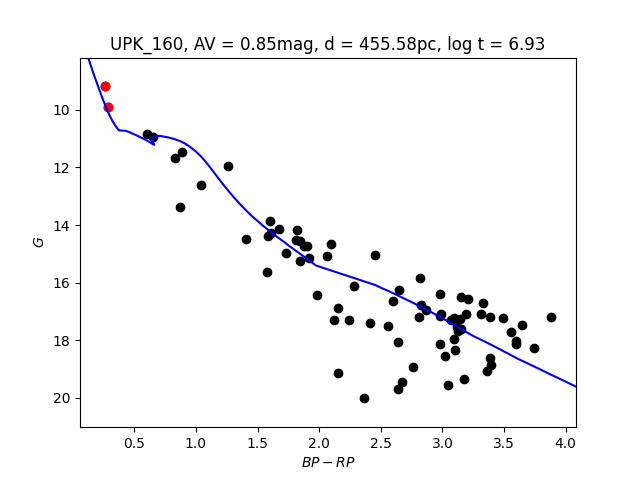}
    \caption{Same as Fig. \ref{fig:cmd_cluster}, but for UPK 160.}
    \label{fig:enter-label}
\end{figure}

\begin{figure}[ht]
    \centering
    \includegraphics[width = \columnwidth]{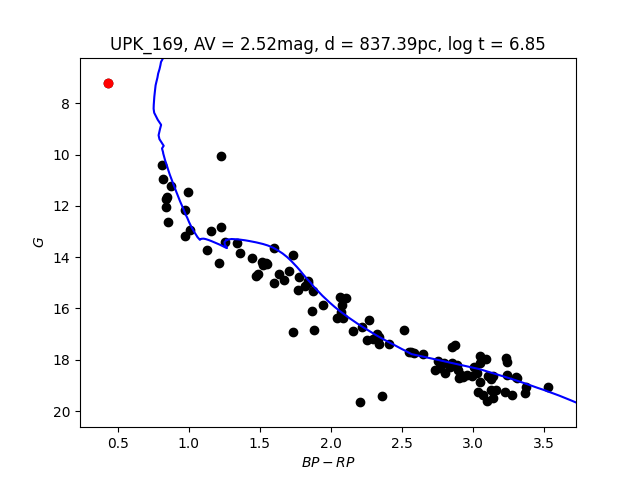}
    \caption{Same as Fig. \ref{fig:cmd_cluster}, but for UPK 169.}
    \label{fig:enter-label}
\end{figure}

\begin{figure}[ht]
    \centering
    \includegraphics[width = \columnwidth]{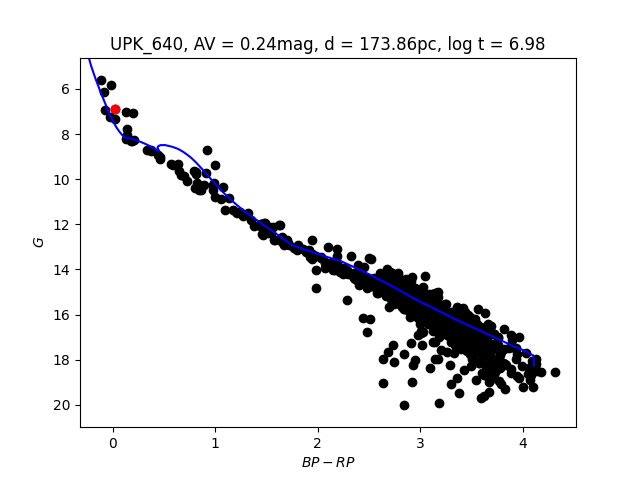}
    \caption{Same as Fig. \ref{fig:cmd_cluster}, but for UPK 640.}
    \label{fig:enter-label}
\end{figure}

\begin{figure}[ht]
    \centering
    \includegraphics[width = \columnwidth]{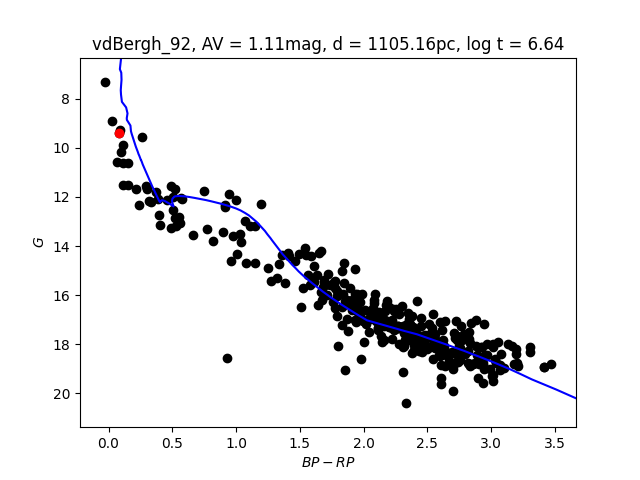}
    \caption{Same as Fig. \ref{fig:cmd_cluster}, but for vdBergh 92.}
    \label{fig:enter-label}
\end{figure}

\clearpage

\section{TESS light curves}\label{appendix_lc_tess}

\begin{figure}[ht]
    \centering
    \includegraphics[width = \columnwidth]{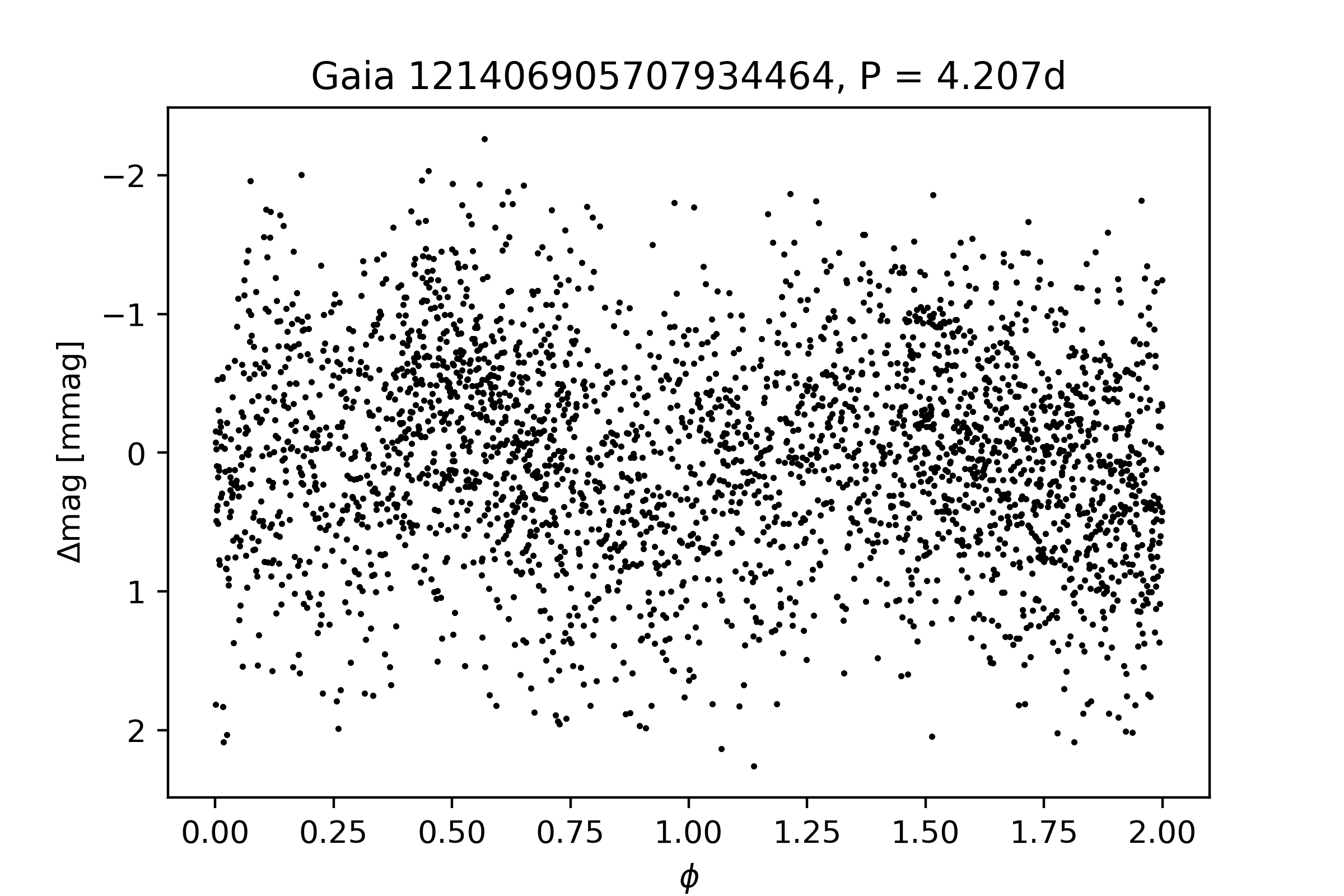}
    \caption{Known Pre-MS CP star without discernable variability in the TESS light curve. In the literature, it is an ACV variable with a period of about 123d.}
    \label{fig:lc_gaia_121406905707934464}
\end{figure}

\begin{figure}[ht]
    \centering
    \includegraphics[width = \columnwidth]{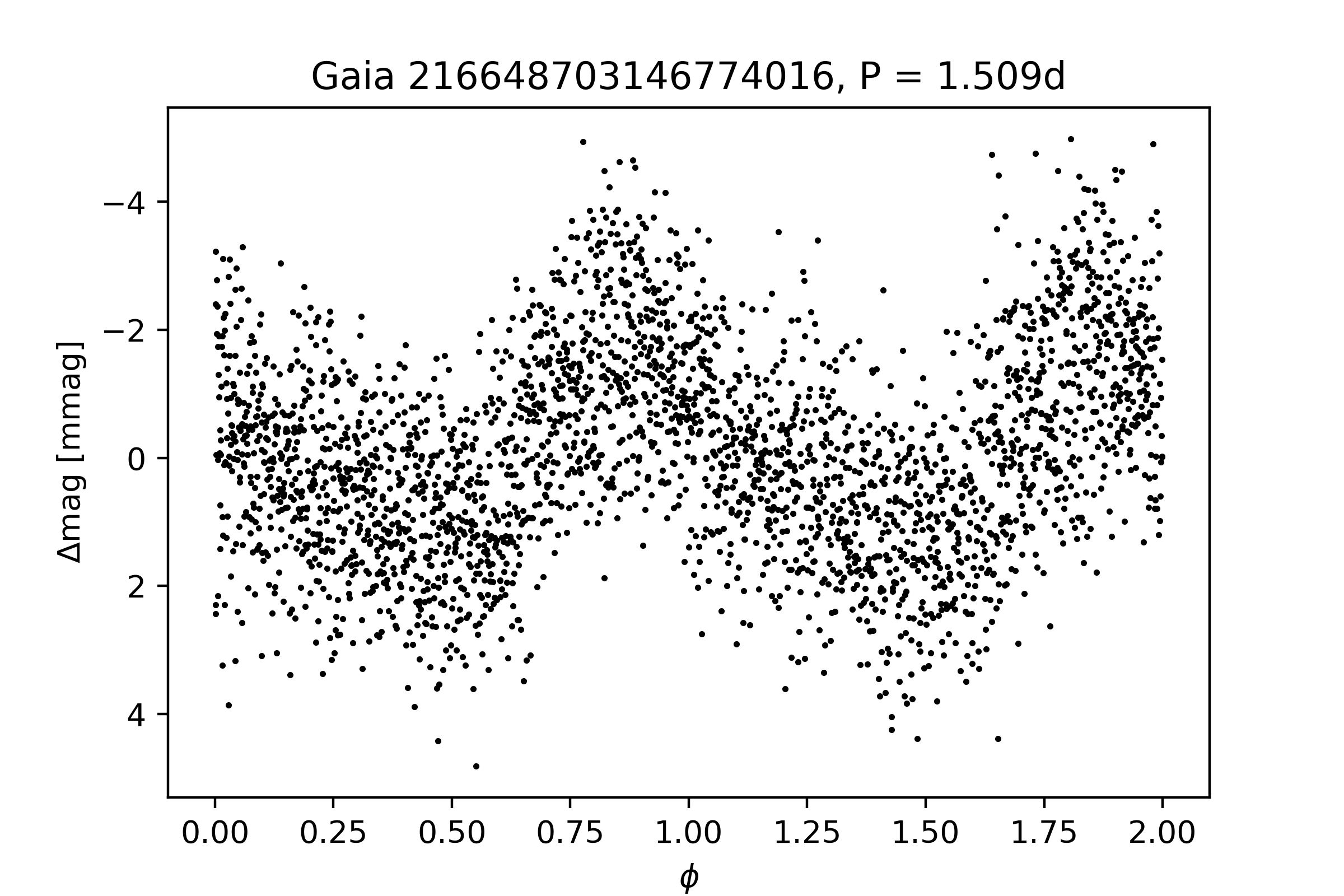}
    \caption{Mixture of two pulsational variability types, GDOR and DSCT in the light curve. The main pulsation with a period of 1.509d comes from the GDOr part.}
    \label{fig:enter-label}
\end{figure}

\begin{figure}[ht]
    \centering
    \includegraphics[width = \columnwidth]{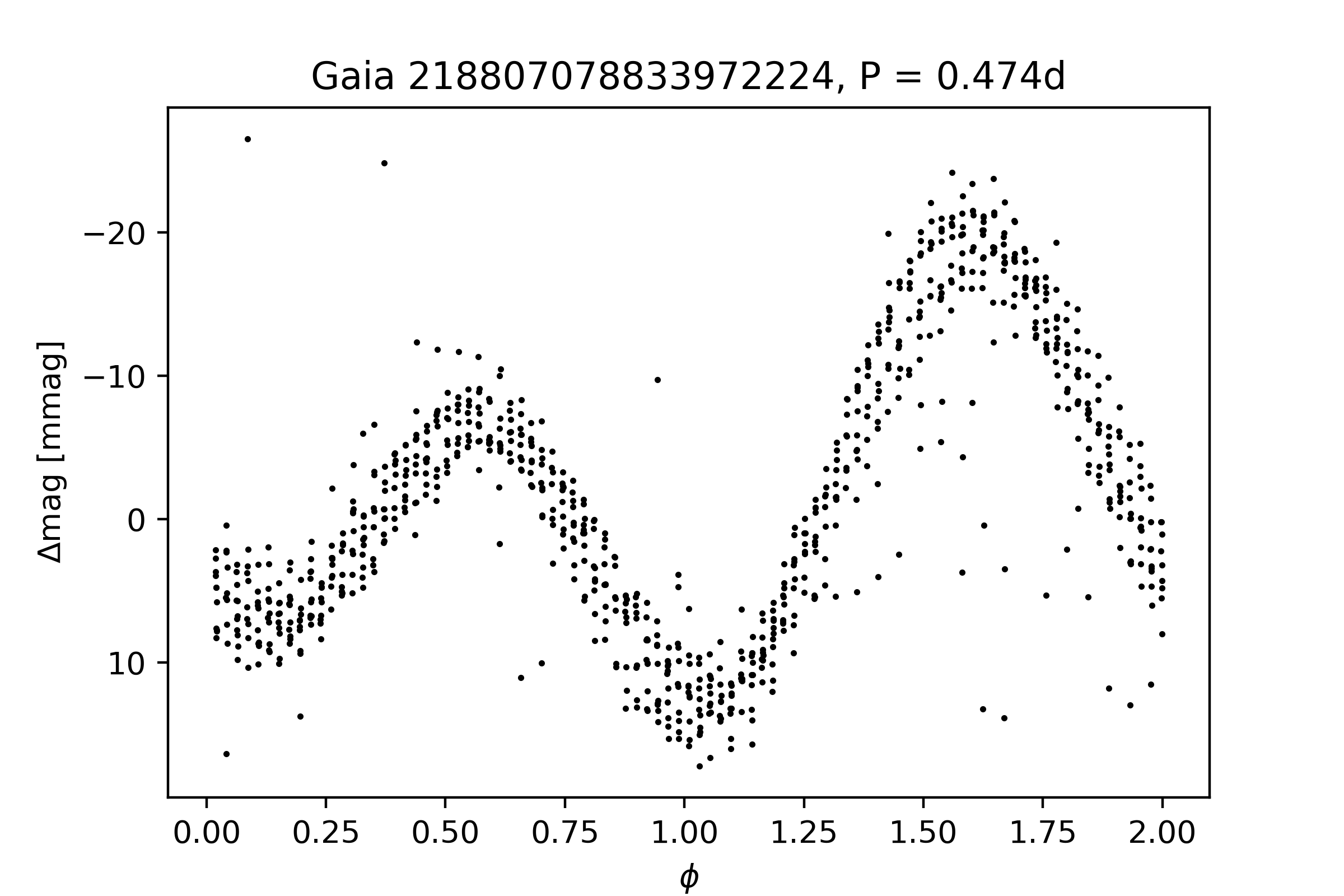}
    \caption{Clear CP star. Based on the light curve and spectrum, we conclude that the variability type is SXARI. The star is folded at half the period of 0.948d.}
    \label{fig:enter-label}
\end{figure}

\begin{figure}[ht]
    \centering
    \includegraphics[width = \columnwidth]{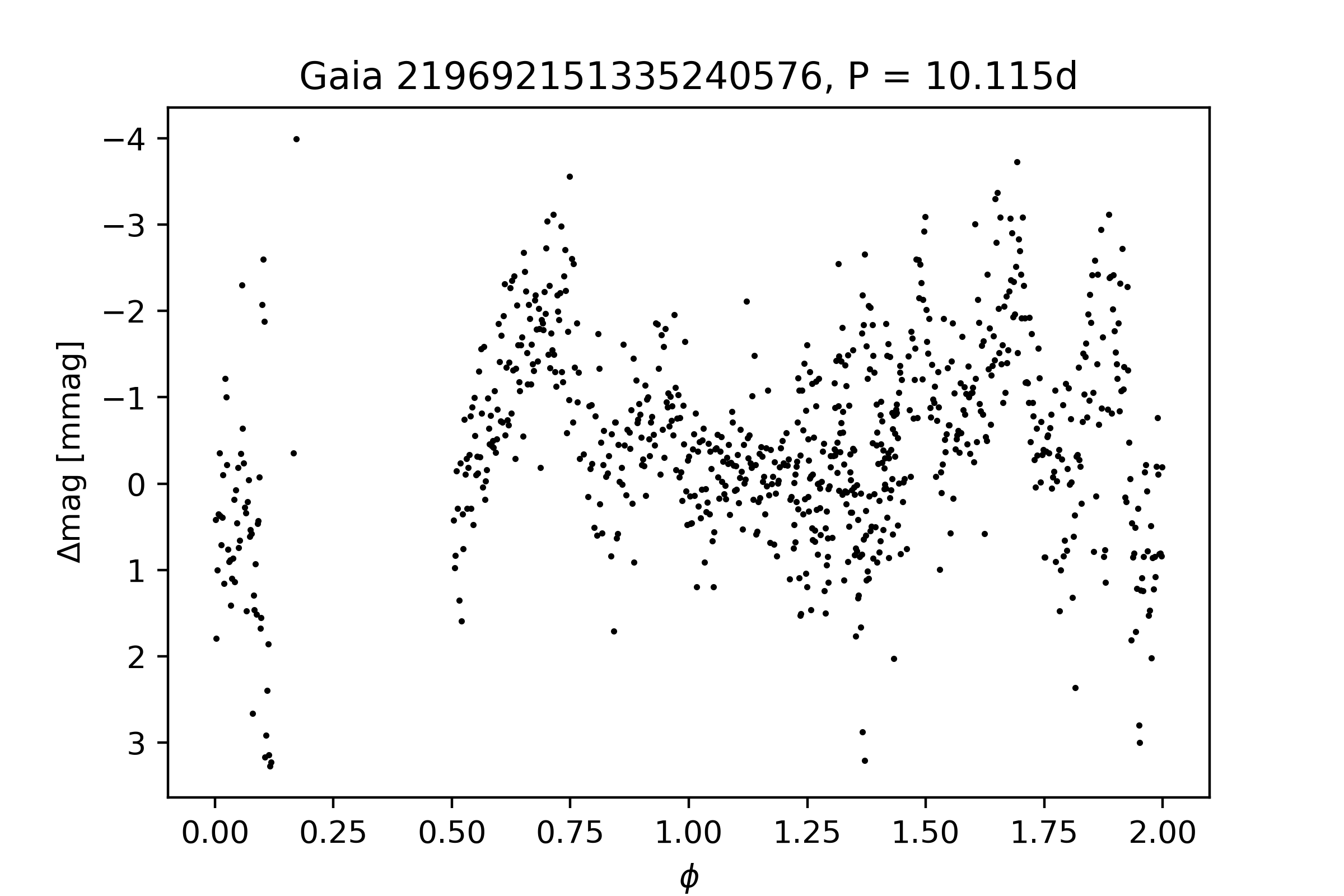}
    \caption{Irregular variability. However, it is possible that some pulsation or rotation feature is present. The low amplitude of the variation prevents us from concluding.}
    \label{fig:enter-label}
\end{figure}

\begin{figure}[ht]
    \centering
    \includegraphics[width = \columnwidth]{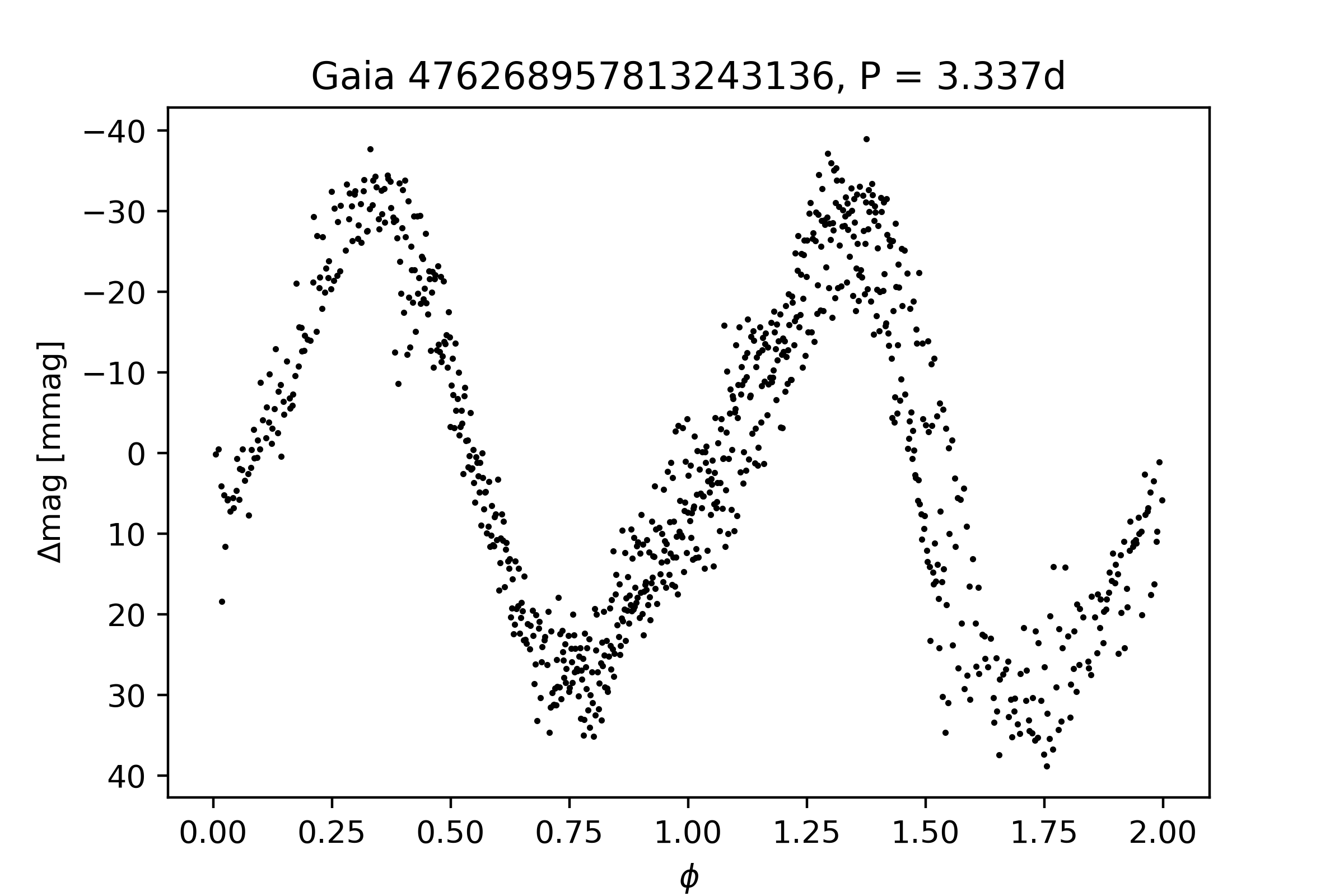}
    \caption{Clearly rotating variable star, probably of the type ACV.}
    \label{fig:enter-label}
\end{figure}

\begin{figure}[ht]
    \centering
    \includegraphics[width = \columnwidth]{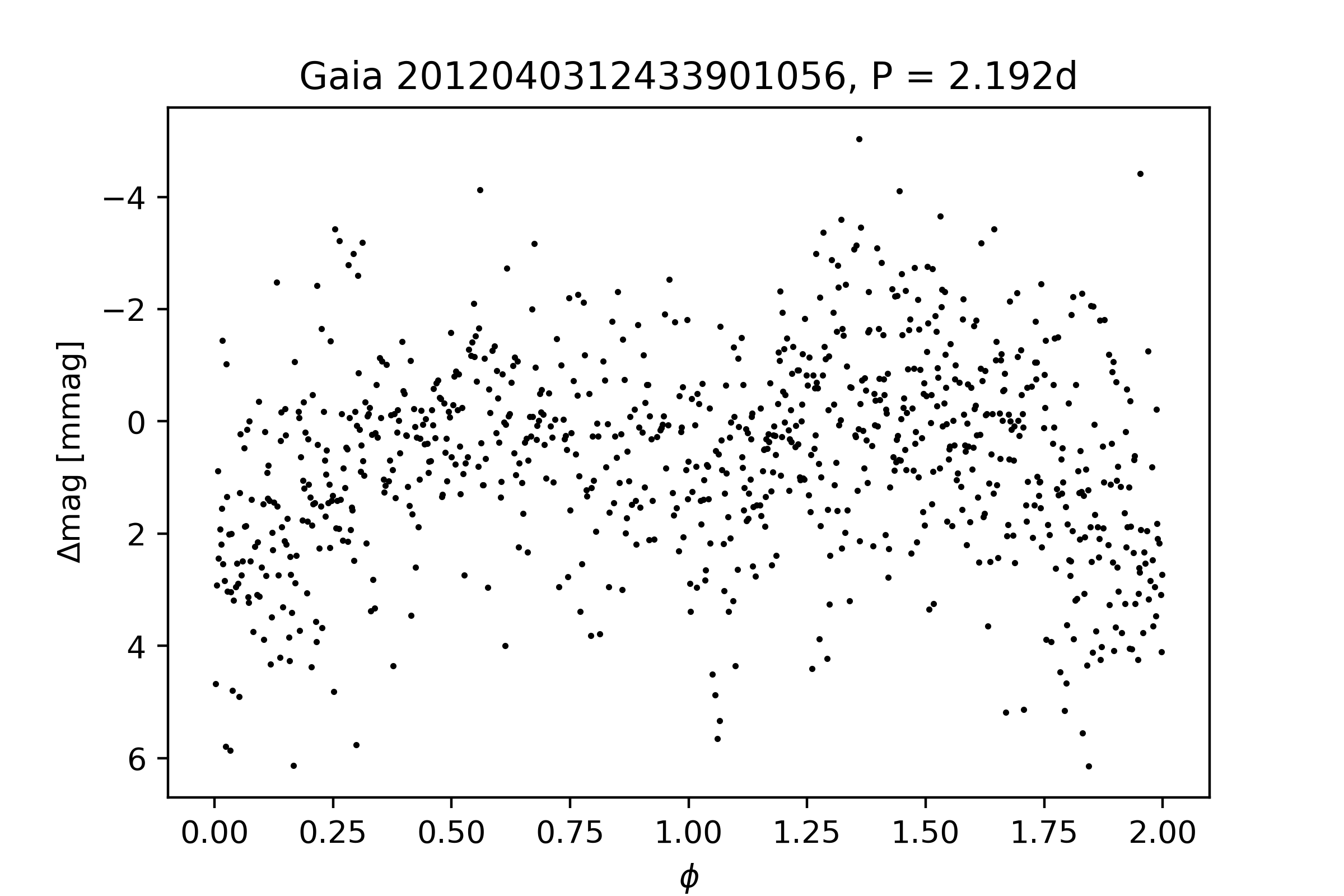}
    \caption{Possible binary star.}
    \label{fig:enter-label}
\end{figure}

\begin{figure}[ht]
    \centering
    \includegraphics[width = \columnwidth]{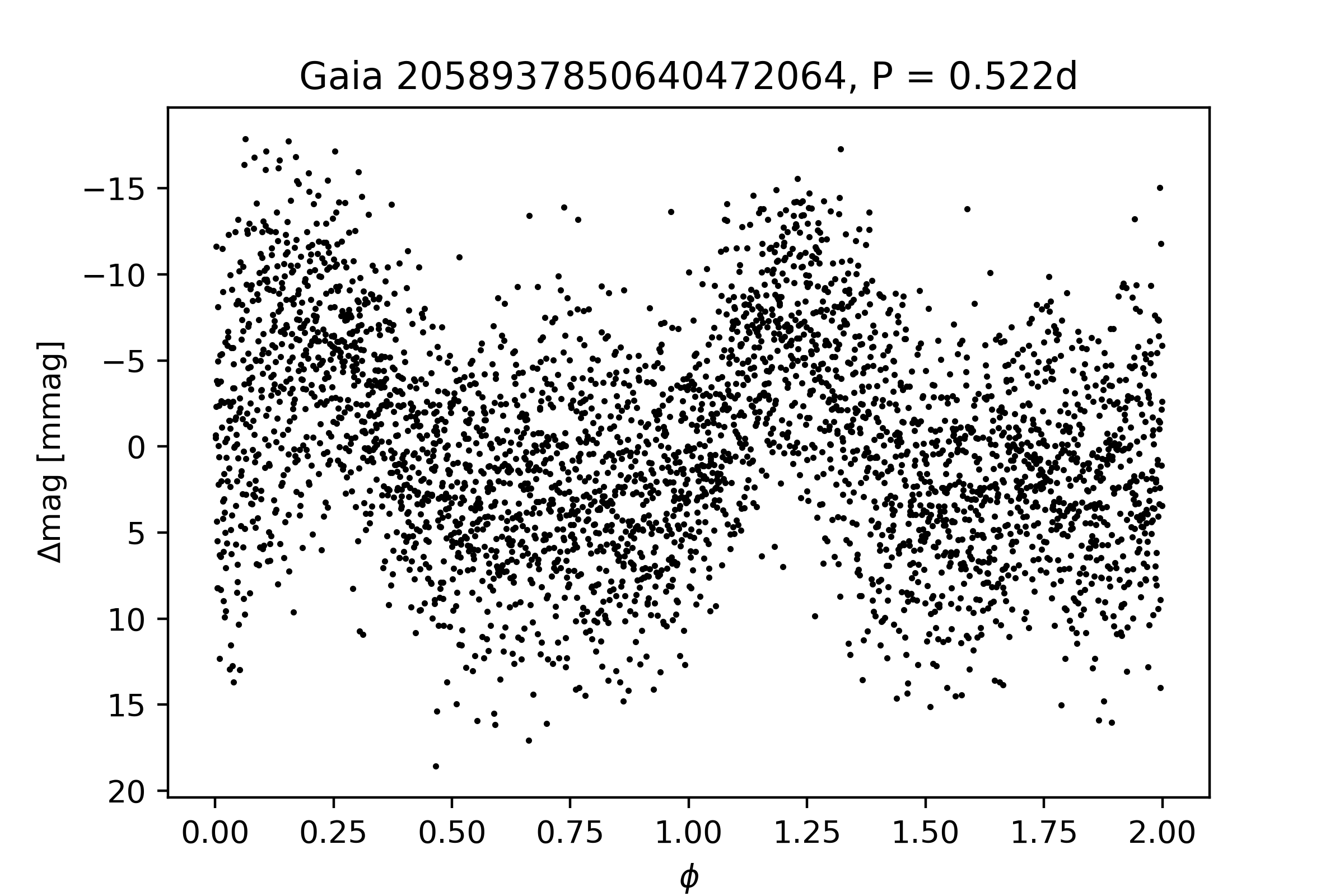}
    \caption{Another ACV variable. A hint of a double wave is visible.}
    \label{fig:enter-label}
\end{figure}

\begin{figure}[ht]
    \centering
    \includegraphics[width = \columnwidth]{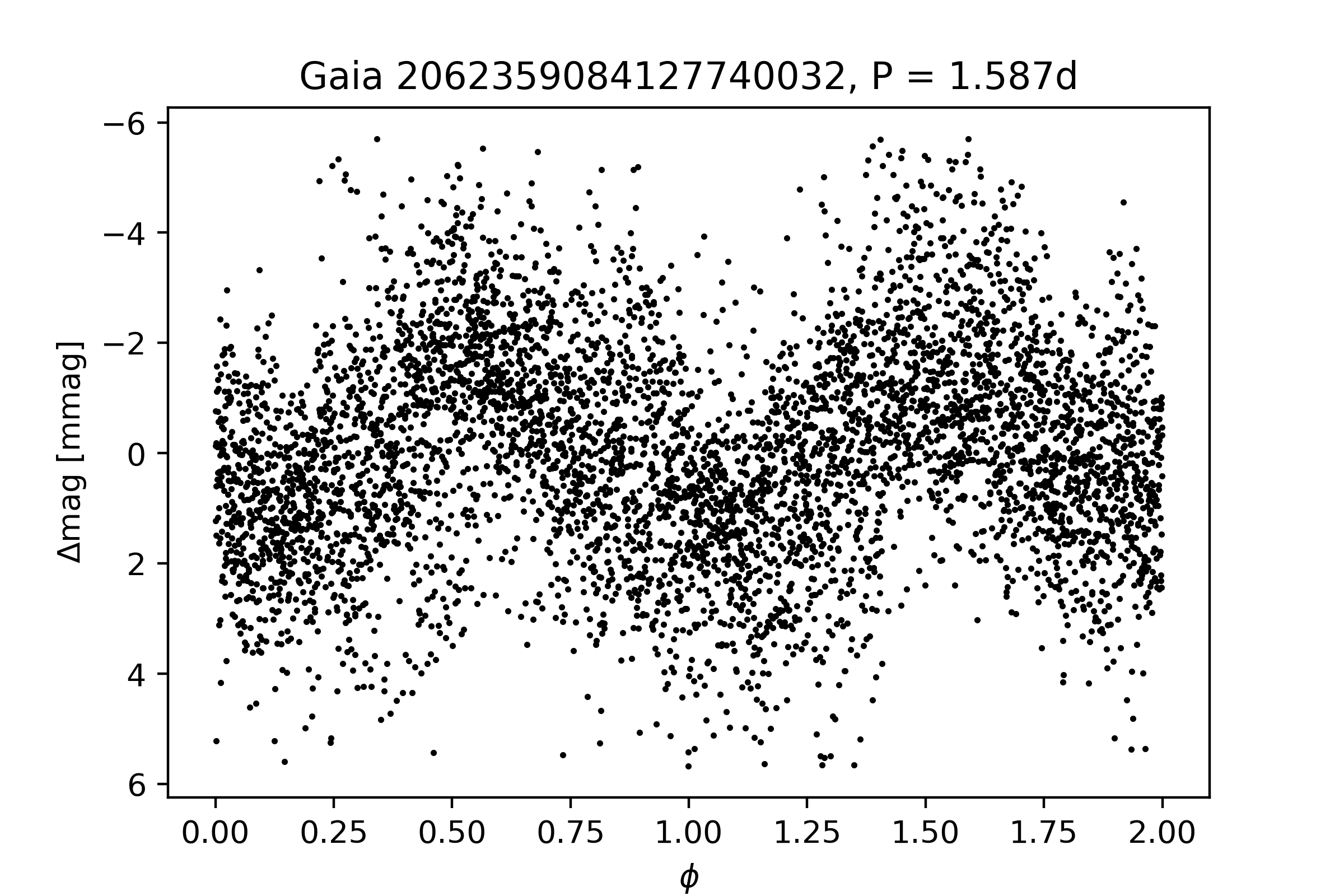}
    \caption{Slowly pulsating B (SPB) star.}
    \label{fig:enter-label}
\end{figure}

\begin{figure}[ht]
    \centering
    \includegraphics[width = \columnwidth]{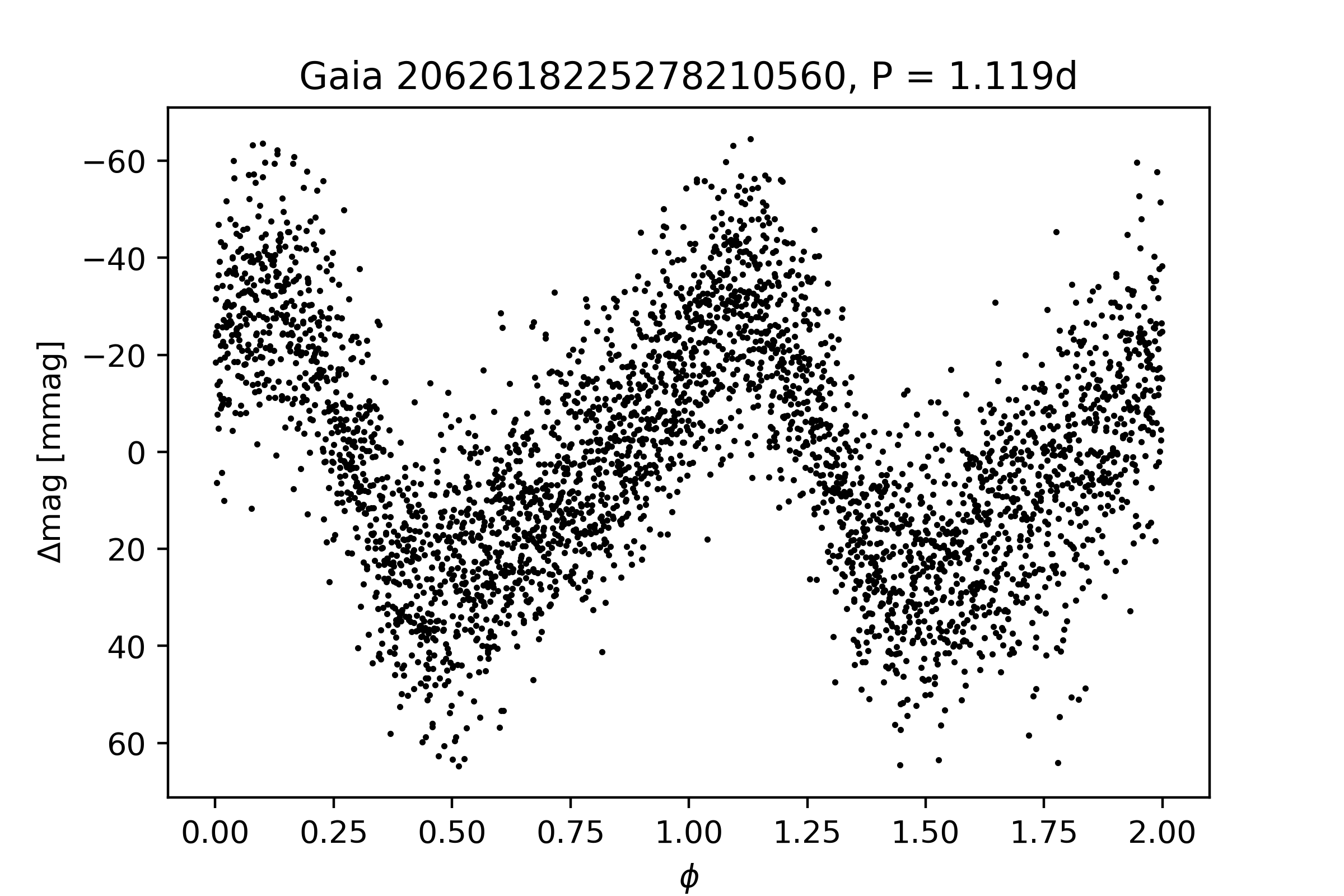}
    \caption{ACV variable}
    \label{fig:enter-label}
\end{figure}

\begin{figure}[ht]
    \centering
    \includegraphics[width = \columnwidth]{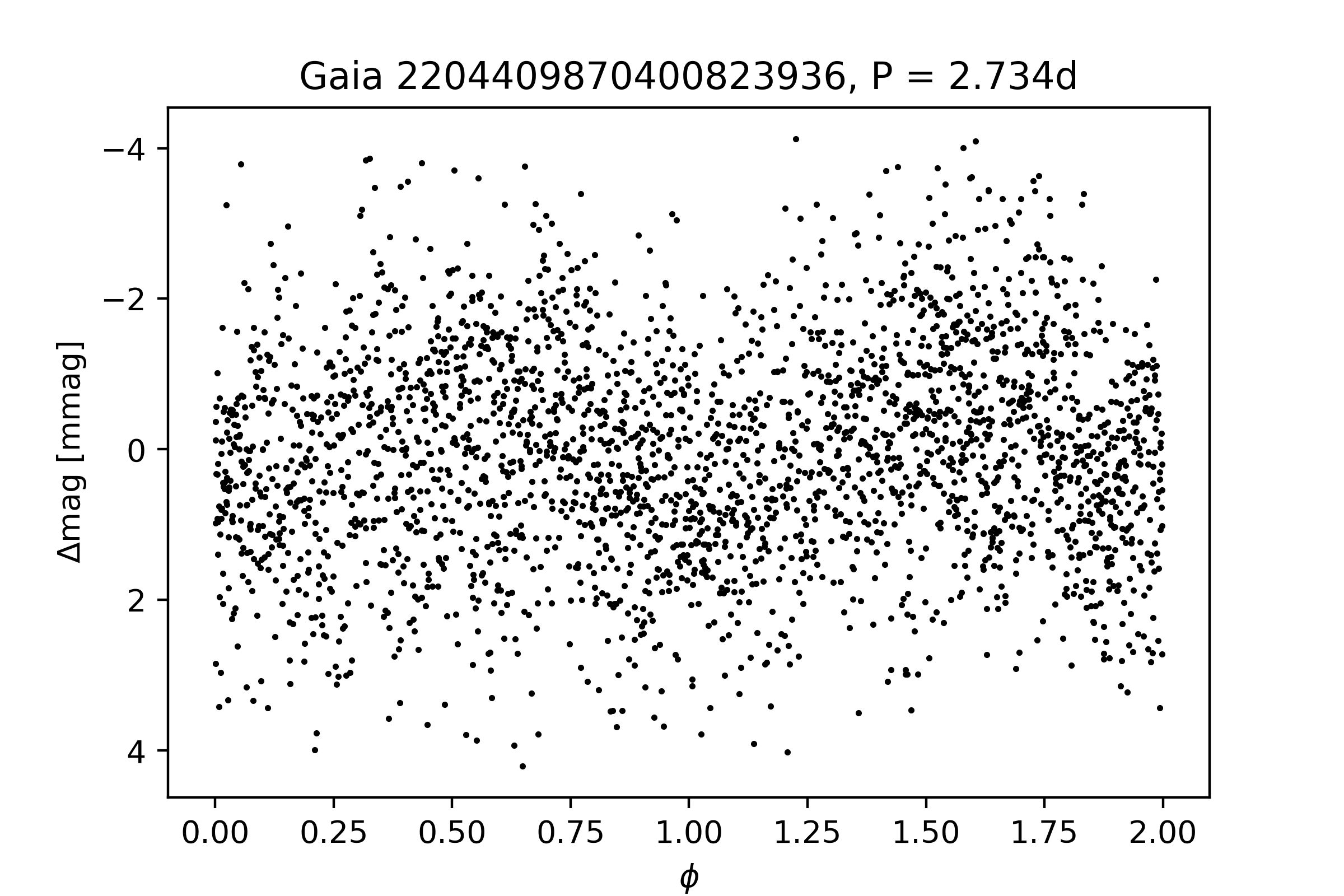}
    \caption{Likely a pulsating star of SPB type.}
    \label{fig:enter-label}
\end{figure}

\begin{figure}[ht]
    \centering
    \includegraphics[width = \columnwidth]{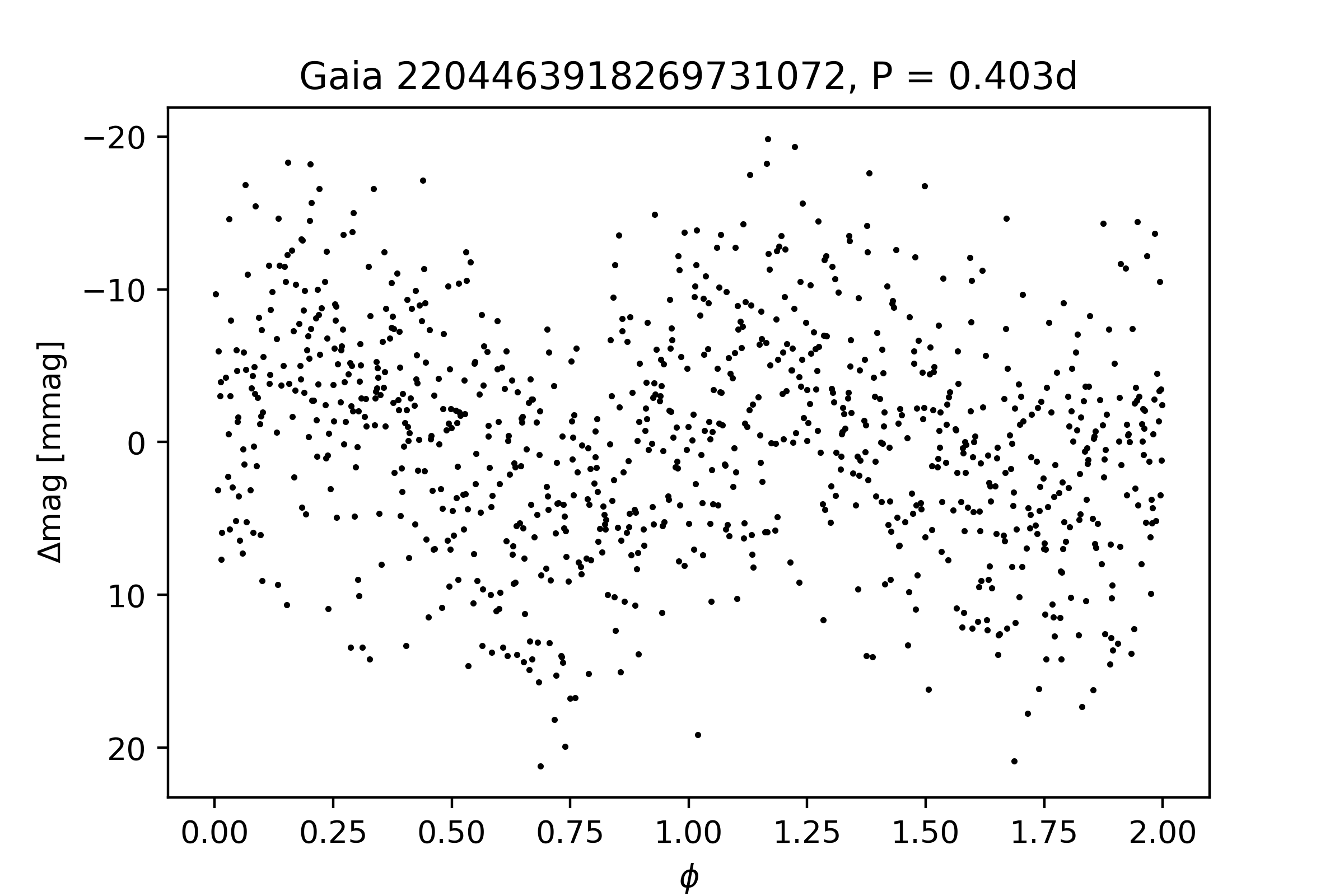}
    \caption{Rotating variable star.}
    \label{fig:enter-label}
\end{figure}

\begin{figure}[ht]
    \centering
    \includegraphics[width = \columnwidth]{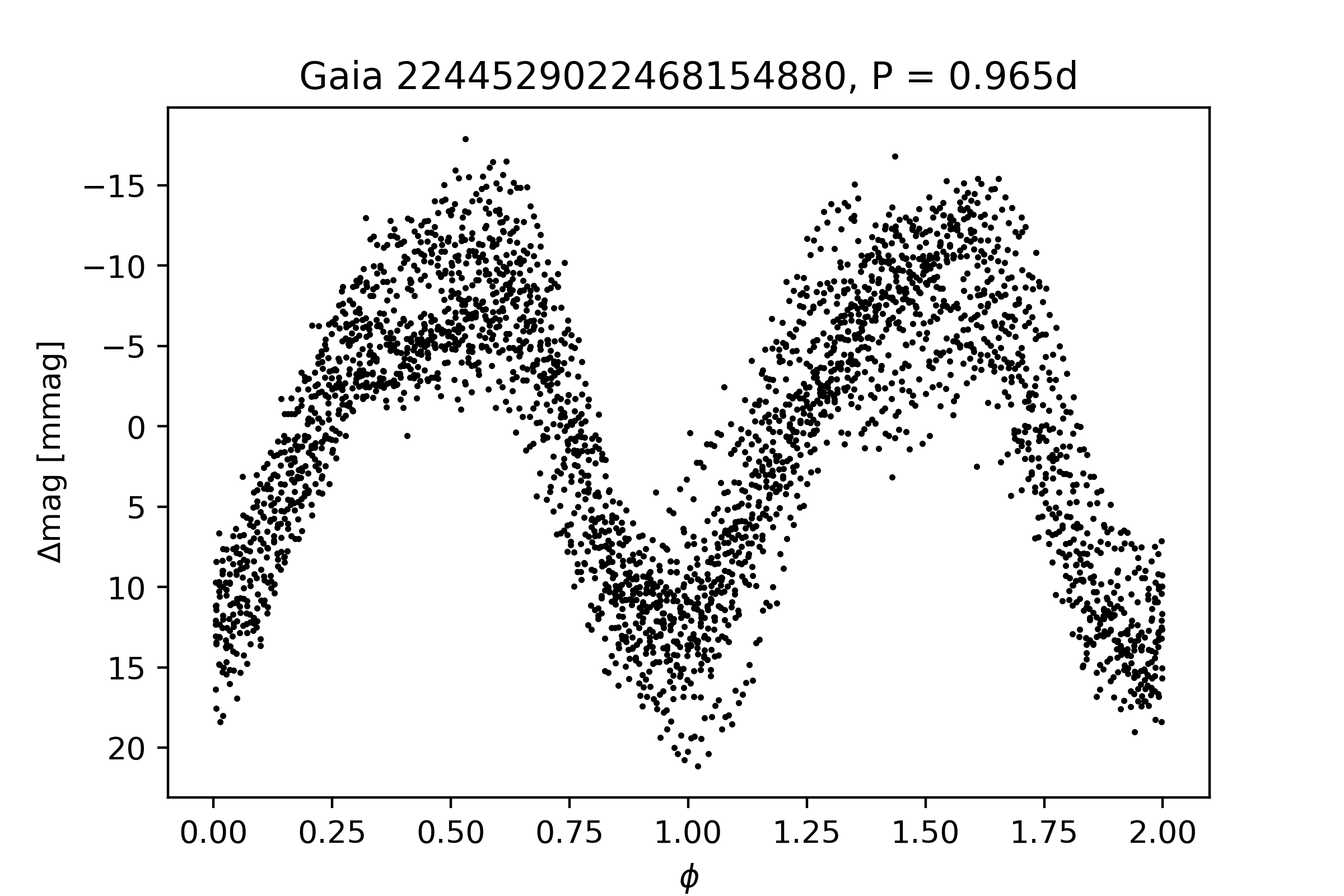}
    \caption{ACV variable.}
    \label{fig:enter-label}
\end{figure}

\begin{figure}[ht]
    \centering
    \includegraphics[width = \columnwidth]{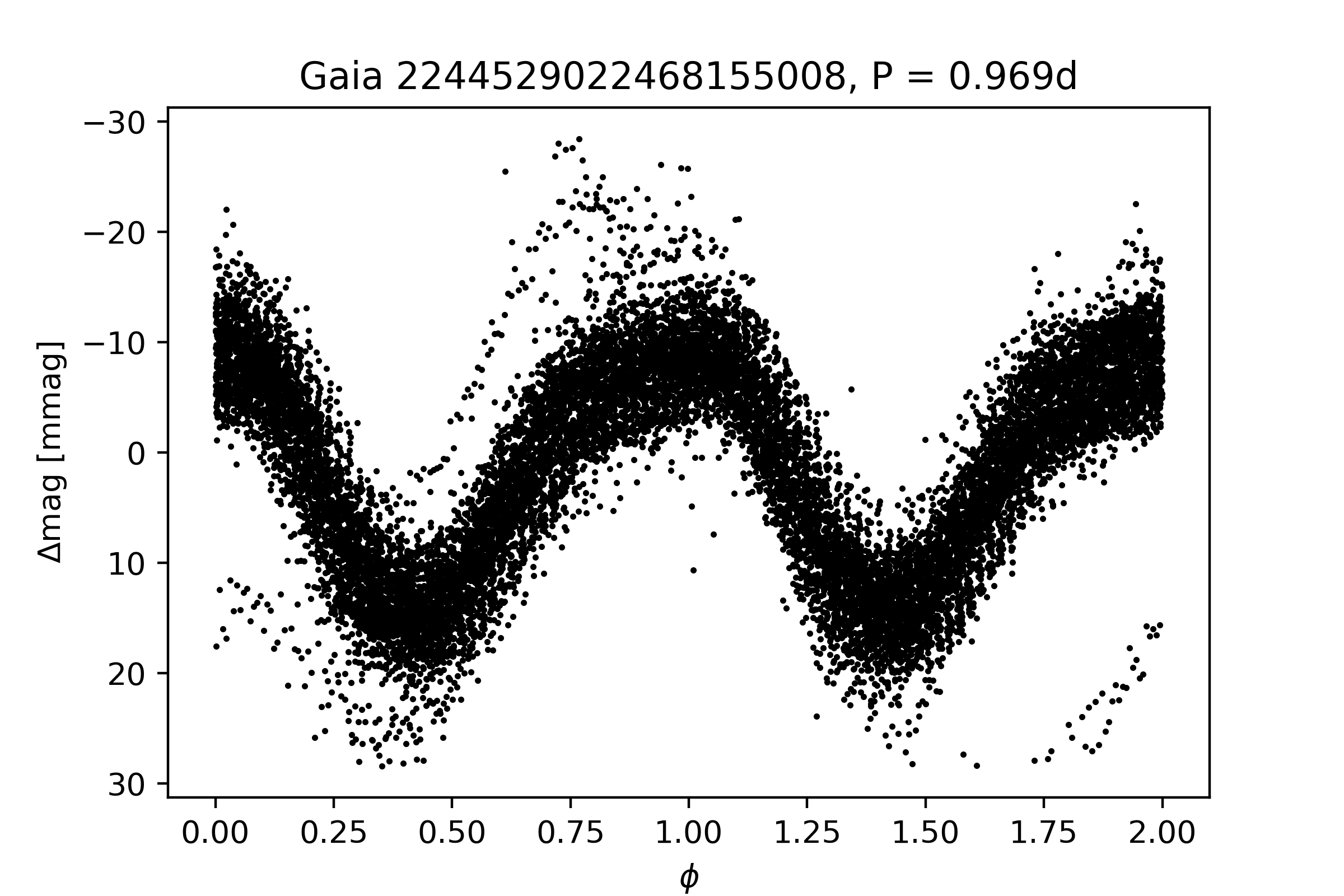}
    \caption{Star belonging to the group of SXARI stars.}
    \label{fig:enter-label}
\end{figure}

\begin{figure}[ht]
    \centering
    \includegraphics[width = \columnwidth]{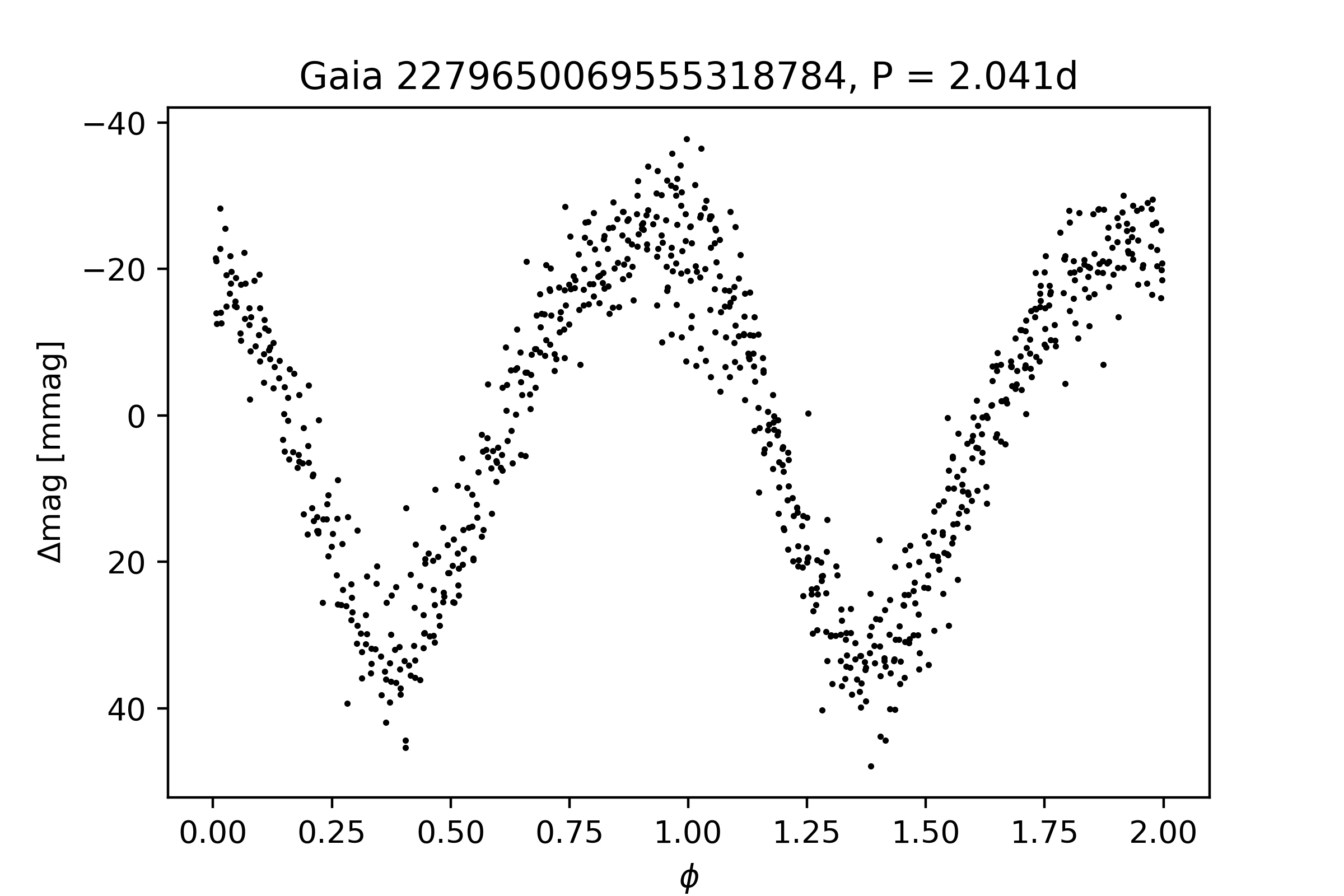}
    \caption{Basic rotation feature with slightly asymmetric sine waves. An ACV variable.}
    \label{fig:enter-label}
\end{figure}

\begin{figure}[ht]
    \centering
    \includegraphics[width = \columnwidth]{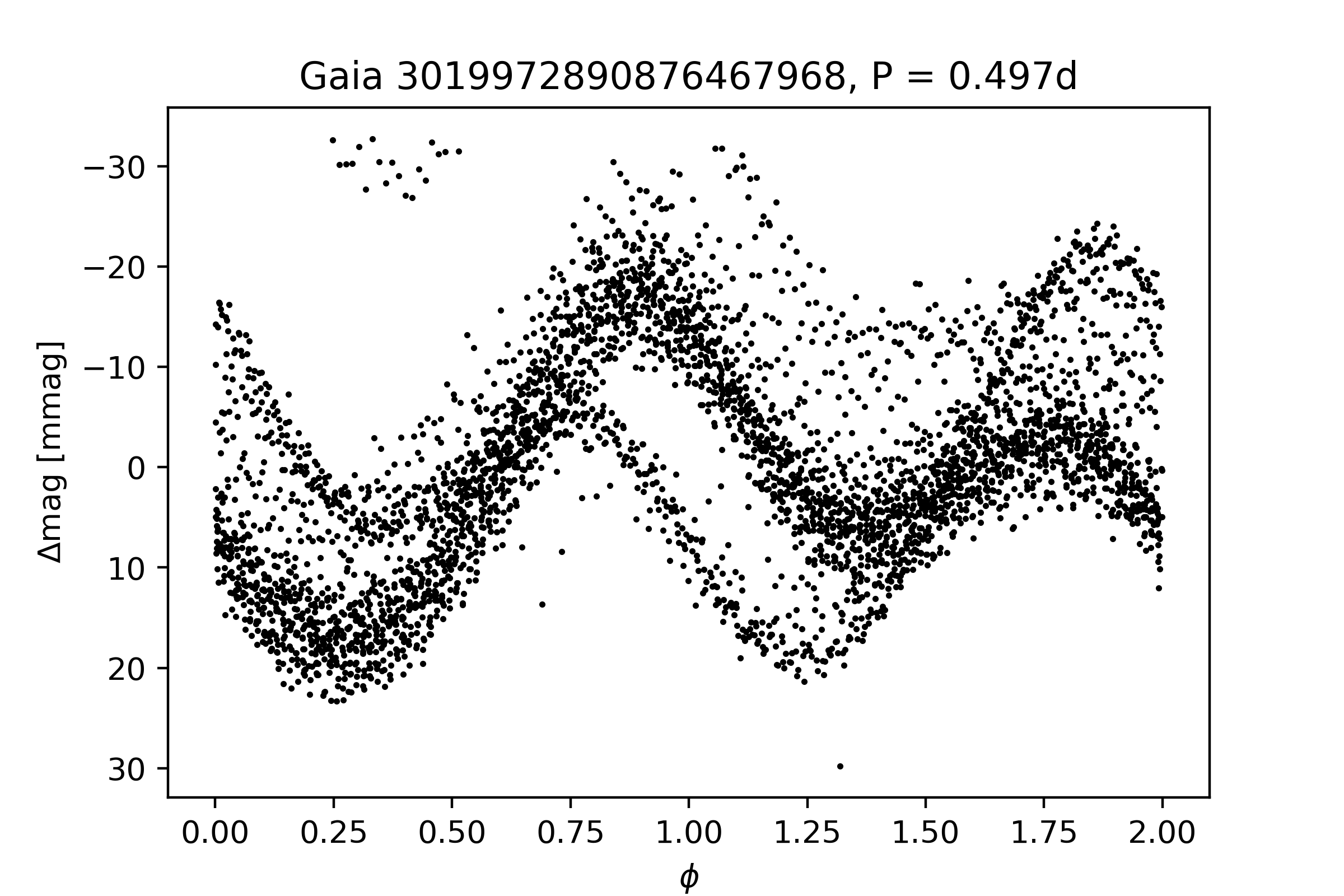}
    \caption{Double wave of this ACV variable, folded at half the period.}
    \label{fig:enter-label}
\end{figure}

\begin{figure}[ht]
    \centering
    \includegraphics[width = \columnwidth]{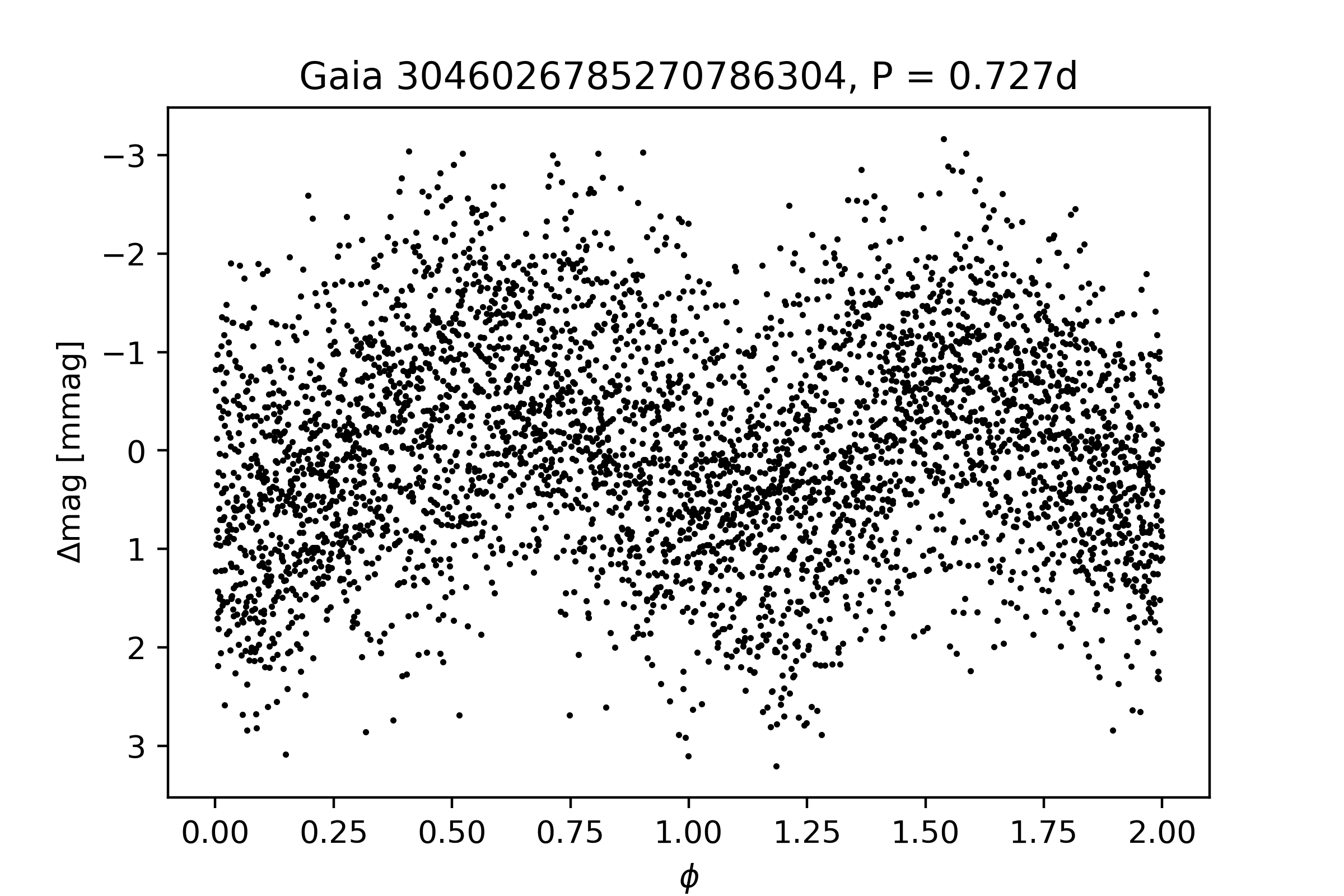}
    \caption{Another SPB star.}
    \label{fig:enter-label}
\end{figure}

\begin{figure}[ht]
    \centering
    \includegraphics[width = \columnwidth]{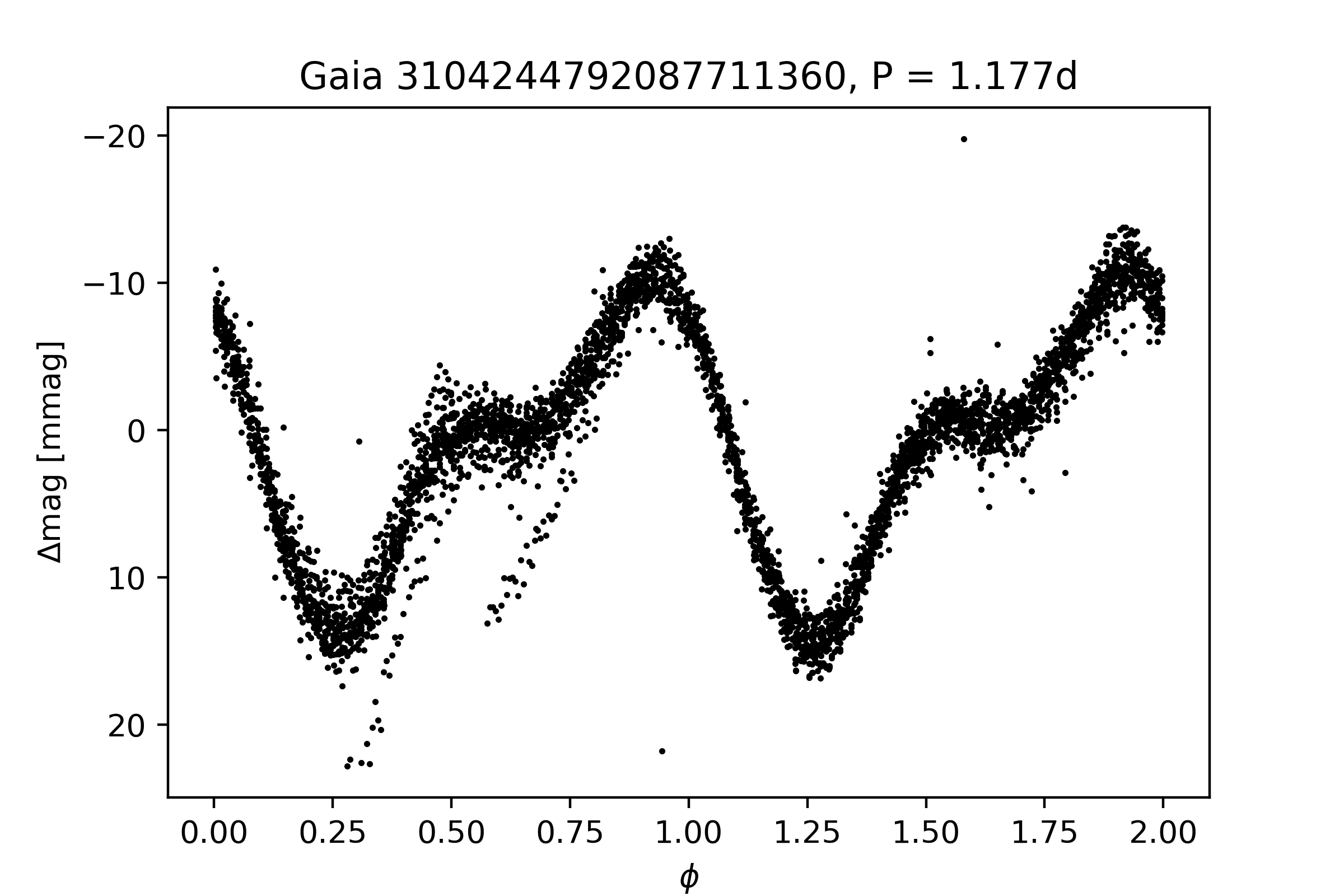}
    \caption{TESS light curve of the known ACV star V682 Mon.}
    \label{fig:blended_star}
\end{figure}

\begin{figure}[ht]
    \centering
    \includegraphics[width = \columnwidth]{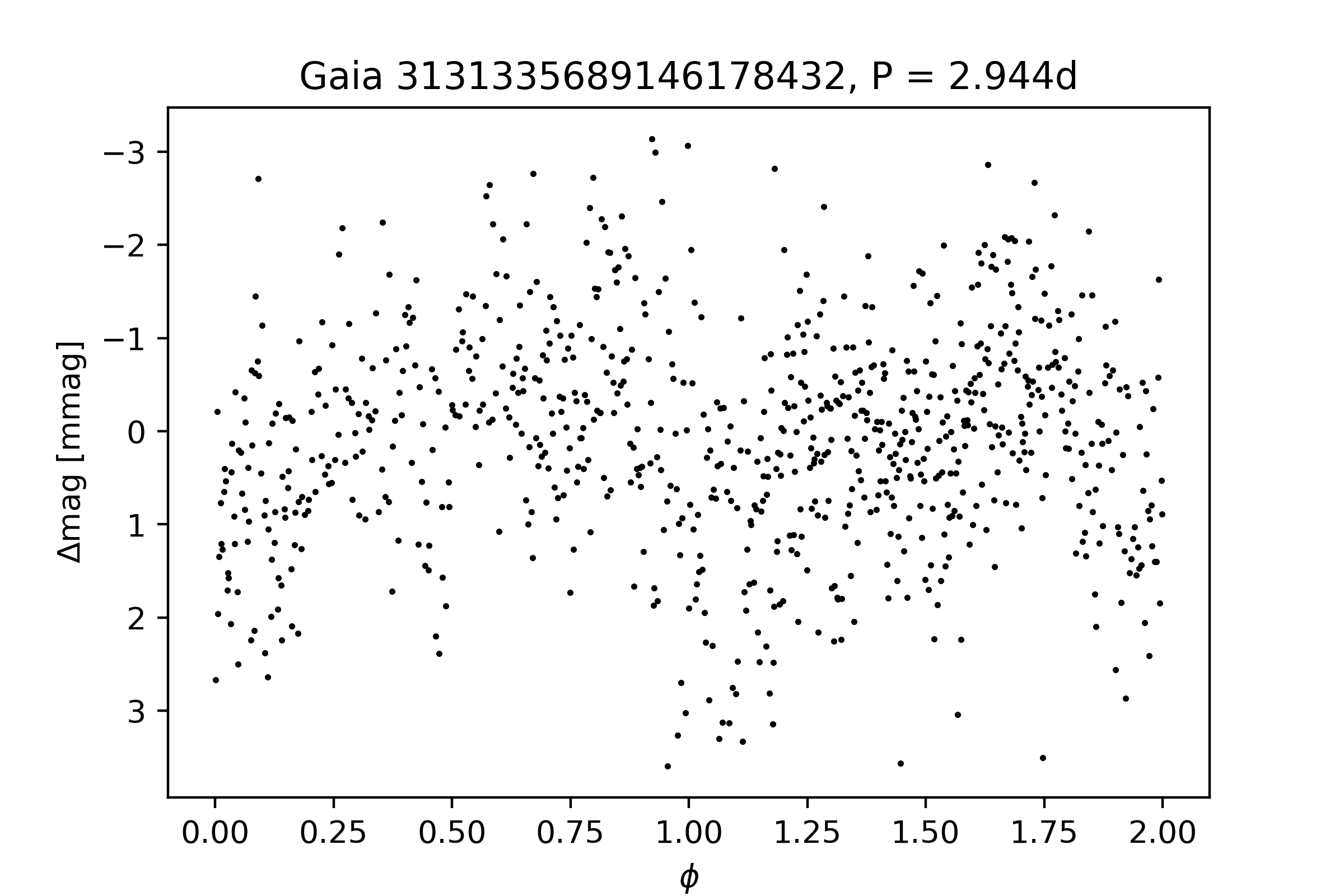}
    \caption{Eclipsing binary star (possibly). However the amplitude of the variability is pretty low.}
    \label{fig:enter-label}
\end{figure}

\begin{figure}[ht]
    \centering
    \includegraphics[width = \columnwidth]{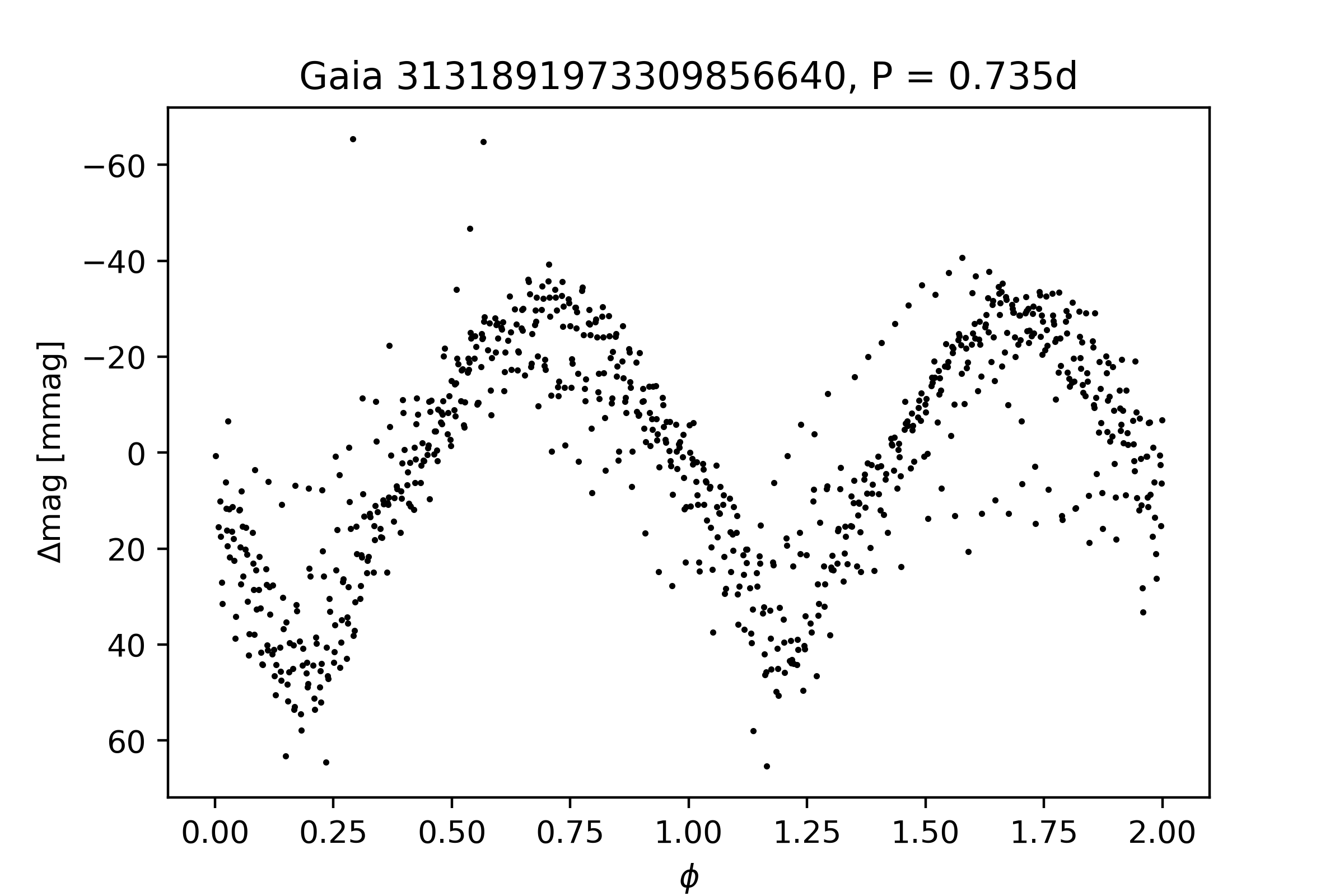}
    \caption{Binary star. The signal likely comes from the close brighter eclipsing variable V649 Mon.}
    \label{fig:enter-label}
\end{figure}

\begin{figure}[ht]
    \centering
    \includegraphics[width = \columnwidth]{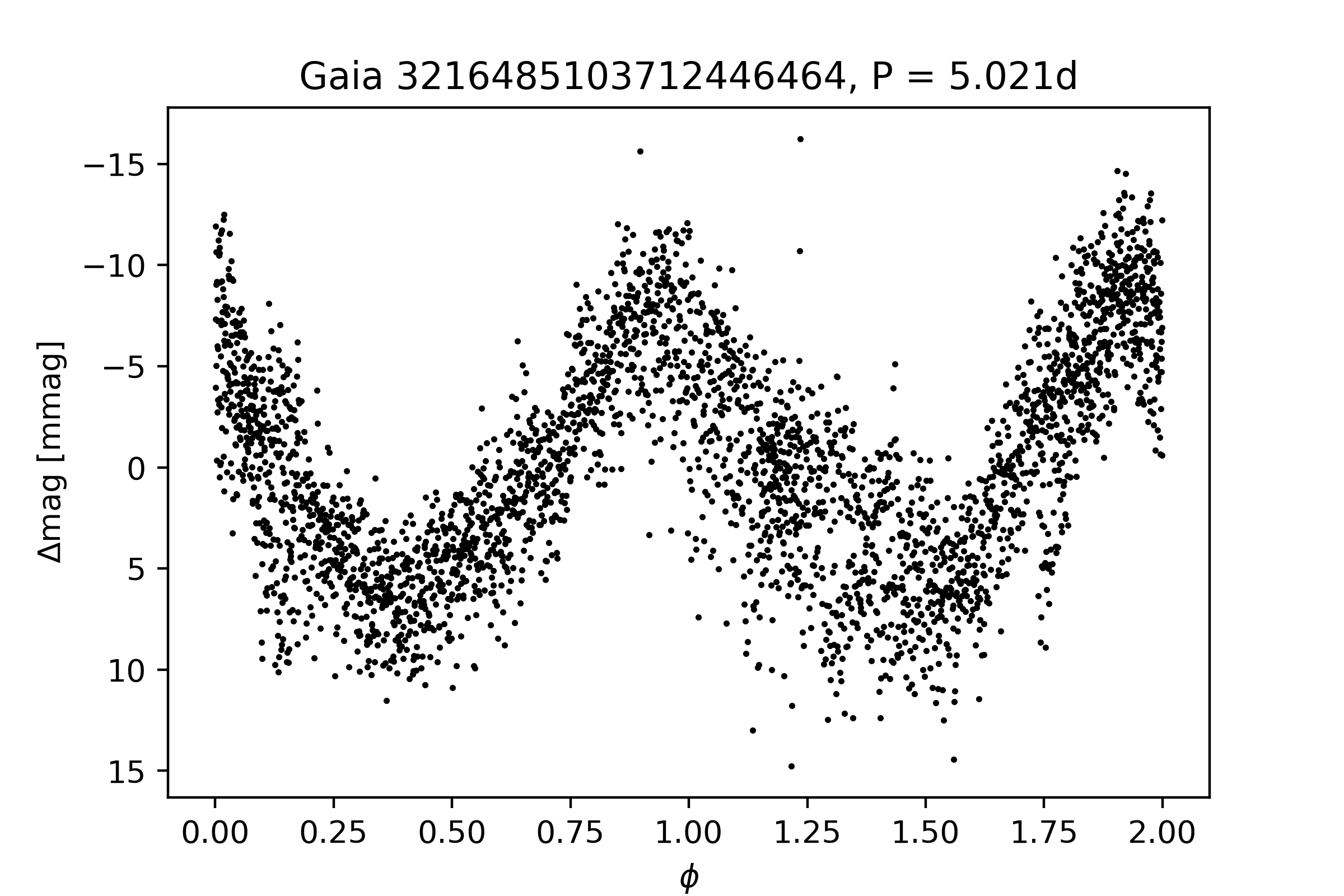}
    \caption{Another ACV variable.}
    \label{fig:enter-label}
\end{figure}

\begin{figure}[ht]
    \centering
    \includegraphics[width = \columnwidth]{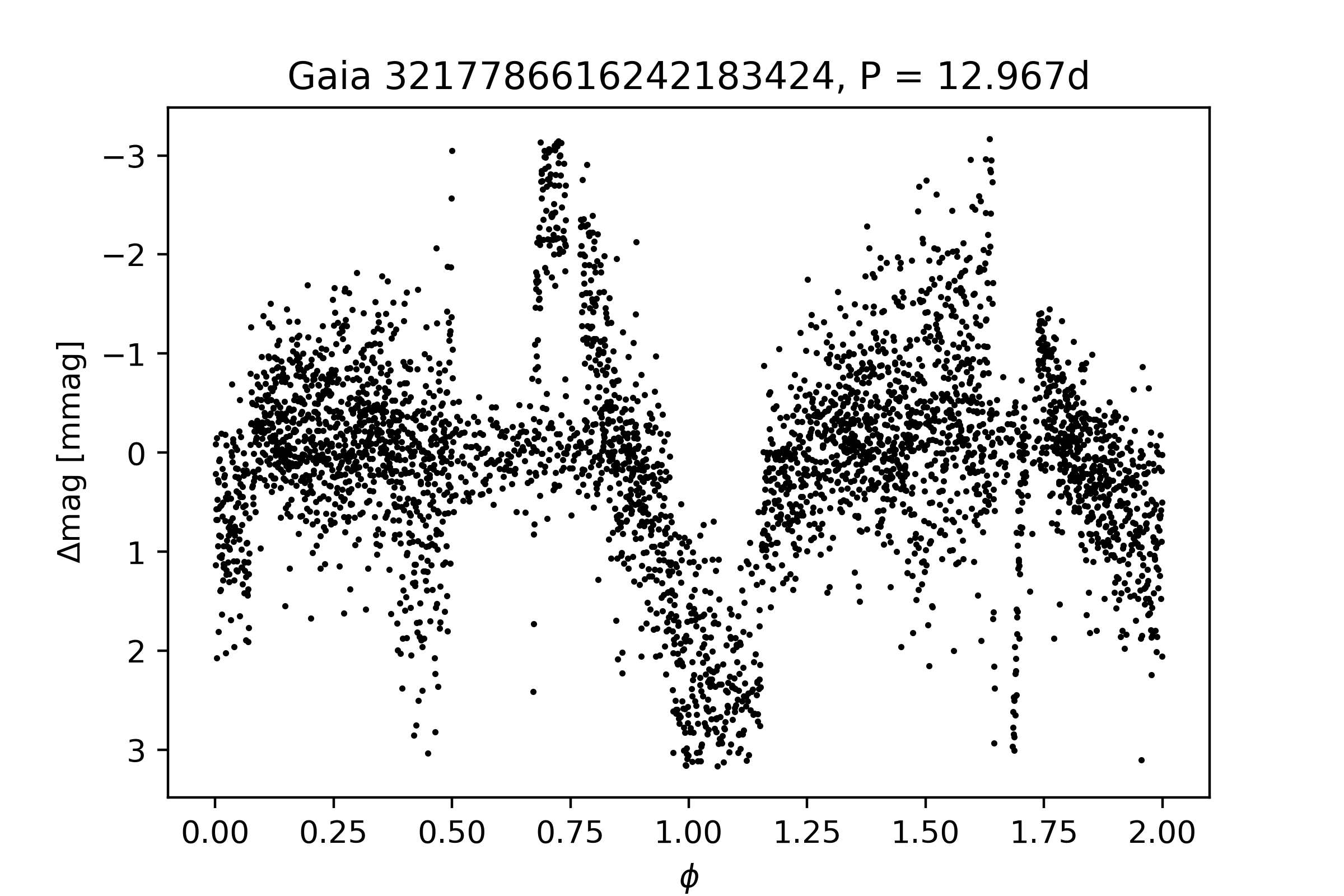}
    \caption{Irregular variable without real variability from the star itself. Likely instrumental features are visible.}
    \label{fig:enter-label}
\end{figure}

\begin{figure}[ht]
    \centering
    \includegraphics[width = \columnwidth]{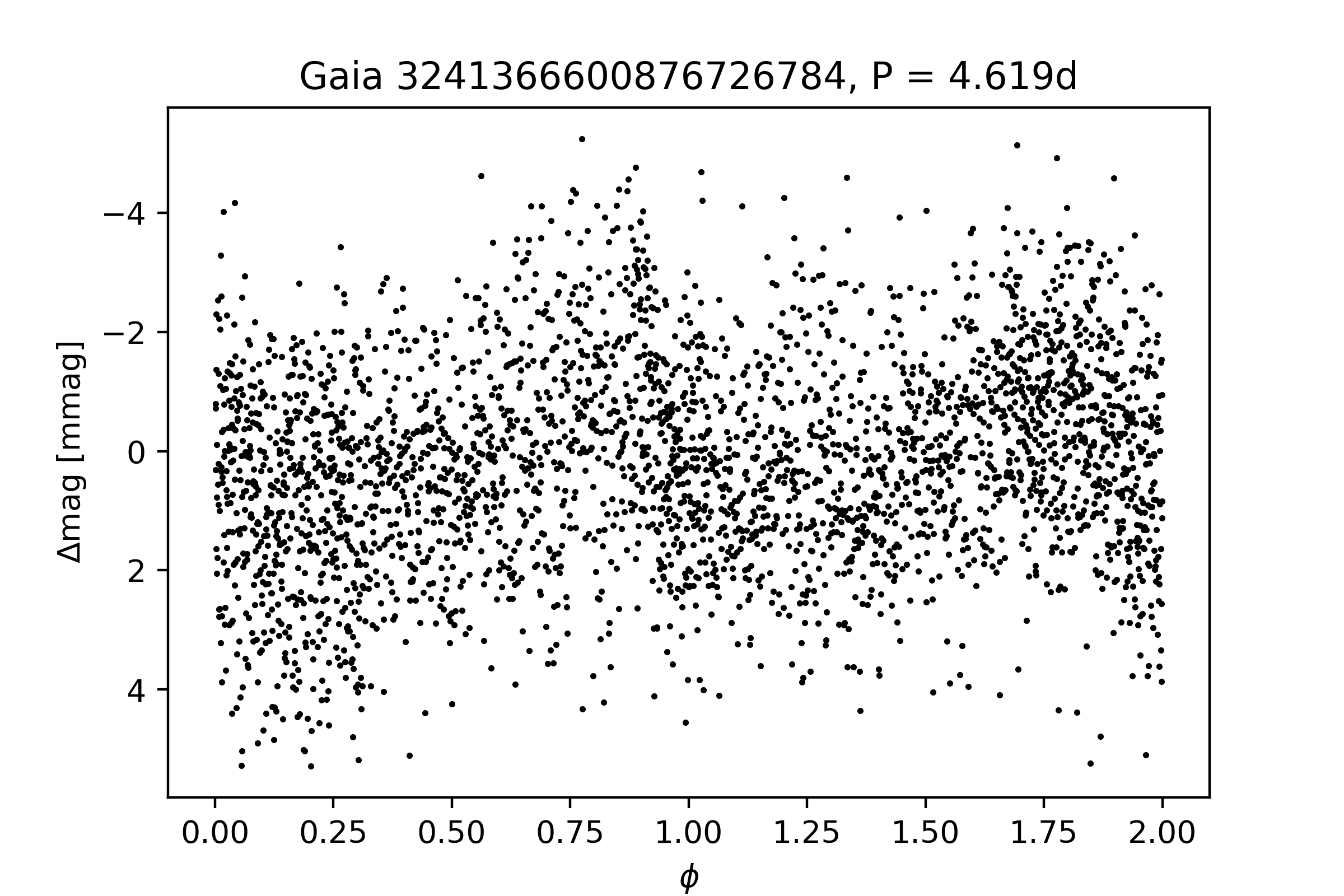}
    \caption{Maybe a rotating variable. The signal is not clear enough to conclude.}
    \label{fig:enter-label}
\end{figure}

\begin{figure}[ht]
    \centering
    \includegraphics[width = \columnwidth]{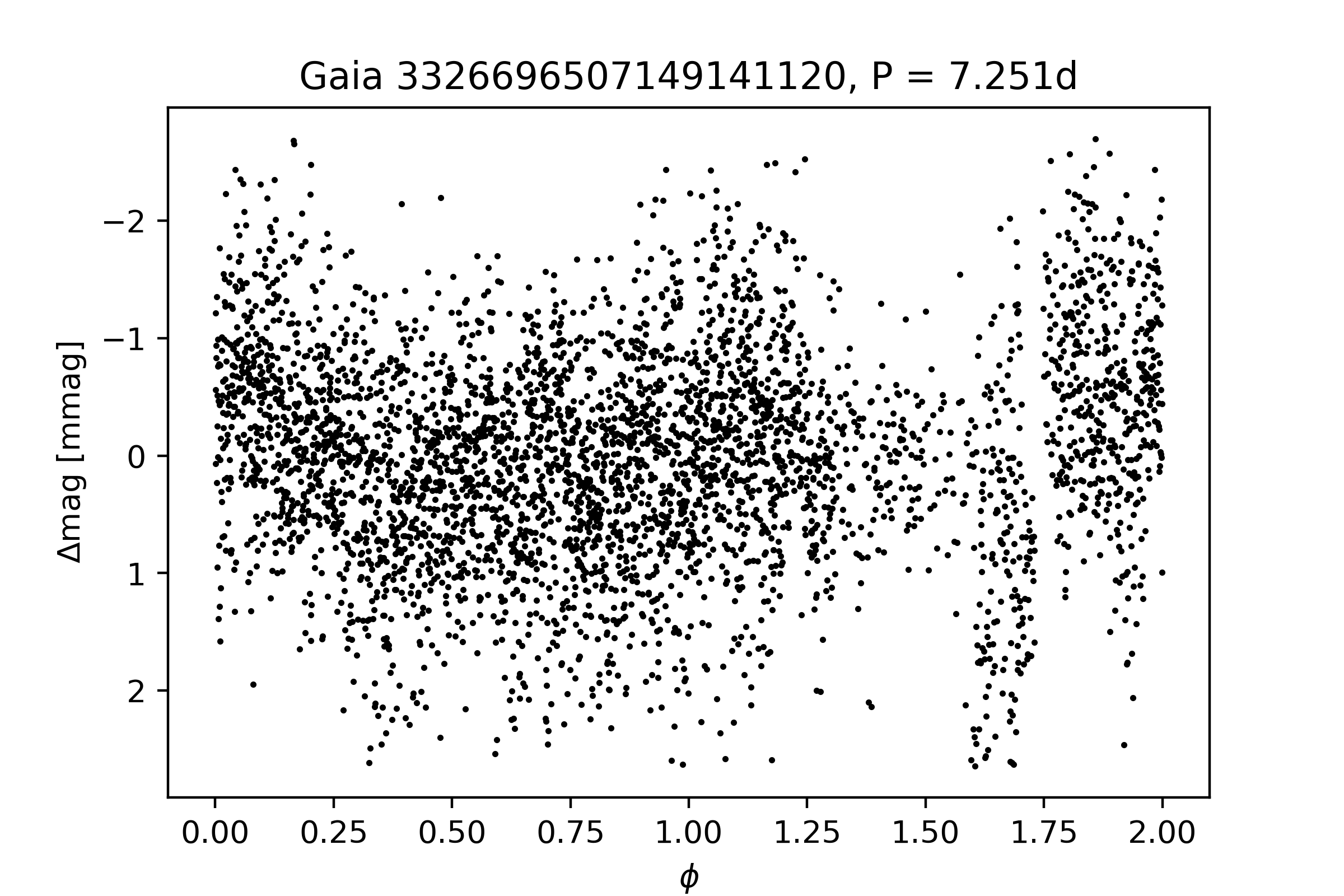}
    \caption{Irregular variability, likely low-amplitude instrumental variations.}
    \label{fig:enter-label}
\end{figure}

\begin{figure}[ht]
    \centering
    \includegraphics[width = \columnwidth]{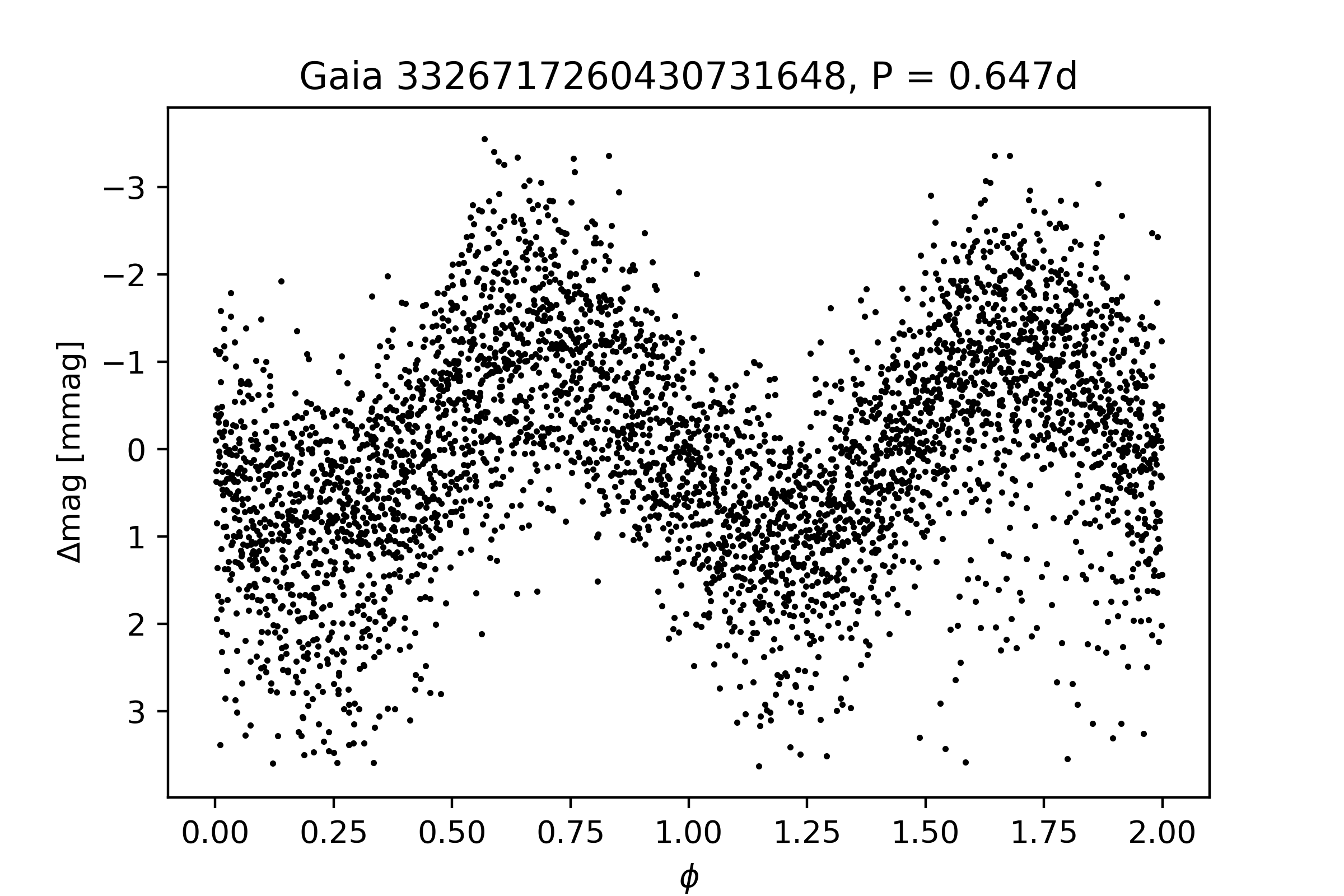}
    \caption{Another SXARI variable.}
    \label{fig:enter-label}
\end{figure}

\begin{figure}[ht]
    \centering
    \includegraphics[width = \columnwidth]{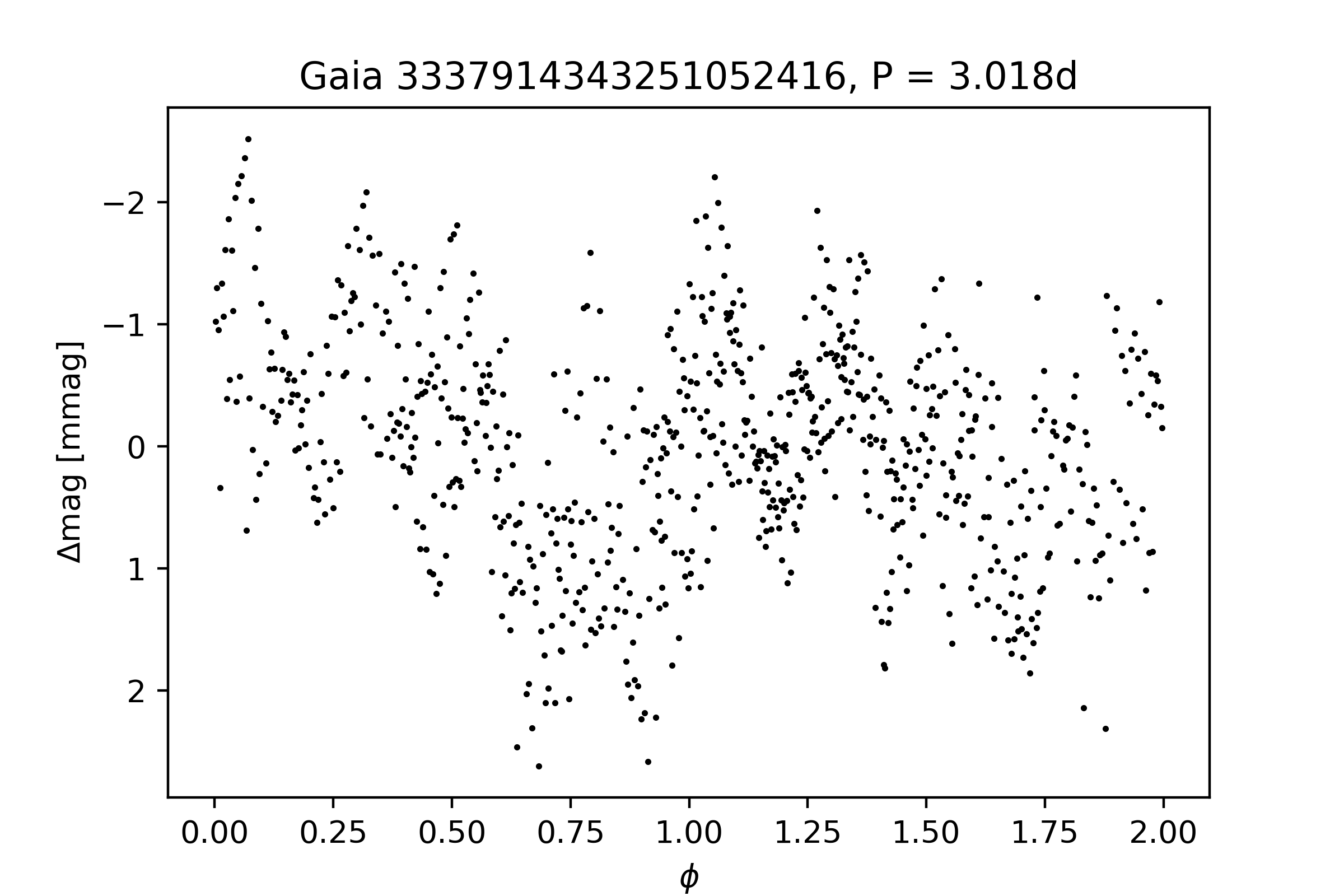}
    \caption{Multiperiodic or semiregular variable with a possible rotation feature (P = 3.018d) and some superimposed pulsational properties.}
    \label{fig:enter-label}
\end{figure}

\begin{figure}[ht]
    \centering
    \includegraphics[width = \columnwidth]{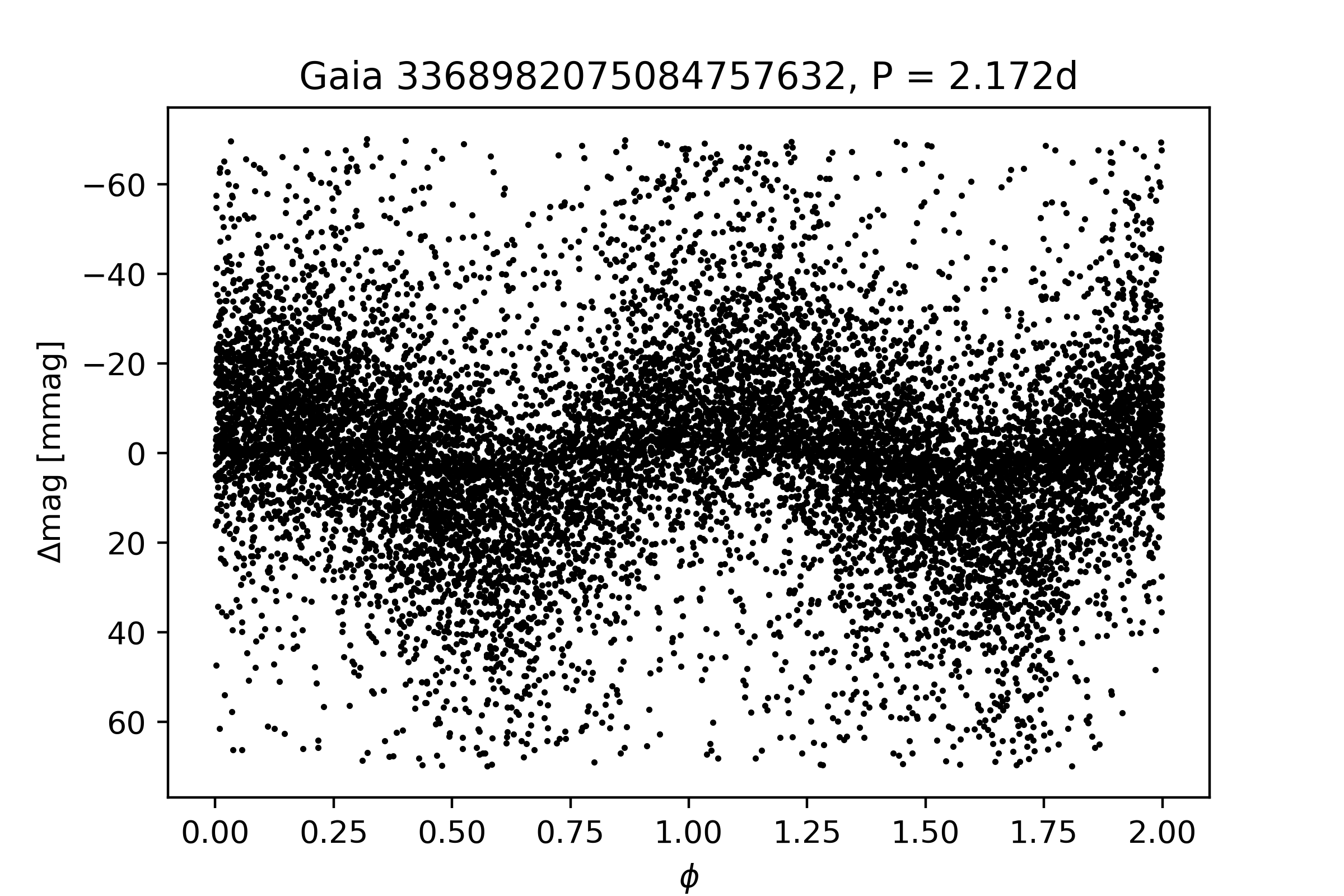}
    \caption{A sinusoidal rotation feature. The amplitude variation comes from observations in different sectors.}
    \label{fig:enter-label}
\end{figure}

\begin{figure}[ht]
    \centering
    \includegraphics[width = \columnwidth]{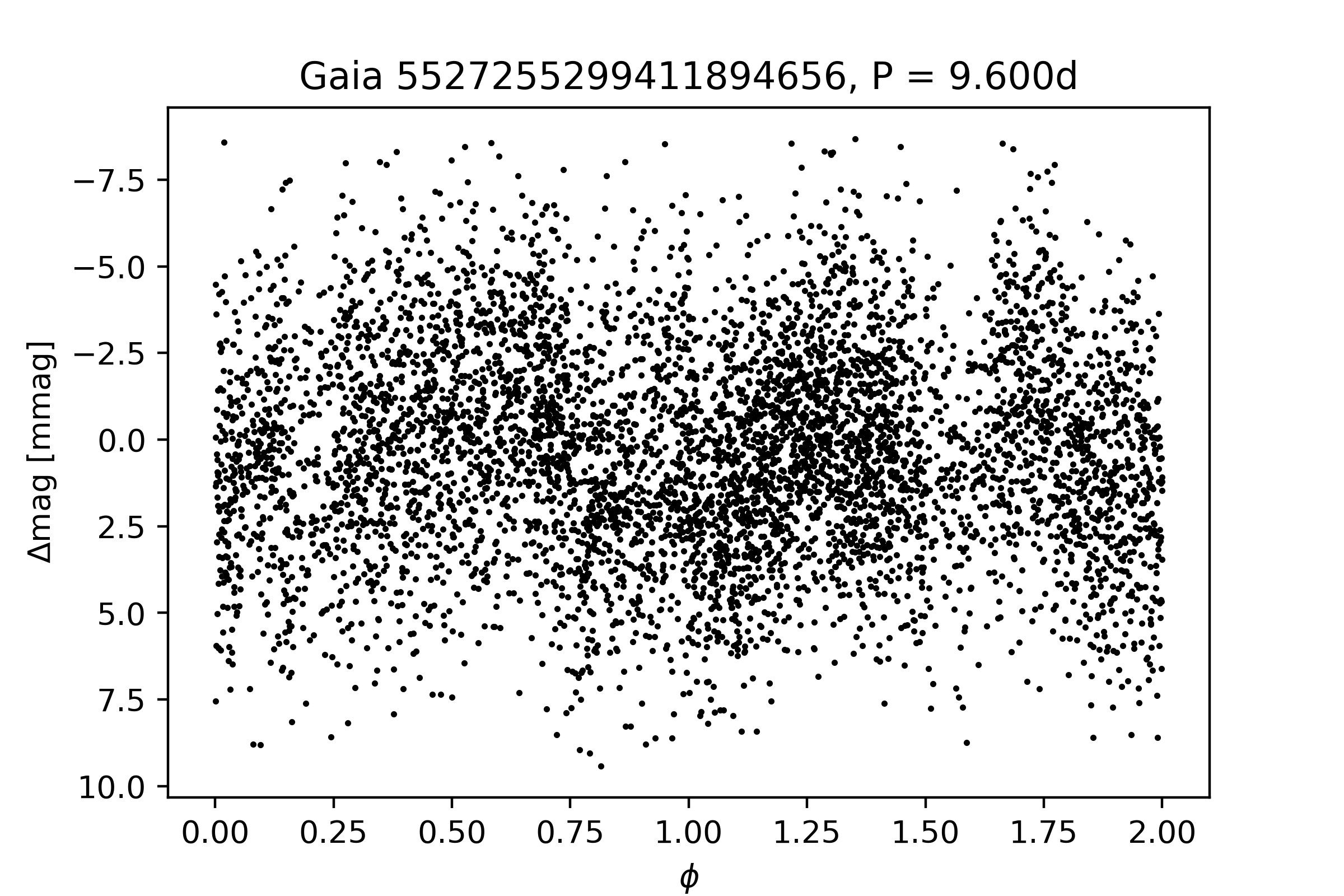}
    \caption{Possibly a rotating star. }
    \label{fig:enter-label}
\end{figure}

\begin{figure}[ht]
    \centering
    \includegraphics[width = \columnwidth]{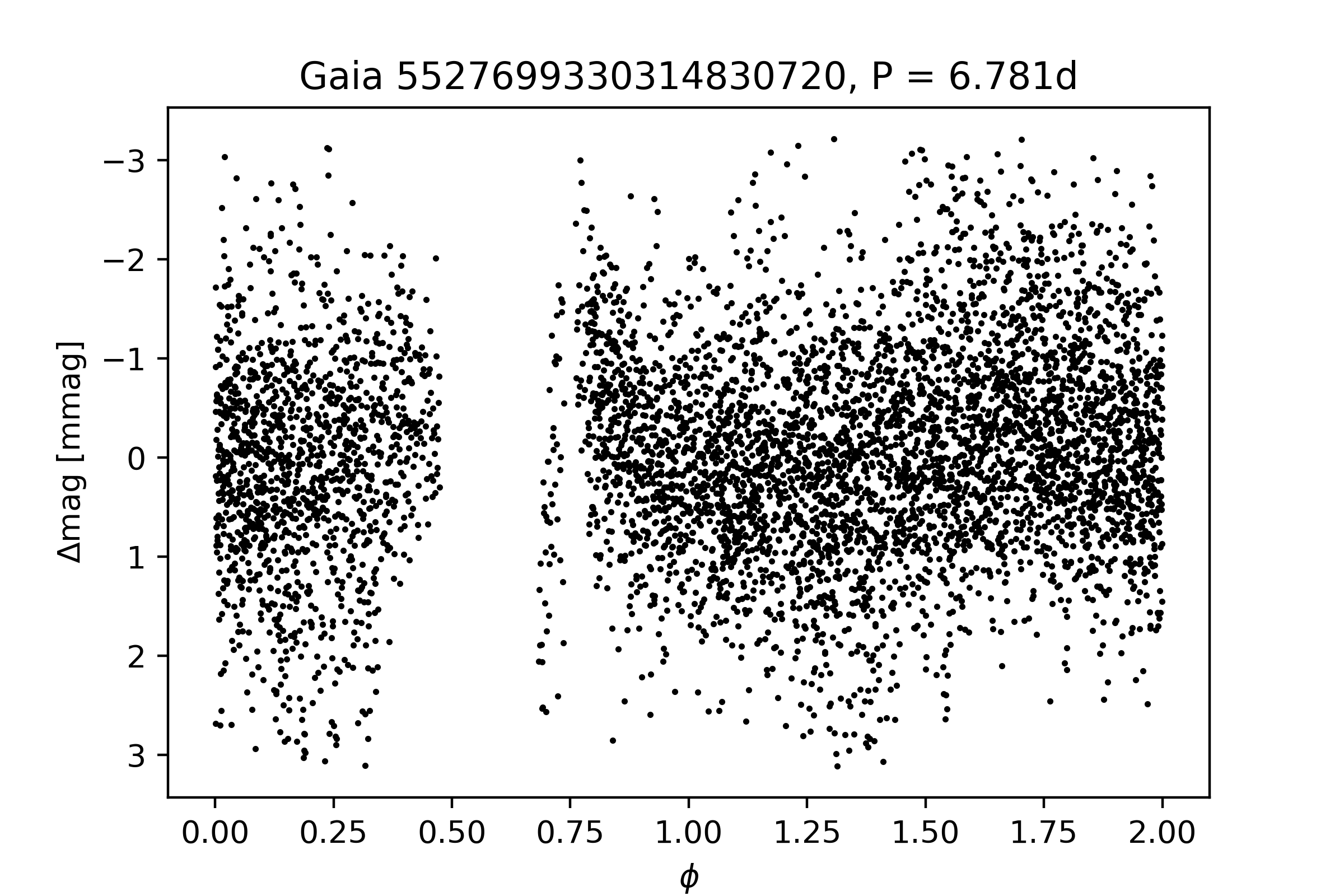}
    \caption{Source with mostly instrumental artefacts that cause a slight variation.}
    \label{fig:enter-label}
\end{figure}

\begin{figure}[ht]
    \centering
    \includegraphics[width = \columnwidth]{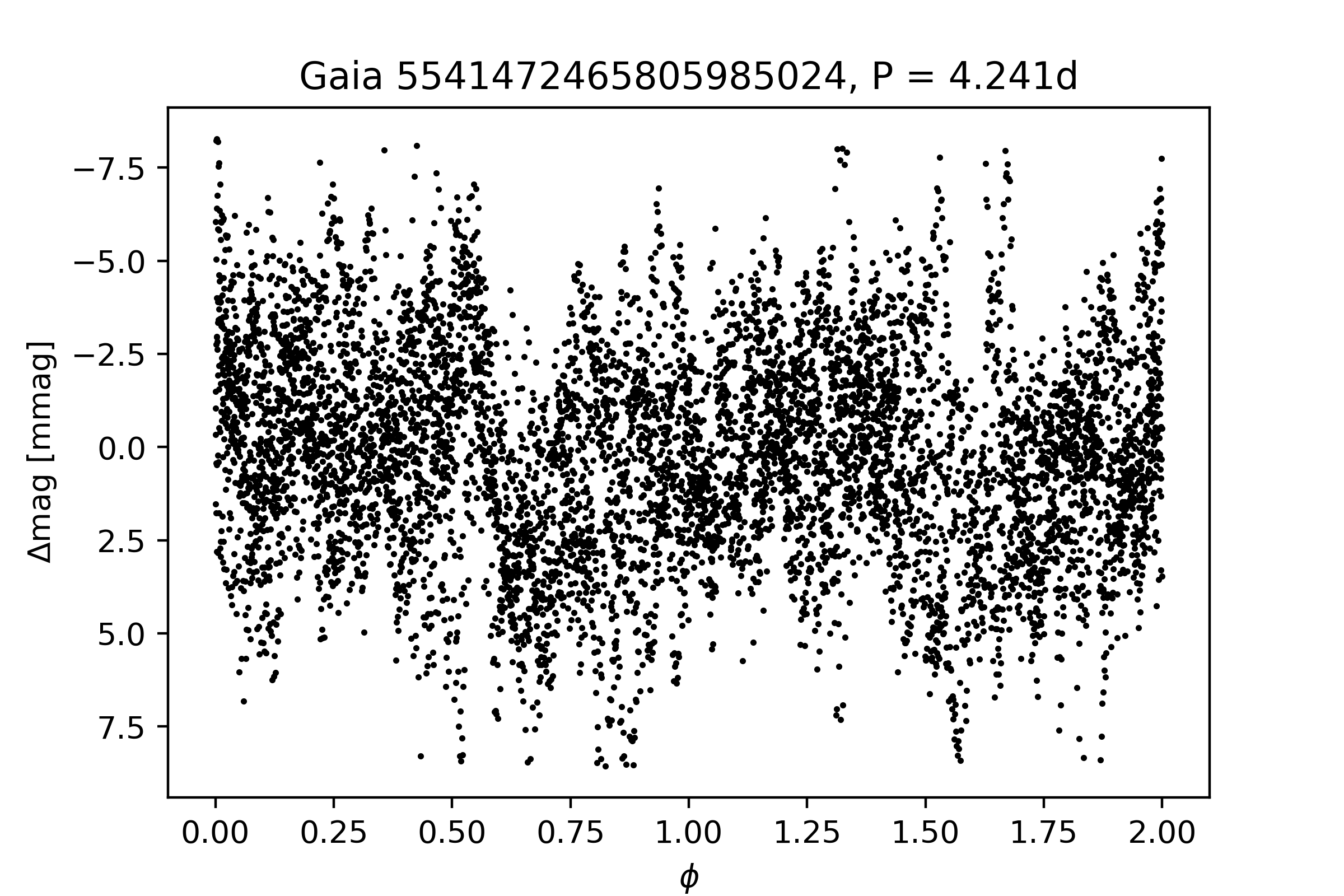}
    \caption{Probable rotation superimposed by pulsations, probably a ROT+GDOR hybrid.}
    \label{fig:rot_gdor}
\end{figure}

\begin{figure}[ht]
    \centering
    \includegraphics[width = \columnwidth]{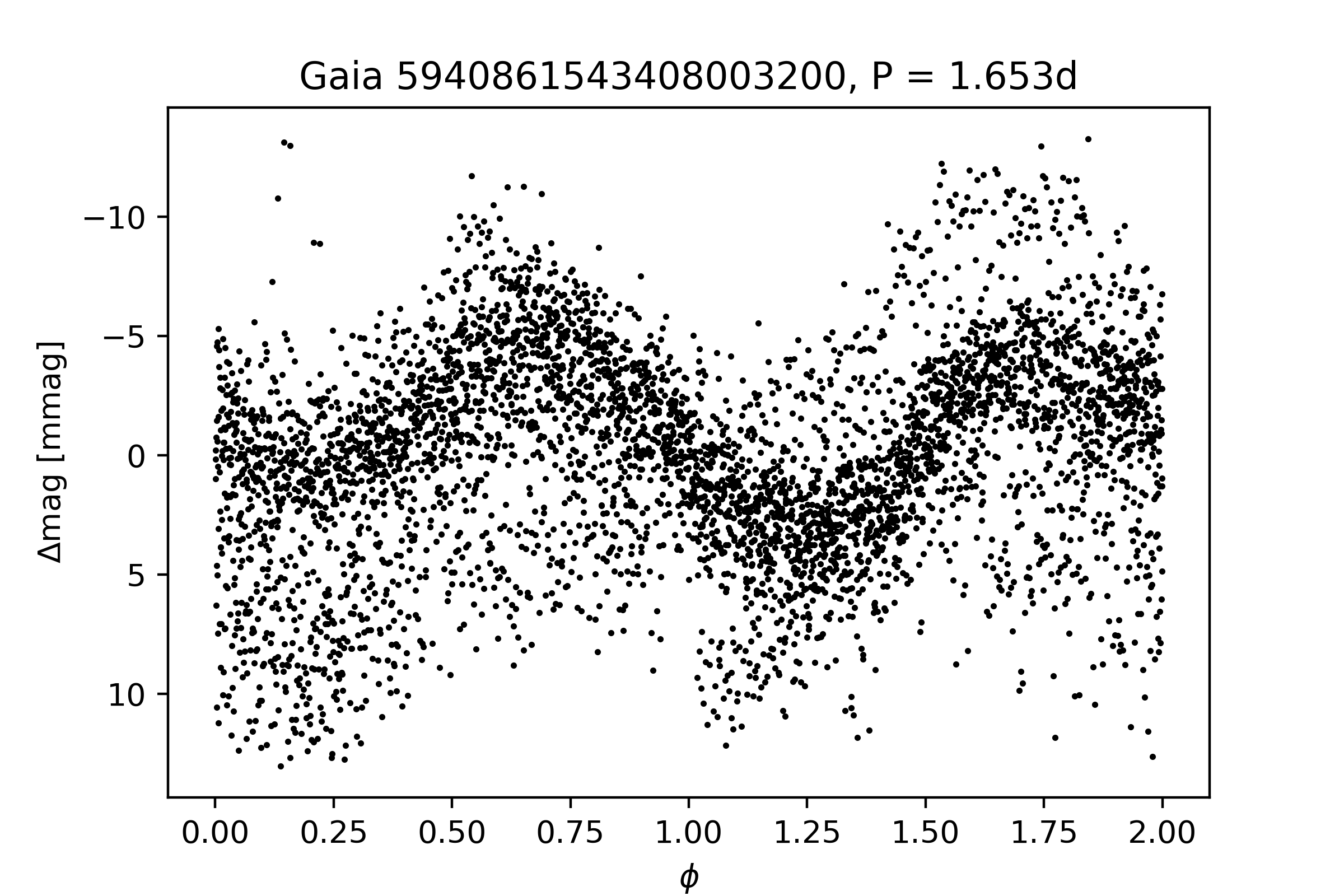}
    \caption{Rotating star with amplitude variations possibly due to chemical spots on the surface. Thus possibly an SXARI variable.}
    \label{fig:enter-label}
\end{figure}

\begin{figure}[ht]
    \centering
    \includegraphics[width = \columnwidth]{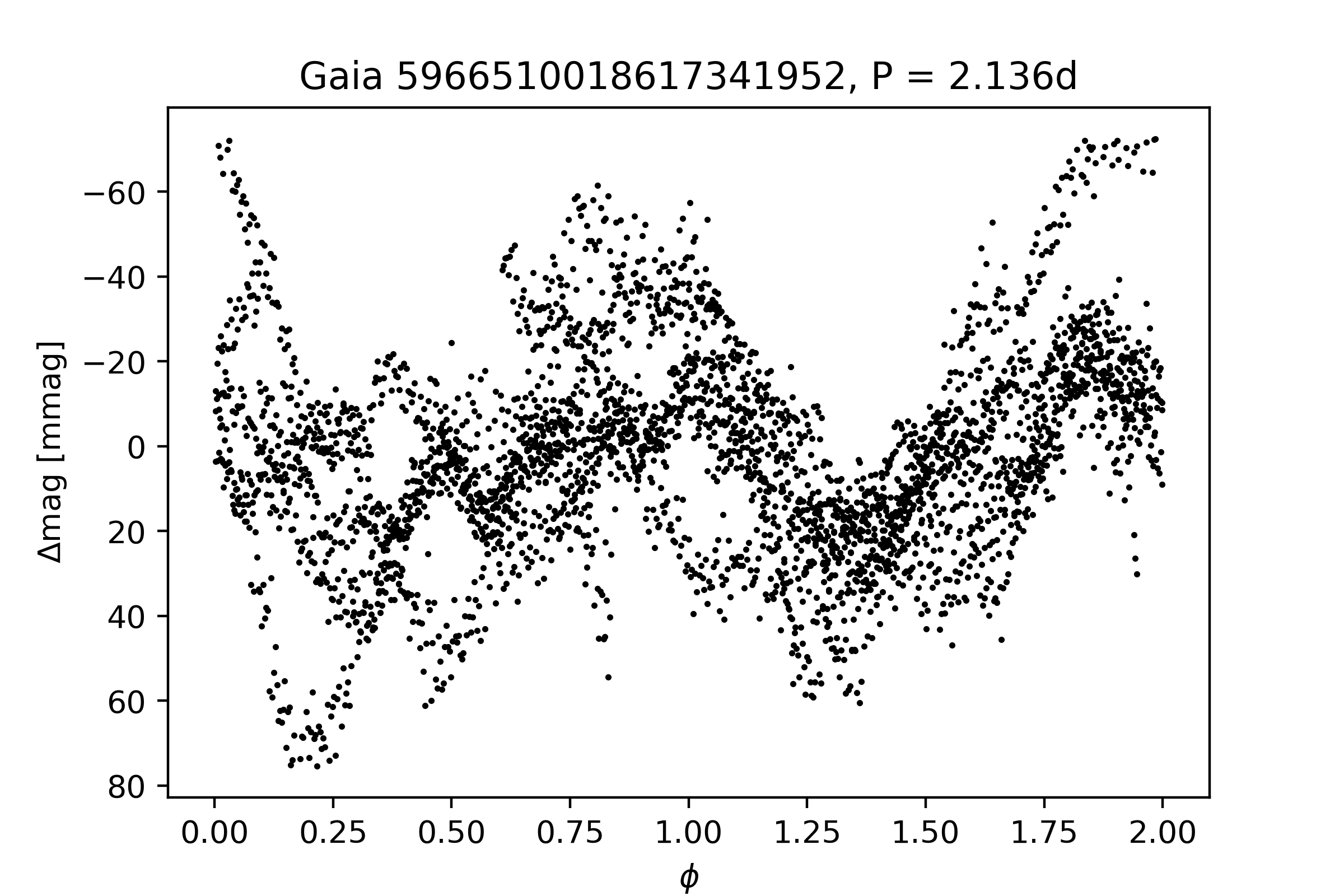}
    \caption{Multiperiodic variable with a basic rotation signal at a period of 2.136d.}
    \label{fig:enter-label}
\end{figure}

\begin{figure}[ht]
    \centering
    \includegraphics[width = \columnwidth]{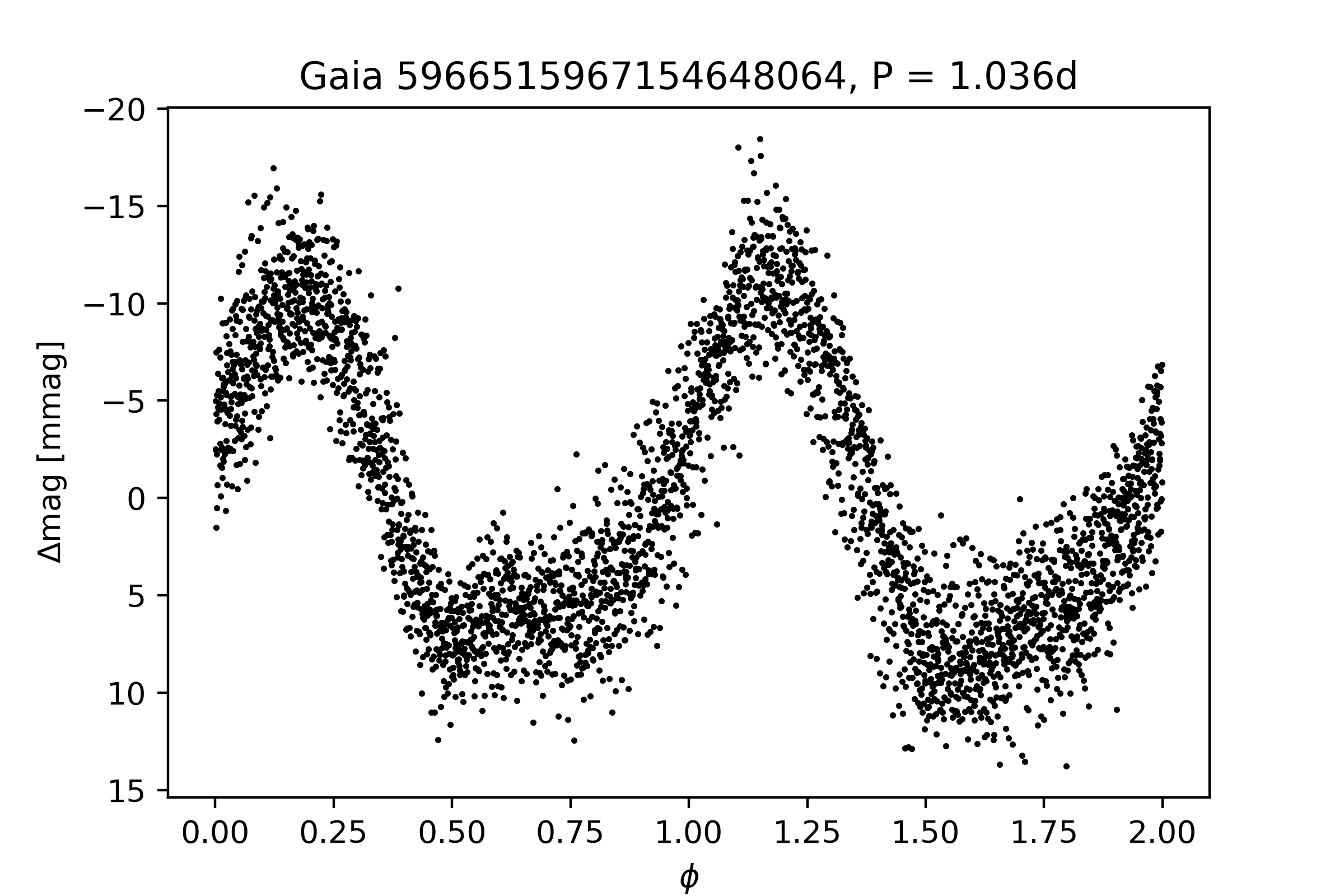}
    \caption{ACV variable previously classified as an SPB.}
    \label{fig:pms_cp_acv}
\end{figure}

\begin{figure}[ht]
    \centering
    \includegraphics[width = \columnwidth]{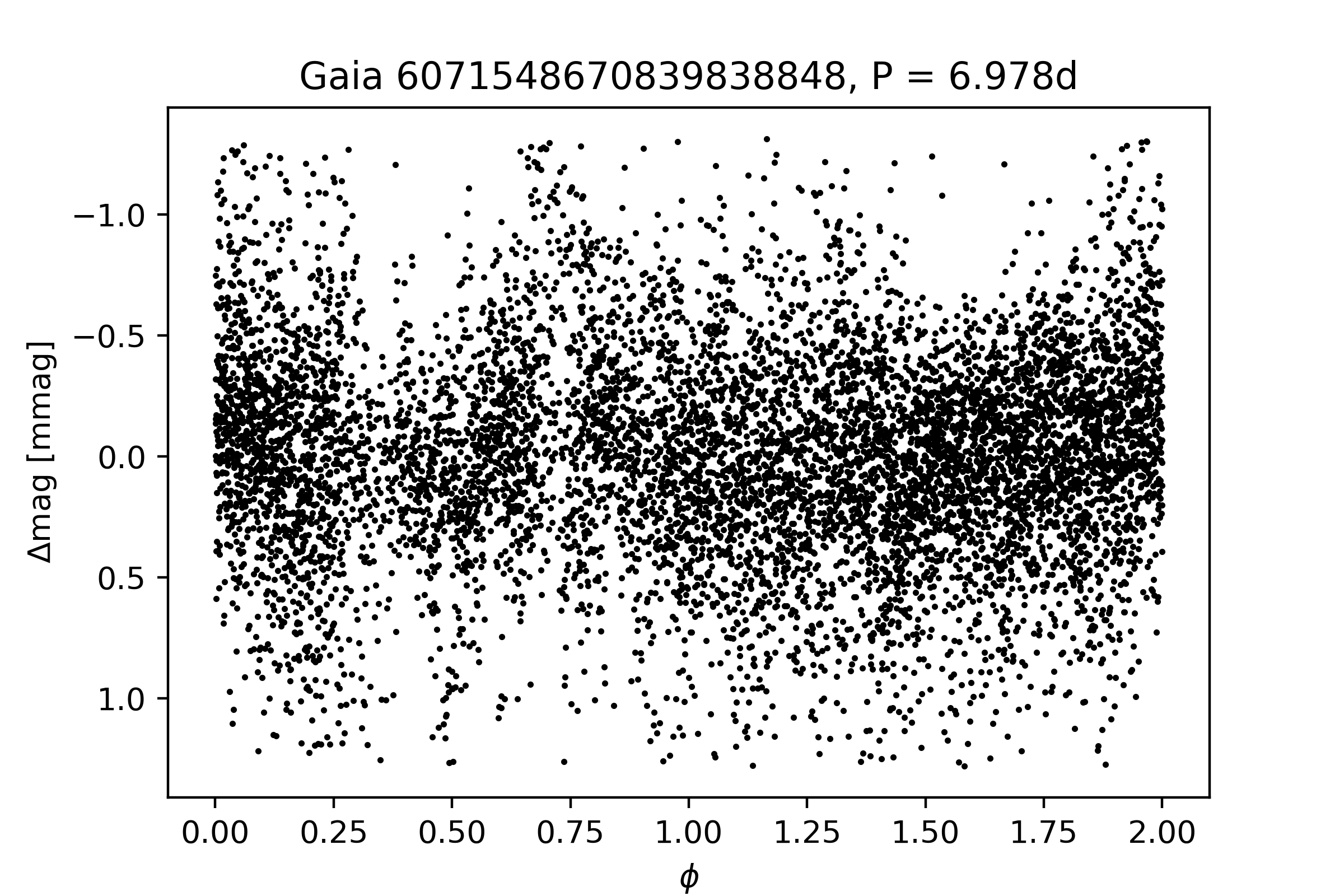}
    \caption{Star without visible variability from the light curve we obtained.}
    \label{fig:enter-label}
\end{figure}

\clearpage

\begin{table*}[]
 \caption{Variability of our sample stars based on the TESS light curves. Entries with a dash mean that no light curve was available via our method.}
     \centering
     \begin{tabular}{c|c|c|c|c|c}
    
     \textbf{Gaia DR3} & \textbf{Type our} & \textbf{P our} & \textbf{P lit.} & \textbf{Reference} & \textbf{Remarks} \\
    \hline
    \hline
    121406905707934464 & VAR & - & 123.3 & (1) &  \\
    216648703146774016 & GDOR + DSCT & 1.509 & 1.52 & (2) &  \\
    218807078833972224 & SXARI & 0.948 & 0.9 & (2) &  \\
    219692151335240576 & VAR & - & - &  &  \\
    476268957813243136 & ROT/ACV & 3.337 & 3.309 & (2) &  \\
    3241266270439474048 & - & - & - &  &  \\
    3241366600876726784 & ROT & 4.619 & - &  &  \\
    3017188622495555328 & - & - & - &  &  \\
    3337914343251052416 & Multiperiodic & - & - &  &  \\
    3217786616242183424 & VAR & - & - &  &  \\
    3216485103712446464 & ACV & 5.021 & - &  &  \\
    3019972890876467968 & ACV & 0.994 & - &  &  \\
    3345191014284099200 & - & - & - &  &  \\
    3368982075084757632 & ROT & 2.172 & 2.172 & (4) &  \\
    3104244792087711360 & ACV & 1.177 & 1.77 & (5) &  \\
    3131335689146178432 & E? & 2.944 & - &  &  \\
    3131891973309856640 & ELL & 0.735 &  &  & possible blend with V649 Mon \\
    3326717260430731648 & SXARI & 0.647 & - &  &  \\
    3326696507149141120 & VAR & - & - &  &  \\
    3046026785270786304 & SPB & 0.727 & - &  &  \\
    5541472465805985024 & ROT+GDOR & 4.241 & - &  &  \\
    5527255299411894656 & VAR & - & - &  &  \\
    5527699330314830720 & VAR & - & - &  &  \\
    6071548670839838848 & - & - & - &  &  \\
    6236109243250504576 & - & - & - &  &  \\
    6244725050721030528 & - & - &  &  &  \\
    5943020743750226048 & - & - & - &  &  \\
    5943020022195591552 & - & - & - &  &  \\
    6017520318720281216 & - & - & - &  &  \\
    5940861543408003200 & SXARI & 1.653 & - &  &  \\
    5966515967154648064 & ACV & 1.036 & 1.04 & (6) &  \\
    5966510018617341952 & Irregular & - & - &  &  \\
    5966508030054972288 & - & - & - &  &  \\
    4065975277785683840 & - & - & - &  &  \\
    4273851209555523328 & - &  &  &  &  \\
    2058937850640472064 & ROT/ACV & 0.522 & 0.521 & (2) &  \\
    2062618225278210560 & ACV & 1.119 & 1.119 & (2) &  \\
    2062359084127740032 & SPB & 1.587 & - &  &  \\
    2244529022468154880 & ACV & 0.965 & - &  &  \\
    2244529022468155008 & SXARI & 0.969 & - &  &  \\
    2203434019472097664 & - & - & - &  &  \\
    2204463918269731072 & ROT & 0.403 & - &  &  \\
    2204409870400823936 & SPB & 2.734 & - &  &  \\
    2279650069555318784 & ACV & 2.036 & 2.038 & (7) &  \\
    2012040312433901056 & E? & 2.197 & - &  &  \\
 
     \end{tabular}
     \tablebib{
     (1)~\cite{2023MNRAS.520.1296P}, (2)~\cite{2018AJ....155...39O},
     (3)~\cite{2017MNRAS.468.2745N}, (4)~\cite{2020ApJS..249...18C},
     (5)~\cite{1998A&AS..127..421C}, (6)~\cite{2013A&A...559A.108M},
     (7)~\cite{2010ApJS..188..473N}
     }
     \label{tab:tess_vars}
 \end{table*}

\clearpage

\section{Spectral types from the literature}

\begin{table*}[h!]
    \centering
    \begin{tabular}{c|c|c|c|c}
    
    \textbf{Gaia} DR3 & \textbf{Name} & \textbf{SpT Literature} & \textbf{Reference} & \textbf{Remarks} \\
    \hline
    \hline
    121406905707934464 & SHI261 & B8 Si He-wk & (1) &   \\
    216648703146774016 & R720 & A0 Si Sr & (2) &   \\
    218807078833972224 & R762 & B9 Si Sr Cr & (2) &   \\
    219692151335240576 & R770 & B8 He-wk & (2) &   \\
    476268957813243136 & R813 & B8  & (2) &  \\
    3241266270439474048 & R1121 & A0 Si & (2) &   \\
    3241366600876726784 & Q9955 & kA3hA3mA7 & (11) &   \\
    3017188622495555328 & R1398 & B9 He-wk & (2) &   \\
    3337914343251052416 & C121 & - & - &   \\
    3217786616242183424 & R1429 & Am & (3) &   \\
    3216485103712446464 & R1483 & B6 He-wk & (3) &   \\
    3019972890876467968 & SHA20504 & B6 IV Si & (4) &   \\
    3345191014284099200 & R1929 & A0 Si & (2) &   \\
    3368982075084757632 & Z583 & - & - &   \\
    3104244792087711360 & R2073 & B9 Si & (2) &   \\
    3131335689146178432 & R2110 & - & - &   \\
    3131891973309856640 & SHA19048 & B5 IV & (5) &   \\
    3326717260430731648 & R2175 & B6 He Var. & (2) &   \\
    3326696507149141120 & R2177 & B7 & (2) &   \\
    3046026785270786304 & R2426 & B2 & (2) &   \\
    5541472465805985024 & R2968 & B0 He & (2) & He-str. (6) \\
    5527255299411894656 & R3038 & A & (2) &   \\
    5527699330314830720 & R3051 & A2 Cr Eu & (2) &   \\
    6071548670839838848 & R4397 & B8 Cr & (7) &   \\
    6236109243250504576 & R5465 & B6 He-wk & (2) &   \\
    6244725050721030528 & R5585 & B9 Si Cr Sr & (2) &   \\
    5943020743750226048 & R5661 & B5 He-r & (2) &   \\
    5943020022195591552 & R5663 & B2 IV-V & (8) &   \\
    6017520318720281216 & R5675 & A0 Si & (2) &   \\
    5940861543408003200 & R5683 & B1 He-r & (9) &   \\
    5966515967154648064 & R5765 & A & (2) & ACV variable (this work)   \\
    5966510018617341952 & R5769 & B8 Si He-wk & (2) &   \\
    5966508030054972288 & R5779 & A & (2) &   \\
    4065975277785683840 & R6205 & B2 He & (2) &   \\
    4273851209555523328 & R6361 & A7 & (2) &   \\
    2058937850640472064 & Z888 & - & - &   \\
    2062618225278210560 & Z905 & - & - &   \\
    2062359084127740032 & R7156 & B8 Si & (2) &   \\
    2244529022468154880 & R7193 & A0 Si & (2) &   \\
    2244529022468155008 & R7192 & B9 Si & (2) &   \\
    2203434019472097664 & R7630 & B1 He & (2) &   \\
    2204463918269731072 & R7657 & A1-A7 & (2) &   \\
    2204409870400823936 & R7690 & B0 He & (2) &   \\
    2279650069555318784 & R7917 & A0 Si & (10) &   \\
    2012040312433901056 & R8060 & . & (2) &   \\
   
    \end{tabular}
    \caption{Spectral types of our candidates from the literature}
    \label{tab:spt_all_lit}

    \tablebib{
    (1)~\cite{2023MNRAS.520.1296P}, (2)~\cite{2009A&A...498..961R}, (3)~\cite{2013AstBu..68..300R}, (4)~\cite{2022ApJS..259...63S},
    (5)~\cite{2022ApJS..259...38Z}, (6)~\cite{2019MNRAS.487.5922G},
    (7)~\cite{2021MNRAS.504.3758P}, (8)~\cite{2020A&A...636A..74T},
    (9)~\cite{2021A&A...652A..31B}, (10)~\cite{2017AJ....153..218C},
    (11)~\cite{2019ApJS..242...13Q}
    }
\end{table*}

\end{appendix}

\end{document}